\documentclass[12pt]{article}
\usepackage{authblk}

\usepackage{lineno,hyperref}
\modulolinenumbers[5]

\usepackage{adjustbox}
\usepackage[utf8]{inputenc}   
\usepackage[T1]{fontenc}      
\usepackage{graphicx}
\usepackage{blindtext}
\usepackage{caption}%
\usepackage{makecell}%
\usepackage{subcaption}
\usepackage{float}
\usepackage{algorithm}
\usepackage{pm-isomath}
\usepackage{amsmath}
\usepackage{amssymb}
\usepackage{comment}%
\usepackage{siunitx}
\usepackage[version=4]{mhchem}
\DeclareSIUnit\angstrom{\text {Å}}
\usepackage{booktabs}
\usepackage{multicol}
\usepackage{subcaption}
\usepackage{multirow}
\usepackage{cleveref}
\usepackage[flushleft]{threeparttable}
\usepackage{tikz} 
\usetikzlibrary{shapes.geometric, arrows} 
\usepackage[section]{placeins} 
\newcommand{\software}[1]{{\fontfamily{qcr}\selectfont {#1}}} 

\usepackage{caption} 
\usepackage{siunitx}
\newcommand{\ntoantin}{\mbox{$n\to \bar{n}$}}

\newcommand{\babar}{\mbox{\slshape B\kern-0.1em{\smaller A}\kern-0.1em B\kern-0.1em{\smaller A\kern-0.2em R} }}

\newcommand{\nbarviasterile}{\mbox{$n \to n'\to\bar{n}$}}

\begin{document}

\vfill

\title{The HIBEAM Instrument at the European Spallation Source}

\author[1,2]{V.~Santoro}
\author[3]{D.~Milstead}
\author[4]{P.~Fierlinger}
\author[5,6,7]{W.~M.~Snow}
\author[8]{J.~Amaral}
\author[9,10]{J.~Barrow}
\author[1]{M.~Bartis}
\author[1]{P.~Bentley}
\author[4]{L.~Björk}
\author[11]{G.~Brooijmans}
\author[12]{L.~Broussard\thanks{Notice: This manuscript has been co-authored by UT-Battelle, LLC, under contract DE-AC05-00OR22725 with the US Department of Energy (DOE). The US government retains and the publisher, by accepting the article for publication, acknowledges that the US government retains a nonexclusive, paid-up, irrevocable, worldwide license to publish or reproduce the published form of this manuscript, or allow others to do so, for US government purposes. DOE will provide public access to these results of federally sponsored research in accordance with the DOE Public Access Plan (https://www.energy.gov/doe-public-access-plan).}}
\author[3]{A.~Burgman}
\author[13,14]{G.~Croci}
\author[1]{N.~de~la~Cour}
\author[1]{D.D.~Di~Julio}
\author[3]{K.~Dunne}
\author[15]{L.~Eklund}
\author[2]{H.~Eriksson}
\author[1]{M.~J.~Ferreira}
\author[1]{U.~Friman-Gayer}
\author[2]{P.~Golubev}
\author[14]{G.~Gorini}
\author[16]{G.P.~Guedes}
\author[2]{V.~Hehl}
\author[17]{A.~Heinz}
\author[2]{H.~Henriksson}
\author[1]{M.~Holl}
\author[18]{Y.~Kamyshkov}
\author[19]{E.~Kemp}
\author[1]{M.~Kickulies}
\author[20]{R.~Kolevatov}
\author[21]{A.~Kozela}
\author[15]{A.~Kupsc}
\author[17]{H.T.~Johansson}
\author[1]{B.~Jönsson}
\author[1]{W.~T.~Lee}
\author[3]{W.~Lejon}
\author[2]{G.~Luzin}
\author[15]{P.~Marciniewski}
\author[1]{J.I.~Marquez Damian}
\author[2,17]{B.~Meirose}
\author[1]{G.~Muhrer}
\author[13]{A.~Muraro}
\author[22]{A.~Nepomuceno}
\author[16]{T.~Nilsson}
\author[2]{A.~Oskarsson}
\author[23]{T.~Pałasz}
\author[2]{R.~Pasechnik}
\author[2]{L.B.~Persson}
\author[13]{O.~Putignano}
\author[21]{K.~Pysz}
\author[8]{T.~Quirino}
\author[2]{B.~Rataj}
\author[2]{A.~Ripa}
\author[15]{J.~Rogers}
\author[13,14]{F.~Scioscioli}
\author[3]{S.~Silverstein}
\author[24]{Y.~V.~Stadnik}
\author[25]{R.~Wagner}
\author[15]{M.~Wolke}
\author[26]{J.~Womersley}
\author[3]{S.C.~Yiu}
\author[1]{L.~Zanini}
\author[2]{L.~\AA strand}

\affil[1]{European Spallation Source ERIC, Partikelgatan 5, 22484 Lund, Sweden}
\affil[2]{Department of Physics, Lund University, Professorsgatan 1, 22363 Lund, Sweden}
\affil[3]{Department of Physics, Stockholm University, 106 91 Stockholm, Sweden}
\affil[4]{Physikdepartment, Technische Universität München, James-Franck-Str. 1, 85748 Garching, Germany}
\affil[5]{Department of Physics, Indiana University, 727 E. Third St., Bloomington, IN, USA, 47405}
\affil[6]{Indiana University Center for Exploration of Energy \& Matter, Bloomington, IN 47408, USA}
\affil[7]{Indiana University Quantum Science and Engineering Center, Bloomington, IN 47408, USA}
\affil[8]{Faculdade de Engenharia, Universidade do Estado do Rio de Janeiro, Rua São Francisco Xavier 524, 20550-900 Rio de Janeiro, RJ, Brazil}
\affil[9]{Massachusetts Institute of Technology, Department of Physics, Cambridge, MA 02139, USA}
\affil[10]{School of Physics and Astronomy, Tel Aviv University, Tel Aviv 69978, Israel}
\affil[11]{Department of Physics, Columbia University, New York, NY 10027, USA}
\affil[12]{Oak Ridge National Laboratory, Oak Ridge, TN 37831, USA}
\affil[13]{Istituto per la Scienza e Tecnologia dei Plasmi, Via Cozzi 53, 20125 Milan, Italy}
\affil[14]{Università degli Studi di Milano-Bicocca, Dipartimento di Fisica G. Occhialini, Piazza della Scienza 3, 20125 Milan, Italy}
\affil[15]{Department of Physics and Astronomy, Uppsala University, Uppsala, Sweden}
\affil[16]{Departamento de Física, Universidade Estadual Feira de Santana (UEFS), 44036-900, Feira de Santana, Bahia, Brazil}
\affil[17]{Department of Physics, Chalmers University of Technology, Gothenburg, Sweden}
\affil[18]{Department of Physics and Astronomy, University of Tennessee, Knoxville, TN 37996, USA}
\affil[19]{Instituto de Física Gleb Wataghin, Universidade Estadual de Campinas (UNICAMP), Av. Sérgio Buarque de Holanda 777, 13083-859 Campinas, SP, Brazil}
\affil[20]{European Spallation Source Consultant, Norway}
\affil[21]{H. Niewodniczański Institute of Nuclear Physics, Polish Academy of Sciences, Krakow, Poland}
\affil[22]{Departamento de Ciências da Natureza, Universidade Federal Fluminense, Rua Recife, 28890-000 Rio das Ostras, RJ, Brazil}
\affil[23]{M. Smoluchowski Institute of Physics, Jagiellonian University, Krakow, Poland}
\affil[24]{School of Physics, The University of Sydney, Sydney, New South Wales 2006, Australia}
\affil[25]{IRAMIS/Laboratoire Léon Brillouin, CEA-CNRS, Université Paris-Saclay, 91191 Gif-sur-Yvette, France}
\affil[26]{University of Edinburgh, Edinburgh, United Kingdom}

\date{April 2025}

\maketitle

\begin{abstract}
The European Spallation Source (ESS) will be the world's brightest neutron source and will open a new intensity frontier in particle physics. The HIBEAM collaboration aims to exploit the unique potential of the ESS with a dedicated ESS instrument for particle physics which offers world-leading capability in a number of areas. The HIBEAM program includes the first search in thirty years for free neutrons converting to antineutrons and searches for sterile neutrons, ultralight axion dark matter and nonzero neutron electric charge. This paper outlines the capabilities, design, infrastructure, and scientific potential of the HIBEAM program, including its dedicated beamline, neutron optical system, magnetic shielding and control, and detectors for neutrons and antineutrons. Additionally, we discuss the long-term scientific exploitation of HIBEAM, which may include measurements of the neutron electric dipole moment and precision studies of neutron decays.
\end{abstract}

\newpage
\tableofcontents
\section{List of acronyms}
{\small
\begin{tabular}{lr}
 \hline
{\bf Acronym/term} & {\bf Meaning}  \\
\toprule
ALP & Axion-like particle \\
BNV & Baryon number violation \\
CAD & Computer aided design \\ 
C.L. & Confidence level \\ 
COMSOL & A finite element analysis and simulation software package \\ 
CDR & Conceptual design report \\ 
DAQ & Data acquisition system \\
EDM & Electric dipole moment \\
ENDF & Evaluated Nuclear Data File \\ 
ESS & European Spallation Source \\
FOM & Figure of merit \\ 
GEANT4 & A Monte Carlo simulation program for GEometry ANd Tracking \\
GEM & Gas electron multiplier \\ 
HPV & Hadronic parity violation \\ 
HIBEAM & High Intensity Baryon Extraction And Measurement collaboration \\
INCL & The Li{\'e}ge Intranuclear Cascade model \\
ILL & Institut Laue Langevin \\ 
LCTPC & Linear Collider Time Projection Chamber Collaboration \\ 
LNV & Lepton number violation \\ 
MCPL & Monte Carlo Particle Lists \\ 
 MCStas &  Monte Carlo simulation of neutron instruments \\
 ML & Machine Learning \\
 NBOA    &  Neutron Beam Optics Assembly \\
 NES & Neutron Extraction System \\
 nEDM@SNS & An experiment to measure the neutron's electric dipole moment at the SNS \\ 
 NNBAR & An experiment to search for neutrons converting to antineutrons at the ESS \\
 ORNL & Oak Ridge National Laboratory  \\
PHITS & Particle and Heavy Ion Transport code System \\
 PMT & Photo-Multiplier Tube \\ 
SAMPA & A Multi-signal Application-specific Integrated Circuit \\ 
SiPM & Silicon photomultiplier \\ 
 SM & Standard Model \\
 SNS & Spallation Neutron Source at ORNL \\ 
 TPC & Time projection chamber \\
 UCN & Ultra-cold neutron \\
 WASA & Wide Angle Shower Apparatus \\ 
\bottomrule
\end{tabular}}
\newpage


\section{Introduction}
The European Spallation Source (ESS) will become the most powerful research facility worldwide for neutron-based studies once it reaches full completion~\cite{Garoby_2017}. With its exceptional capabilities, including a higher useful flux of neutrons compared to any existing research reactor and an unprecedented level of neutron brightness, the ESS surpasses any currently available neutron source and opens a new intensity frontier in particle physics. 

Although the ESS is presently constructing 15 instruments,  22 instruments are ultimately foreseen to fully achieve the scientific objectives set out in the ESS statutes~\cite{ESS-statutes}. In addition to neutron scattering studies for materials and life sciences, the ESS has a dedicated mandate for fundamental physics research. A prioritisation exercise identified the absence of a dedicated beamline for particle physics as being a missing scientific capability of the highest importance~\cite{ess-gap}. In this context, taking advantage of ESS's unique scientific potential, the High-Intensity Baryon Extraction and Measurement (HIBEAM) collaboration is designing a beamline and related infrastructure for a world-leading particle physics research program. 

The original focus of the HIBEAM project, as outlined in Refs.~\cite{Addazi:2020nlz,Abele_2023}, comprised searches for the violation of baryon number ($\mathcal{B}$) via  high-sensitivity searches for neutron conversions to antineutrons and/or sterile neutrons~\cite{zurab1}. This provides the first competitive search for free neutrons converting to anti-neutrons in over thirty years. The HIBEAM discovery potential/sensitivity exceeds that of the previous search at the Institut Laue Langevin (ILL)~\cite{Baldo-Ceolin:1994hzw} by around an order of magnitude\footnote{Searches have been performed with bound neutrons in large-mass detectors~\cite{Homestake,KGF,NUSEX,IMB,Kamiokande,Frejus,Soudan-2,Aharmim:2017jna,Abe:2011ky,Gustafson:2015qyo}. A recent analysis by Super-Kamiokande~\cite{Abe:2011ky,Gustafson:2015qyo,Super-Kamiokande:2020bov} has placed competitive limits on this process. However, in nuclei, the $n \rightarrow \bar{n}$ conversion rate is suppressed due to the energy difference between neutrons and antineutrons in the nuclear potential, which breaks degeneracy and affects the oscillation probability. This introduces a strong model dependence, making it challenging to directly compare limits obtained from free and bound neutron searches.
}. 

In addition to neutron oscillations, the presently designed HIBEAM instrument can be utilized for direct searches for other phenomena, such as ultralight axion-like particles (ALPs), with exceptionally high sensitivity~\cite{PhysRevLett.133.181001}. A similarly high-sensitive search for a nonzero neutron electric charge can be made. 

For the various searches, HIBEAM can deliver at least an order of magnitude improvement in sensitivity compared with previous work. The work addresses key open questions in modern physics, including baryogenesis and the nature of dark matter, and the possible falsification of the Standard Model (SM) beyond the neutrino sector. Unsurprisingly, the HIBEAM program corresponds to `essential scientific activities' in the 2020 Update to the European Particle Physics Strategy~\cite{EuropeanStrategyGroup:2020pow}.  

This paper focuses on the design of HIBEAM, which requires the largest possible neutron flux, as described above. However, modifications to this design—such as adjustments to the neutron guide—could enable an even broader program, including measurements of the neutron electric dipole moment (EDM), neutron beta decay, and parity violation searches. These possibilities are discussed in this paper.

Additionally, it is noteworthy that the flexible beamline design permits conducting measurements and searches with neutron guides different from the one considered here, even before the full program is completed with the currently planned guide.




This paper is organised as follows. A brief motivation for the HIBEAM program including relevant phenomenology is given in Section~\ref{sec:motivation}. An overview of the ESS is provided in Section~\ref{sec:ess}. This is followed by a description of the principles behind each of the searches within the HIBEAM program and the expected sensitivities in Section~\ref{sec:hibeamoverview}. A short discussion on possible activities beyond the program is also included in this Section. A detailed description of the HIBEAM beamline, including neutron extraction and focusing, magnetic control, biological shielding, and the radiation profile of the beamline is given in Section~\ref{sec:hibeambeamline}. The suite of detectors used at HIBEAM is then described in Section~\ref{sec:det}. Both Sections~\ref{sec:hibeambeamline} and~\ref{sec:det} also include brief descriptions of prototype work. A summary and discussion of future plans is given in Section~\ref{sec:summary}. 


\section{Theoretical background}\label{sec:motivation}
This section describes the motivation for the HIBEAM experimental program together with relevant phenomenology which guides the design of the HIBEAM beamline.  

\subsection{Neutron conversions to antineutrons and/or sterile neutrons}

The observation of neutron conversions to anti-neutrons~\cite{Phillips:2014fgb} and/or to sterile neutrons at HIBEAM would be of fundamental significance. The HIBEAM searches exploit unique channels for baryon number violation (BNV). Unlike single proton decay in which $\mathcal{B}$ and $\mathcal{L}$ must be simultaneously violated to conserve angular momentum, in these channels $\mathcal{B}$ is the only hitherto conserved quantity which is violated. Furthermore, owing to the experimental challenges involved, relatively few searches for free neutron-antineutron conversions $(\Delta \mathcal{B}=2, \Delta \mathcal{L}=0)$ have been performed, compared with, for example, the leptonic equivalent process of neutrinoless double beta decay~\cite{Dolinski_2019} $(\Delta \mathcal{B}=0, \Delta \mathcal{L}=2)$. 

While the need for `blue sky' exploration and high-sensitivity testing of conservation laws strongly motivates the HIBEAM program, there also exists a number of theoretical arguments, summarised below, outlining why BNV is to be expected. Furthermore, as discussed, neutron conversions can arise as a phenomenon addressing a specific problem, such as baryogenesis, as well as being coupled to other signals of new physics such as neutrinoless double beta decay. 

\begin{itemize}
\item The mechanism by which the observed matter-antimatter asymmetry in the Universe came about is not understood. However, it is known, as one of the so-called Sakharov conditions~\cite{Sakharov:1967dj}, that BNV is required for baryogenesis. Neutron conversions feature  in a number of scenarios of baryogenesis~\cite{Babu:2006xc,Mohapatra:1980qe,Berezhiani:2015uya,Dev:2015uca,Allahverdi:2017edd}.
\item The conservation of $\mathcal{B}$ corresponds, like lepton number ($\mathcal{L}$), to an accidental symmetry in the SM. 
For theories extending the SM,  BNV and lepton number violation (LNV) therefore tend to occur generically. 
\item Observable low-scale BNV via neutron conversions features in a number of extensions of the SM, such as scenarios of extra dimensions~\cite{Nussinov:2001rb}, branes \cite{Dvali:1999gf}, and supersymmetry~\cite{Barbier:2004ez,Dutta:2005af,Calibbi:2016ukt}. 

\item Sterile neutrons can belong to a dark/hidden sector of particles which interact gravitationally and not via the gauge forces of the SM~\cite{
Lanfranchi:2020crw,Petraki:2007gq,Berezhiani:2018eds,Hostert:2022ntu}. Such a sector can provide an explanation for Dark Matter (DM). As meta-stable and electrically uncharged objects which can be copiously produced and studied, neutrons offer an attractive portal to a dark sector.
\item There exists a symbiosis between neutron-antineutron conversions and other key observables for new physics sought experimentally. 
The conversion of a neutron to antineutron ($\Delta \mathcal{B}=2$, $\Delta \mathcal{L}=0$) is the baryonic equivalent of neutrinoless double beta decay ($\Delta \mathcal{B}=0$, $\Delta \mathcal{L}=2$). Both processes feature in unification scenarios and theories of neutrino masses~\cite{Mohapatra:2009wp,Dev:2015uca,Allahverdi:2017edd,Berezhiani:2015afa}. Furthermore, neutron-antineutron conversion, neutrinoless double beta decay, and single proton decay are theoretically linked via the electroweak sphaleron interaction~\cite{Mohapatra:2014yla,PhysRevD.79.015017}, which is a fundamental non-perturbative feature of the Standard Model. The sphaleron process can be written as: \begin{equation} Q Q Q Q Q Q ~ Q Q Q L ~ L L \end{equation} as shown in~\cite{Mohapatra_2009}, where the first term (the six-quark operator) corresponds to baryon number violation ($\Delta \mathcal{B} = 2$) as in neutron-antineutron oscillations, the second term (the four-fermion operator) corresponds to proton decay ($\Delta \mathcal{B} = 1, \Delta \mathcal{L} = 1$), and the last term represents lepton number violation (LNV) ($\Delta \mathcal{L} = 2$), implying a low-energy process such as neutrinoless double beta decay. If two of these processes were observed experimentally, then the existence of the third would be strongly suggested within many unification scenarios and extensions of the Standard Model~\cite{Mohapatra:2014yla}.
\end{itemize}

\subsubsection{Phenomenology of neutron conversions and search principles}
HIBEAM is a unique facility that can probe the full range of neutron mixing possibilities between neutrons, antineutrons and sterile states ($n'$)~\cite{Phillips:2014fgb,zurab4}. 
The Hamiltonian for neutrons in a $B$-field is given in  Equation~\ref{eq:nbarmatrix}. 

\begin{equation}
    \label{eq:nbarmatrix}
    \hat{\mathcal{H}}=\left(\begin{array}{cccc}
    m_n + \vec{\mu}_n \vec{B} & \varepsilon_{n\bar{n}} & \alpha_{nn'} & \alpha_{n\bar{n}'}  \\
    \varepsilon_{n\bar{n}} & m_n -\vec{\mu}_n \vec{B} & \alpha_{n\bar{n}'}  & \alpha_{nn'}  \\
    \alpha_{nn'} & \alpha_{n\bar{n}'}  & m_{n'} +\vec{\mu}_{n'} \vec{B}' & \varepsilon_{n\bar{n}}  \\
    \alpha_{n\bar{n}'}  & \alpha_{nn'} & \varepsilon_{n\bar{n}} & m_{n'} - \vec{\mu}_{n'} \vec{B}'
    \end{array}\right)
\end{equation}
The terms $\varepsilon_{n\bar{n}}$ , $\alpha_{nn'}$  and $\alpha_{n\bar{n}'}$
are  the $n \bar{n}$ Majorana 
mass mixing parameter, and mass mixing parameters for $n n{'}$ and for $n \bar{n}{'}$, respectively\footnote{A more general expression for the Hamiltonian allows for the possibility of a transition magnetic moment (TMM)~\cite{Berezhiani:2018qqw}. This is not included here though a search for a TMM can be performed with HIBEAM.}. In a minimal approach, the neutron mass ($m_n$) and magnetic moment ($\mu_n$) are assumed to the same as that in the sterile sector. The magnetic moments for particle and anti-particle are opposite for both the visible and sterile sectors. Magnetic fields, $B$ and $B'$ exist in the visible and sterile sectors, respectively.  

Many different processes are possible.  
Equation~\ref{eq:free} gives the process probabilities as a function of propagation time $t$, in the quasi-free regime,    
for free neutrons to antineutrons ($P_{n\bar{n}}$). Probabilities are also given for the following sterile neutron processes:\footnote{These processes can involve sterile neutrons and sterile antineutrons. For simplicity, only the sterile neutron mode is written here. 
Physical limits and sensitivities are not dependent on this simplification. In the event of a discovery, the ability to perform a range of searches, including neutron-antineutron transitions via sterile states,  helps pin down the various contributions from sterile neutrons and sterile antineutrons.} 
neutron-antineutron transformation ($P_{n\rightarrow  n' \rightarrow \bar{n}}$), neutron regeneration ($P_{n\rightarrow  n' \rightarrow n}$) and neutron disappearance  ($P_{n\rightarrow  n'}$).
The characteristic oscillation times for each process ($\tau_{n\bar n}, \tau_{n\rightarrow n'}, \tau_{n\rightarrow  n'}, \tau_{\bar{n}\rightarrow  n'}$) are the reciprocal of each of the amplitudes. The terms $t_1$ and $t_2$ denote the propagation times before and after conversion. 

\begin{equation}
\label{eq:free}
 P_{n\bar{n}}(t)
 =\frac{t^2}{\tau^2_{n\bar n}}  ;
 \hspace{0.25cm}
P_{n\rightarrow  n' \rightarrow \bar{n}}=\frac{t_1^2\cdot t_2^2}{\tau_{n\rightarrow n'}^2\tau_{\bar{n} \rightarrow {n}'}^2}; \hspace{0.25cm}  
 P_{n\rightarrow  n' \rightarrow n}=\frac{t_1^2\cdot t_2^2}{\tau_{n\rightarrow  n'}^4};  \hspace{0.25cm} P_{n\rightarrow  n'}=\frac{t^2}{\tau_{n\rightarrow  n'}^2}   
 \end{equation}

There are several notable features of neutron conversion phenomenology that guide experimental searches.

\begin{itemize}
    \item The probability of observing a process in the quasi-free limit is enhanced by a power of increasing propagation time, motivating a long beamline and a sample of slow neutrons. The discovery potential of a future experiment is often best quantified by an expression involving the propagation time raised to a power rather than the propagation time.     
    \item Outside of the quasi-free limit,  the above probabilities become massively suppressed. This requires very tight control of the magnetic field experienced during propagation. To achieve the quasi-free condition in a beam experiment for neutron-antineutron conversions requires $B<5-10$~nT~\cite{Davis:2016uyk}. For transitions to sterile neutron states, the magnetic fields seen by the visible ($B$) and sterile states ($B'$) combine in such a way that the transition probability has a resonant-like behaviour as the visible field is varied~\cite{zurab2}.  
    \item A sterile magnetic field  ${B'}$ can be generated in a number of ways. For example, by hypothetical ionization and flow of gravitationally captured dark material in and around the Earth \cite{zurab2}. Such an accumulation could occur due to ionized gas clouds of sterile atoms captured by the Earth e.g. due to photon--sterile photon kinetic mixing; present experimental and cosmological limits on such mixing 
\cite{vigo}  and geophysical limits \cite{igna} still allow the presence of a relevant amount of sterile material at the Earth for magnetic fields less than several gauss~\cite{zurab10}.  
\end{itemize}

The HIBEAM sensitivities to the conversions of free neutrons to antineutrons and neutron-sterile neutron processes are given in Sections~\ref{nnsearches} and~\ref{sec:sterileres}, respectively. 

\subsection{Axions}
Axions are one of the leading candidates to explain cosmic DM, the mysterious form of matter that makes up about five-sixths of the total matter content of the Universe \cite{particle_data_group_review_2022}, while also providing the leading explanation for the strong $\mathcal{CP}$ problem of quantum chromodynamics \cite{Kim_axions_review_2010}. 
Searches for axion DM have mainly focused on the axion's electromagnetic coupling to photons \cite{adams_axion_2022}. 

The HIBEAM neutron beamline provides a sensitive experimental setup to probe ultralight (sub-eV mass) axion DM via its coupling to neutron spins. 
The coupling of a Galactic axion DM field to the axial-vector current of the neutron would cause spin-polarised neutrons to precess about the direction of the axion DM momentum in the laboratory frame of reference~\cite{Flambaum_Patras_2013,stadnik_axion-induced_2014}. 
An ``axion wind'' spin-precession effect on the Larmor precession frequency of neutrons in the presence of an ordinary magnetic field can be observed using Ramsey's method of separated oscillating fields \cite{1950PhRv...78..695R}, whereby the phase of the neutrons accumulates an additional time-varying phase due to the interaction of the neutrons with the axion DM field.  The sensitivity of HIBEAM to ultralight axions is given in Section~\ref{ultraaxion}. 

\subsection{Neutron electric charge}
 It is not always widely appreciated that the SM does not predict electric charge quantization or the electrical neutrality of the neutron~\cite{RFoot_1993,PhysRevD.49.3617}. An additional free parameter~\cite{PhysRevD.48.4481} could allow for a small but nonzero charge in apparently neutral particles, such as neutrons, neutrinos, and atoms~\cite{PhysRevLett.100.120407}, which must therefore be determined experimentally~\cite{PhysRevD.44.3706}. The very stringent limit on the neutron charge $q_{n} < 1.8 \times 10^{-21}{e}$ makes it clear that this is yet another Standard Model parameter that requires (extreme) fine-tuning. This circumstance obviously favors extensions of the Standard Model which lead to electric-charge quantization and $q_{n}=0$ naturally, including higher dimensions \cite{Klein:1926fj}, superstrings \cite{green2012superstring,JacquesDistler_2008}, magnetic monopoles \cite{Dirac:1931kp} and Grand
Unified Theories (GUTs) \cite{PhysRevD.10.275,PhysRevLett.32.438,OKUN1984115}. Note that a nonzero value for $q_{n}$ would eliminate the possibility of neutron-antineutron
oscillations. As discussed in Section~\ref{qneutron}, HIBEAM can be configured with a deflecting electric field over a long propagation distance such that sensitivity to a nonzero neutron electric charge is provided.

\subsection{Other potential areas of exploitation}

In addition to the aforementioned searches, HIBEAM offers opportunities for a broader range of fundamental physics investigations.

A measurement of a nonzero neutron electric dipole moment (EDM) would be of profound significance. It would provide direct evidence of CP violation beyond that predicted by the Standard Model (SM), fulfilling one of the Sakharov conditions for baryogenesis. Additionally, it would impose constraints on the QCD $\theta$ parameter and various beyond-SM scenarios~\cite{Pospelov:2005pr,Engel:2013lsa,Chupp:2017rkp}. Section~\ref{edm} discusses the feasibility of conducting an EDM search at HIBEAM, leveraging existing preparatory work for a similar search planned at Oak Ridge National Laboratory (ORNL)~\cite{Ahmed_2019}.

HIBEAM can also be used to study parity violation in hadronic interactions, shedding light on the interplay between the weak and strong nuclear forces~\cite{Snow:2016zyq,deVries:2020iea}. Investigations of nucleon-nucleon (NN) weak interactions provide a unique testing ground for models of low-energy QCD, including effective field theory~\cite{Zhu:2004vw} and lattice gauge theory~\cite{Wasem:2011tp}.

Additionally, HIBEAM could play a role in high-precision studies of neutron beta decay, which serve as sensitive tests of the SM and probes of potential new physics processes~\cite{Abele:2008zz,Falkowski:2020pma,Baessler:2014gia}. This field encompasses a variety of precision observables and is crucial for refining the determination of the CKM matrix element $V_{ud}$, a fundamental parameter of the electroweak sector.

\section{The European Spallation source} \label{sec:ess}
The ESS facility and how it relates to neutron oscillation projects is described in Refs.~\cite{Addazi:2020nlz,Backman_2022}.  ESS is a spallation neutron source.  It uses a linear accelerator to accelerate  protons on to a neutron production target; in order to deliver the exceptional neutron flux specified in its design requirements, ESS is installing the world's most powerful proton accelerator. The proton beam is pulsed at a repetition rate of 14 Hz, with each pulse lasting 2.86 ms. Through acceleration, the proton beam reaches an energy of 2 GeV, powered by a current of 62.5 mA.
ESS is presently committed to delivering 2 MW power on target by 2028, with the underlying infrastructure in place to allow a straightforward upgrade to 5 MW.

\begin{figure}[tb]
\centering
    \includegraphics[width=0.7\textwidth]{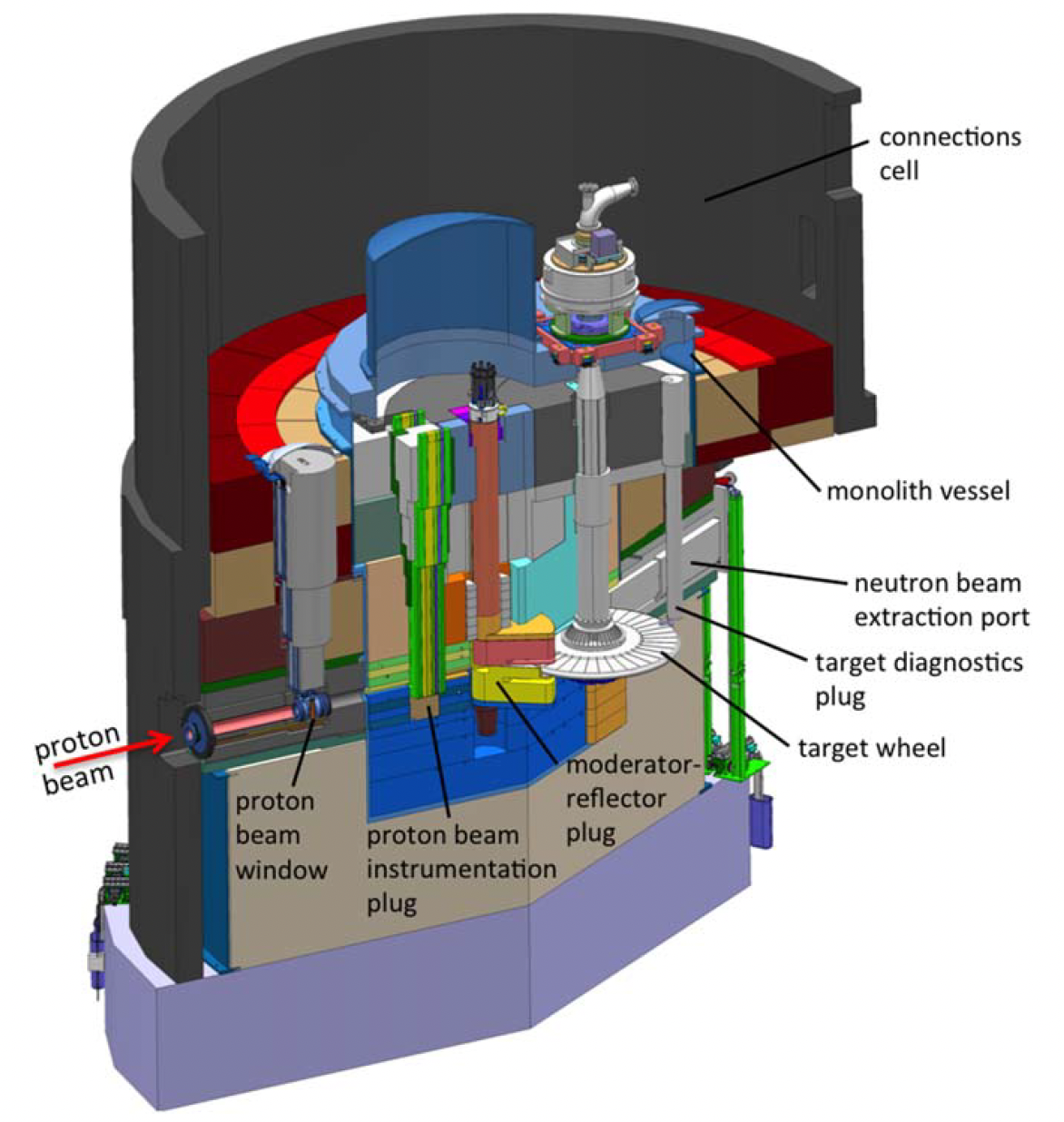}
     \caption{The ESS target monolith.}
        \label{target}
       \end{figure}

\begin{figure}[ht]
\centering
               \includegraphics[width=0.8\textwidth]{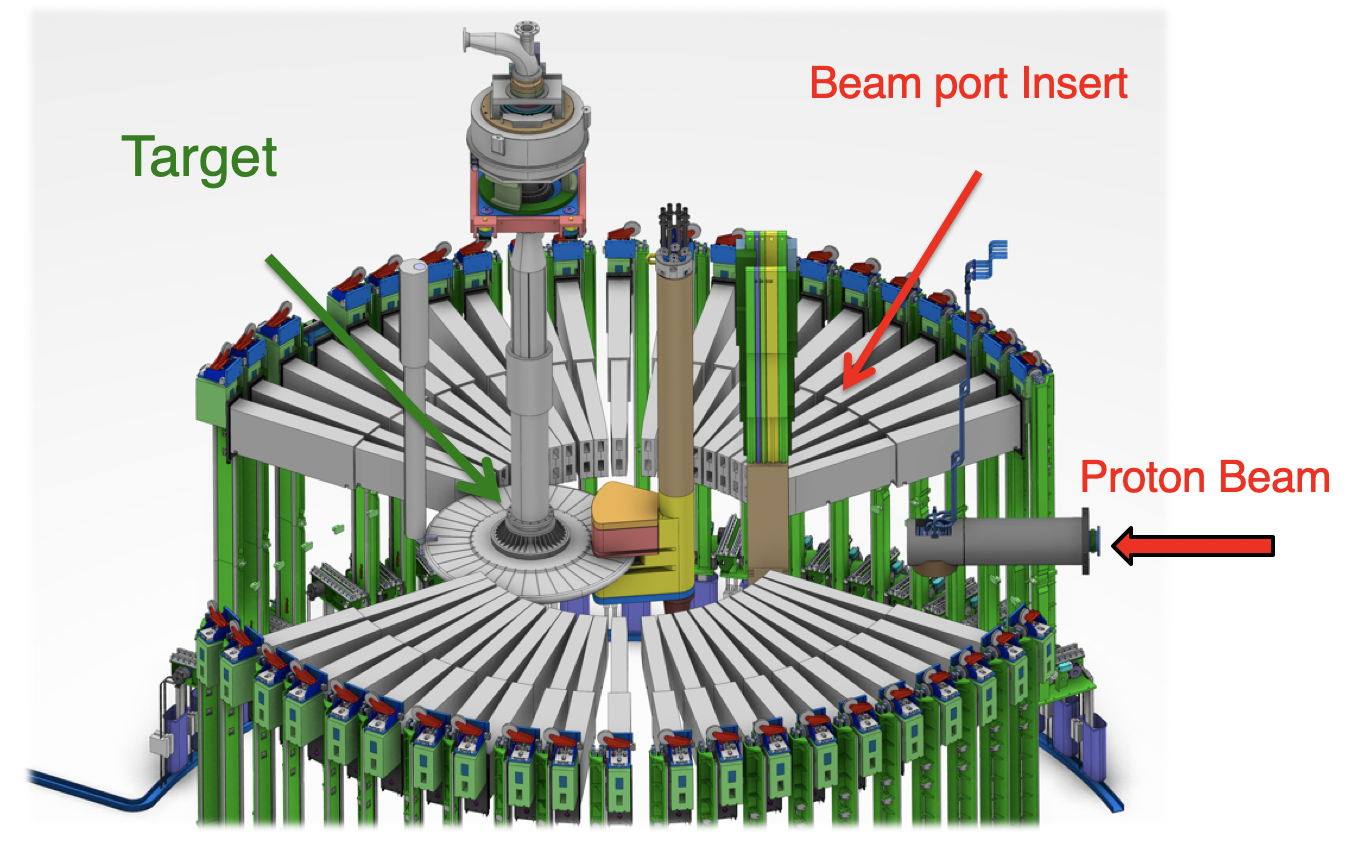}
                \caption{Zoom in the target area. The 42 beam ports surrounding the target are clearly visible, as well as the ESS moderator positioned above the spallation target. }
    \label{essmonolith}
\end{figure}

Once the proton beam reaches its ultimate energy, it collides with a rotating tungsten target, resulting in spallation and the production of primarily evaporation neutrons at a kinetic energy of roughly 2 MeV. 
The spallation neutrons undergo moderation within the neutron moderators contained within the moderator-reflector plug, as shown in Figure~\ref{essmonolith}. Initially, the ESS will be equipped with only a single compact low-dimensional moderator located above the spallation target, which has been designed to deliver brightest neutron beams for condensed matter experiments~\cite{zanini_design_2019}. The target and moderator-reflector system is located within a shielding and cooling configuration referred to as the monolith. Figures~\ref{target} and ~\ref{essmonolith} illustrate the proton beam, the target, and the monolith structure.
       
As shown in Figure~\ref{essmonolith}, the ESS possesses a total of 42 beam ports. These beam ports play a crucial role in the facility as they serve as the neutron extraction systems, which are responsible for transporting neutrons from the target to the instrument area. Surrounding the monolith is a shielding structure known as the bunker~\cite{bunkerpaper}. It serves as a radiation protection shield, enclosing the ESS monolith and shielding the instrument area from high levels of ionizing radiation produced during operation. The bunker’s roof and walls are constructed from heavy magnetite concrete for effective radiation attenuation. Within the bunker, neutron beamlines are equipped with neutron guides and instrument-specific components such as choppers and shutters.

\section{HIBEAM: beamline overview and search sensitivities}\label{sec:hibeamoverview}
\begin{figure}[tb]
	\centering
	\includegraphics[width=\textwidth]{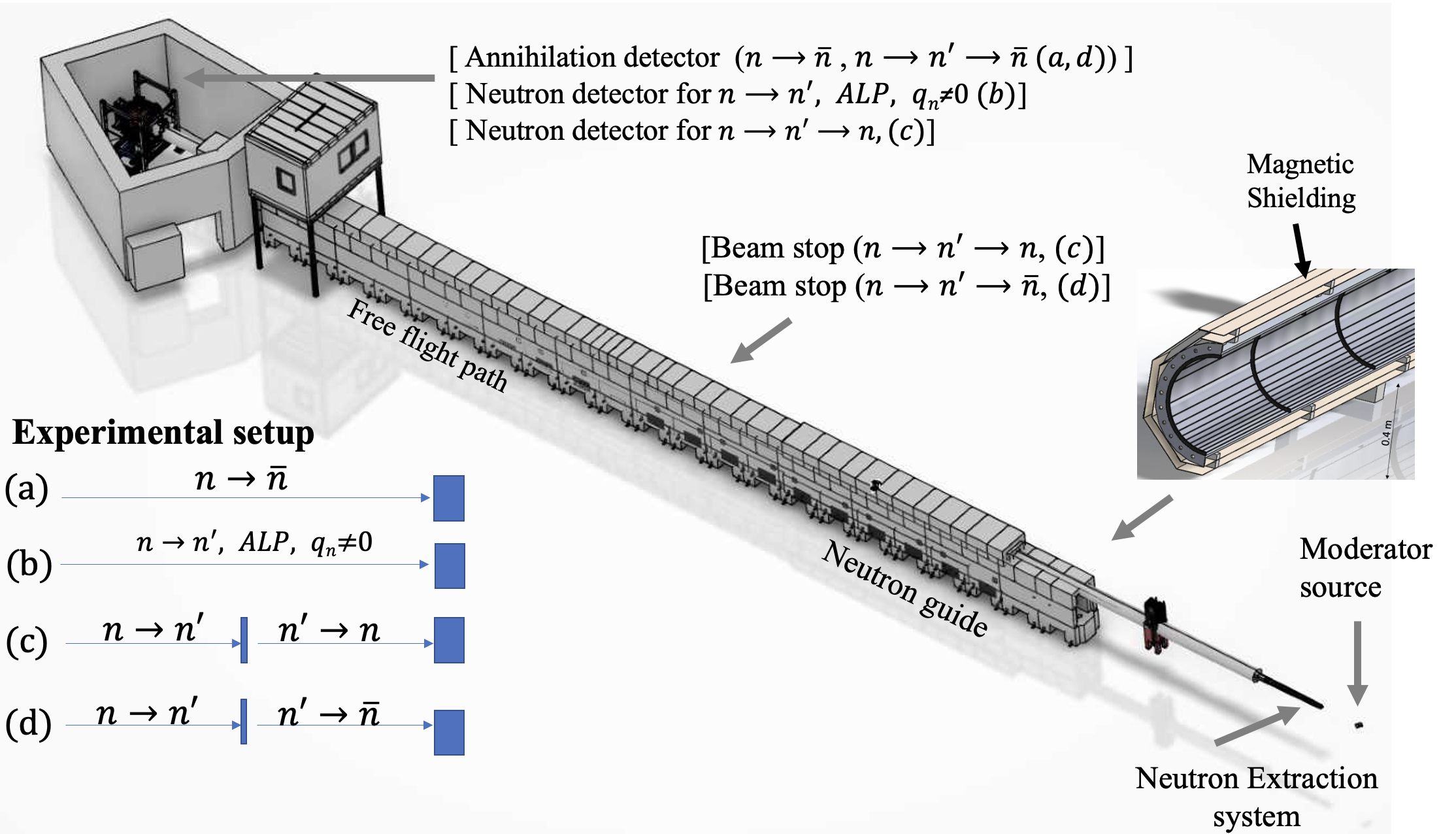}
	\caption{A CAD model of the HIBEAM beamline designed for the proposed searches. Neutrons emitted from the moderator are directed through a neutron guide into a flight path region and then into a detector area. The guide and flight path are magnetically controlled. Different configurations for the searches for sterile neutrons, ALPs, and nonzero neutron charge ($q_{n}\neq 0$) are indicated. The principles of these searches are illustrated in the bottom left of the figure. The right-side inset displays the vacuum pipe and magnetic shielding.}
	\label{beamlineprinciple}
\end{figure}

This Section gives an overview of the principles of the HIBEAM searches, as illustrated by the HIBEAM CAD model in Figure~\ref{beamlineprinciple} (a-d). This is then followed by the search sensitivities and a comparison to previous work.  Detailed information on the geometry of the beamline, beam extraction, neutron optics, neutron propagation, magnetic shielding, and the beamline's radiation profile is given in Section~\ref{sec:hibeambeamline}. The suite of detectors is described in Section~\ref{sec:det}. 

For the free neutron-antineutron search (a), neutrons pass through a neutron guide into a free propagation volume in which the magnetic field is less than $5-10$~nT,  eventually impinging upon a thin ($100$ $\mu$m) carbon target. The target is surrounded by a detector capable of observing the products of any antineutron-nucleon annihilation taking place in the target. 

For the sterile neutron searches, neutrons propagate in magnetic fields of fixed intervals such that the conversion to a sterile state is not suppressed.  Searches for sterile neutrons via the anomalous disappearance of neutron flux take place in configuration (b). A beam-stop is used in configuration (c) for the regeneration search. The incoming neutron flux is thus absorbed but a sterile neutron could pass through prior to converting back to a neutron which is then detected. Neutrons converting to antineutrons via sterile neutron states can also be sought with this approach (d). 

The search for ALPs and a nonzero neutron electric  would take place in mode (b) albeit with different field to those used for the sterile neutron searches. The configurations for the searches for ALPs and neutron charge are described in detail in Ref.~\cite{PhysRevLett.133.181001} and Section~\ref{qneutron}, respectively.




\subsection{Search for free neutron to antineutron conversions at the HIBEAM beamline} 
\label{nnsearches}

Previous searches for free $n\rightarrow \bar{n}$ oscillations were conducted at the Triga Mark II reactor at the University of Pavia~\cite{Bressi:1989zd,Bressi:1990zx} and at the Institut Laue-Langevin (ILL)~\cite{Fidecaro:1985cm,oldnnbar1}. The ILL search~\cite{oldnnbar1}, though performed almost 30 years ago, still provides the most stringent limit for the free neutron oscillation time of about $8.6 \times 10^7$ seconds. No new searches for free $n\rightarrow \bar{n}$ oscillations have been performed since then, despite theoretical interest and several proposals~\cite{Gudkov:2021wvn,Gudkov_2020,PhysRevLett.122.221802,Kronfeld:2016} to improve the experimental sensitivity. These experiments require both an intense neutron source and a team with diverse expertise in magnetic shielding, particle physics detectors, and slow neutron optics. 

The experiment requires a focused beam of free neutrons propagating through a sufficiently field-free (or ``quasi-free'') region~\cite{Phillips:2014fgb} toward a detector capable of identifying antineutron annihilation events. This detector, known as the {\it annihilation} detector, is designed to capture and analyze the annihilation of antineutrons within a thin target. The resulting final state consists of charged pions and photons with an invariant mass of up to $\sim 1.8$~GeV, which are then observed and recorded.

A significantly more advanced antineutron detector system is now feasible compared to the one used in the previous ILL experiment. The ILL detector primarily relied on streamer tubes and scintillators, which had limitations in resolution and efficiency. HIBEAM will have a substantially improved detector, including a high-precision crystal calorimeter and Time Projection Chamber (TPC) tracking (see Section \ref{sec:annihilationdetector}). It is therefore expected to be able to match the ILL efficiency of $\sim 50$\% with full background suppression. Moreover, as shown for the NNBAR Conceptual Design Report (CDR), the use of a more advanced detector than that available at the ILL, together with more advanced analysis techniques, such as the use of machine learning (ML), can potentially lead to antineutron signal efficiencies of greater than 80\%~\cite{Santoro:2024lvc}.

The most appropriate proxy for the discovery potential for a free $n\rightarrow {\bar{n}}$ search is given by the figure of merit ($FOM$) defined in Equation~\ref{eq:fomnnbar}. This quantity is directly proportional to the number of antineutrons that hit the foil and are therefore potentially observable. For a background-free search, the oscillation time sensitivity varies as the square root of the $FOM$.   

\begin{equation}
    FOM=\sum_i N_{n_{i}} \cdot t_{n_{i}}^2 \sim \langle N_n\cdot t_n^2 \rangle, 
    \label{eq:fomnnbar}
\end{equation}

The quantity $N_{n_{i}}$ is the number of neutrons per unit time reaching the annihilation detector after $t_{n_{i}}$ seconds of flight through a magnetically protected, quasi-free vacuum region. The probability of a conversion is, therefore, proportional to the number of neutrons multiplied by the square of the transit time. A high-precision search therefore requires a large flux of cold neutrons (neutrons with energy below 0.025 eV) which are allowed to propagate freely over a long time to allow conversions to antineutrons. These conditions are satisfied at the ESS. 

The $FOM$ achieved in the ILL search was 
$1.5 \times 10^{9}~n\cdot s$. The cold neutron flux available at the ESS, combined with advancements made in neutron focusing in the last 30 years allow this value to be more than doubled (see Section~\ref{sec:optics}). 

Taken together, improvements in anti-neutron detection, and in neutron flux and transmission, can lead to an improvement on the discovery sensitivity for free $n\rightarrow \bar{n}$ by a factor of 10, assuming 3-4 years of running time for the experiment\footnote{The ILL experiment ran for a year.}.

\subsection{Searches for sterile neutrons}\label{sec:sterileres}
These search modes require scans over different magnetic field configurations. 
Following the procedure in~\cite{Berezhiani:2017azg,Addazi:2020nlz}, the sensitivity in oscillation times can be estimated for  the HIBEAM beamline for magnetic fields scans in the up-down direction with step size 2~mG (200~nT) between $\pm 2$~G ($\pm$200~$\mu$T)~\cite{Addazi:2020nlz}. 
Dedicated 3-D magnetic scans with high statistics would then be used to investigate any candidate signal. 

A sensitivity at 95\% Confidence Level on $\tau_{n \rightarrow n'}$ of $97$~s for the proposed neutron regeneration experiment with the ESS power of 5MW can be achieved for a running period of around a year for a magnetic field region up to 2~G (200~$\mu$T) and a background rate of 0.1 neutron/second. For the disappearance mode, this increases to $185$~s. These sensitivities typically outstrip earlier limits~\cite{ban,serb1,altarev,bodek,serb2,zurab11,berbiondi,edm,Ban:2023cja} and can achieve order-of-magnitude improvements for various magnetic field regions, in particular the higher field region in which limits are extremely weak. The search for neutron-antineutron conversions via sterile neutrons provides a sensitivity on ${{\tau_{n\rightarrow n'}\tau_{\bar{n} \rightarrow {n'}}}}$. Unlike the other searches, an effective background-free search can be done, as the ILL experiment showed. Background-free conditions, i.e. an expected number of background events $n_b < 1$, leads to a sensitivity on $\sqrt{{\tau_{n\rightarrow n'}\tau_{\bar{n} \rightarrow {n'}}}}$ of  $230$~s. 

\begin{figure}[H]
  \begin{center}
    \includegraphics[width=0.75\textwidth]{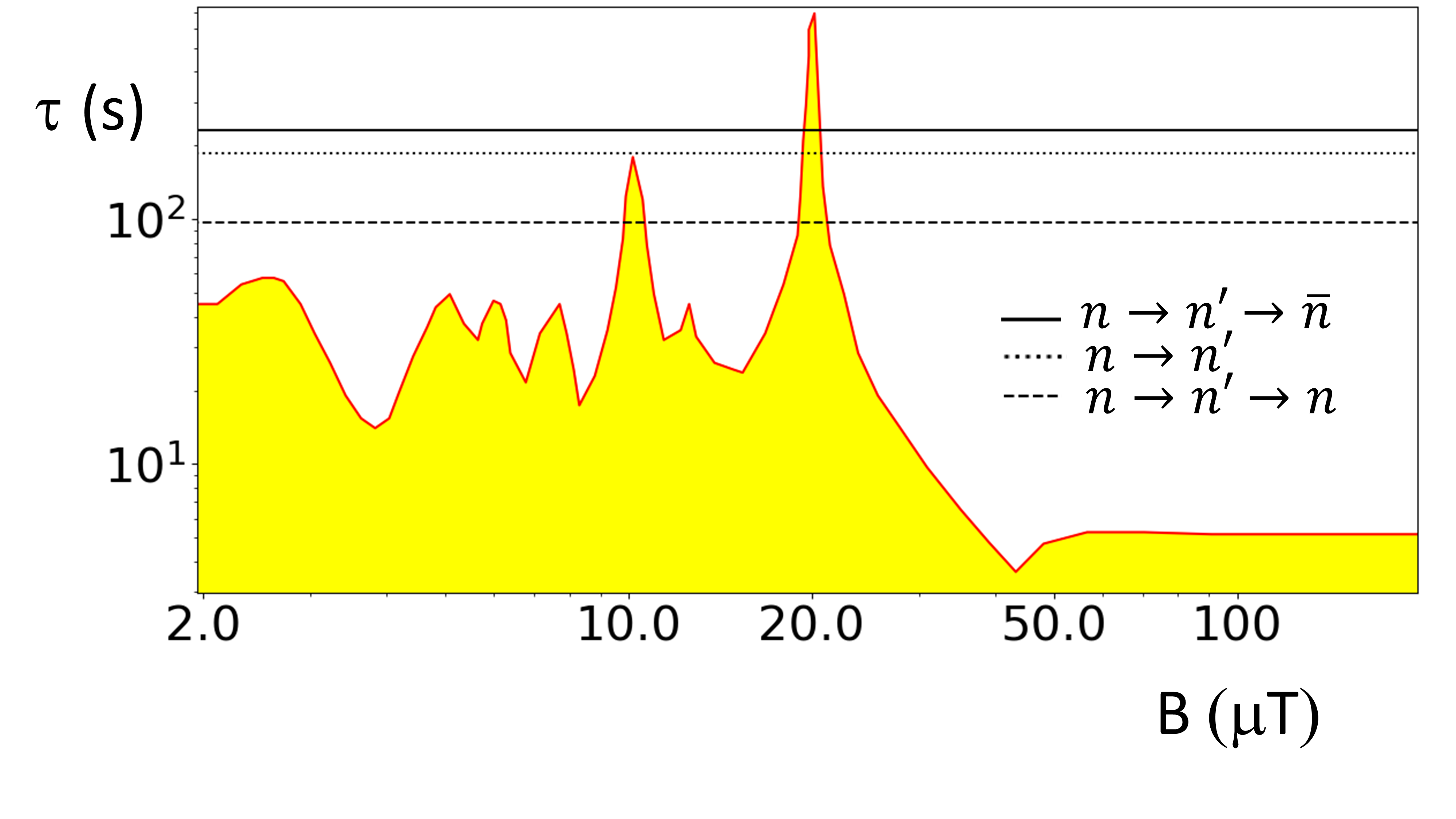}
  \caption{Excluded regions (shaded/yellow) of $\tau_{nn'}$ by experiments with ultra-cold neutrons for magnetic fields up to around  $\pm 2$~G~\cite{ban,serb1,altarev,bodek,serb2,zurab11,berbiondi,edm,Ban:2023cja}. The sensitivity of the HIBEAM program for $\tau_{nn'}$ from regeneration and disappearance experiments is shown, as is the HIBEAM sensitivity for $\sqrt{{\tau_{n\rightarrow n'}\tau_{\bar{n} \rightarrow {n'}}}}$ together with earlier limits on $\tau_{nn'}$. }
  \label{fig:limsen}
 \end{center}
 \end{figure}

The HIBEAM sensitivities are shown in Figure~\ref{fig:limsen} and compared to earlier limits. The large HIBEAM sensitivity gain is robust to ESS power scenarios. The oscillation time sensitivities typically decline by around 20\% for a linac power of 2MW.

\begin{figure}[H]
\begin{center}
\includegraphics[width=.8\textwidth]{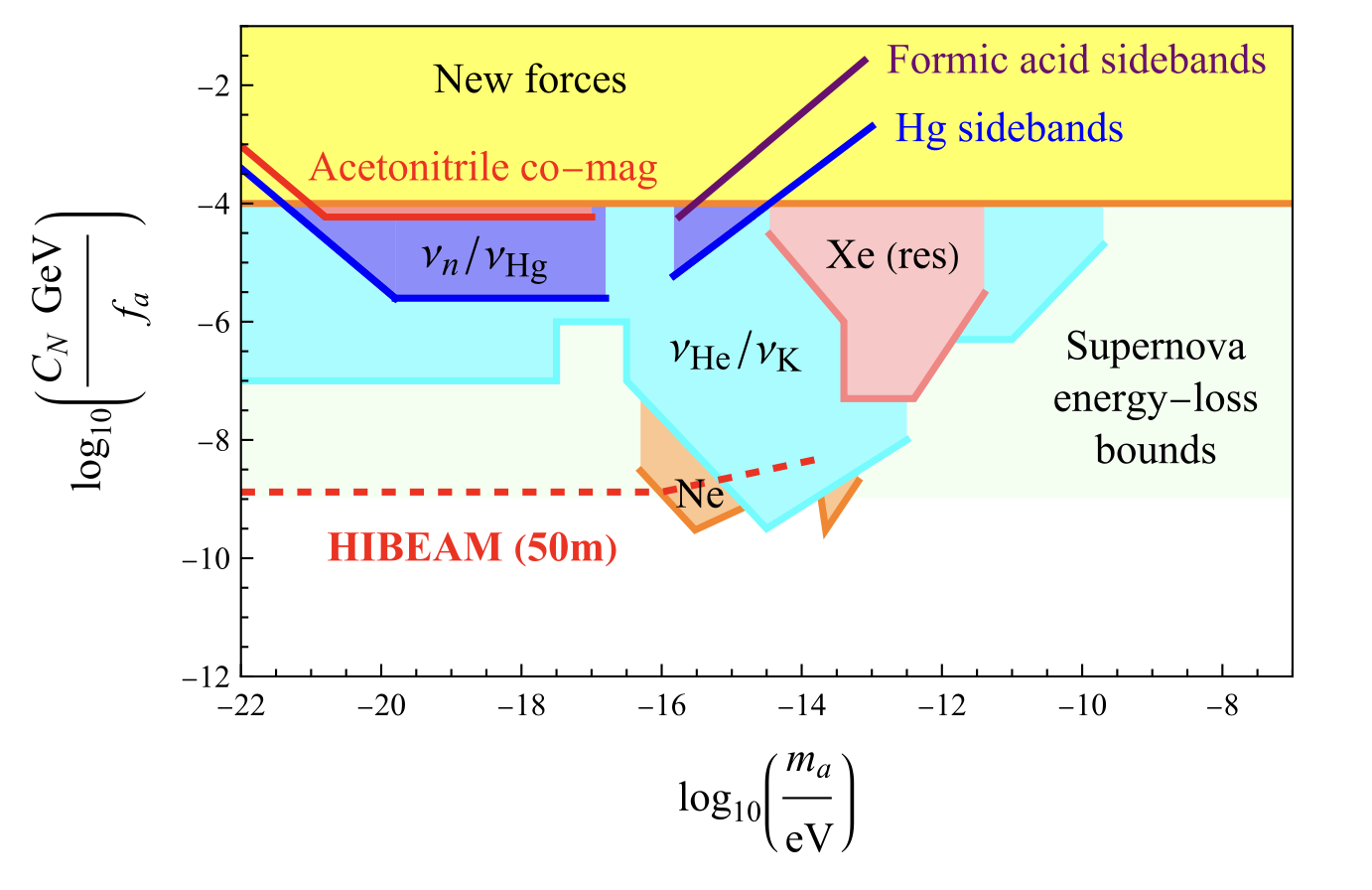}
\caption{The projected sensitivity of a 50\,m scale Ramsey neutron-beam experiment, utilizing the HIBEAM neutron beamline at the ESS (represented by the dashed red line), to the coupling strength of axion dark matter with a neutron is shown as a function of the axion mass $m_a$, assuming one year of operation. The cyan, blue, pink, red, and purple regions represent areas of the parameter space that have already been explored by magnetometry-based searches for time-varying spin-precession effects induced by axion dark matter. The yellow region indicates the parameter space excluded by a magnetometry-based search for spin-dependent forces mediated by virtual axion exchange. The pale green region outlines constraints derived from astrophysical observations of supernovae, which are model-dependent and may be evaded altogether. 
For further details, see Ref.~\cite{PhysRevLett.133.181001}.
} 
\label{fig:axion}
\end{center}
\end{figure}

\subsection{Searches for ultralight axion dark matter}
\label{ultraaxion}

The HIBEAM neutron beamline can also be utilized to search for ultralight axion dark matter as described in~\cite{PhysRevLett.133.181001}. The sensitivity of a search using a 50m magnetic control beamline (see Section~\ref{sec:magin}) is shown in Figure~\ref{fig:axion}. This sensitivity is presented as a function of the coupling strength of axion dark matter with a neutron, depending on the axion mass $m_a$, assuming that axions account for the observed density of dark matter in our local Galactic region. The coupling strength is parameterized by the combination $f_a/C_N$, where $f_a$ is the axion decay constant and $C_N$ is a model-dependent dimensionless parameter~\cite{PhysRevLett.133.181001}. One year of run time is assumed.

The HIBEAM search extends the sensitivity by up to $2-3$ orders of magnitude compared with other direct laboratory searches based on magnetometry. 
Such searches have been made for time-varying spin-precession effects induced by axion dark matter~\cite{abel_search_2017,wu_search_2019,garcon_constraints_2019,bloch_2020_axion,jiang_search_2021,bloch_new_2022,bloch_constraints_2023,abel_search_2023,Romalis_axion_2023}. 
Bounds from astrophysical observations of supernovae \cite{carenza_improved_2019} are also shown. 
These are subject to model-dependent assumptions and may be evaded altogether \cite{Blum_2020_supernova}.

\subsection{Searches for a nonzero-neutron electric charge} 
\label{qneutron}

\begin{figure}[ht]
 \begin{center}
    \includegraphics[width=0.8\textwidth]{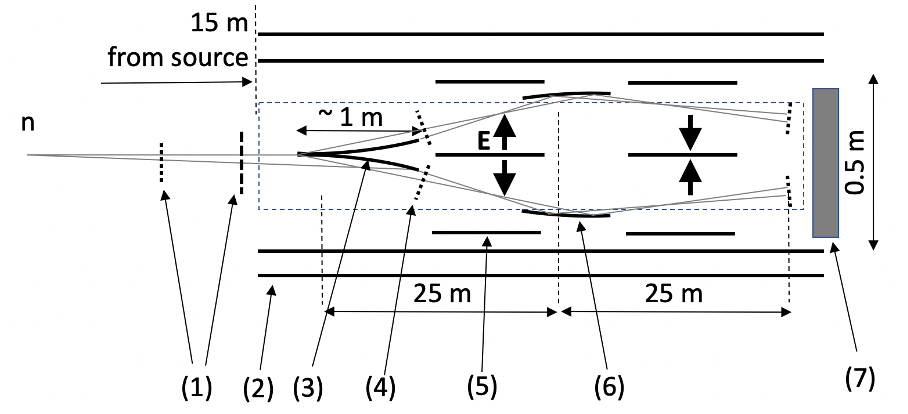}
 \end{center}
  \caption{Layout of the neutron charge measurement setup within the proposed HIBEAM beamline: (1) collimators for the neutron beam, (2) magnetic shielding, (3) bent neutron optics, (4) slit patterns, (5) electric field plates, (6) bent neutron optics, and (7) neutron detectors.
 }
  \label{fig:neutronchargeexperiment}
\end{figure}

With a setup similar to that of the ALP search, the HIBEAM instrument can also be used to search for a nonzero neutron electric charge. The slightly modified experimental setup is shown in Fig.~\ref{fig:neutronchargeexperiment} and is based on the best previous neutron beam measurement \cite{PhysRevD.37.3107}. A collimated, intense neutron beam passes through a vertically aligned slit pattern composed of absorber material with strips spaced a few tens of micrometers apart. The beam is then imaged using a system of two achromatic mirrors, which act as lenses, onto a second pattern, followed by a neutron detector. If neutrons possess a nonzero charge, an electric field applied transverse to the beam will deflect them, leading to a measurable change in the transmitted intensity.
The inverting optics generated by the double mirror geometry allow for the application of oppositely-directed electric fields to cancel first order systematics. The deflection of the neutrons due to an electric field follows $ y = \frac{q_n  E  L^2}{2 m v^2} $ with $y$ the transverse deflection, $E$ the electric field, $q_n$ the charge of the neutron, $L$ the free flight path, $m$ neutron mass and $v$ velocity. Using $q_n \sim 10^{-21} e$, $E = 6 \times 10^6$~V$/$m and 600~m$/$s neutron velocity, a deflection of $\sim$~0.1~nm is expected for 10 m experiment length. By aligning the double-slit pattern one can set the operating point of the device to lie at the point of steepest transmission change due to deflection. In the previous experiments, the change in count rate was 923~n$/$($\mu$ms) on top of a total rate of $3 \times 10^4$~n$/$s. The uncertainty of the measurement scales with $\sigma_y = \omega / \sqrt{N}$, $\omega$ the beam divergence and $N$ the total number of neutrons.
The sensitivity improvements with HIBEAM for this measurement arise from multiple factors. At 15 meters from the source, where the magnetically controlled section of the beamline begins, HIBEAM will achieve a flux of 10$^{12}$ n/s with ESS operating at 2 MW power, with a divergence of about 2 mrad from the installed collimation, compared with $3 \times 10^4$~n$/$s through a slit with 0.3 mm width and 20 cm height and an allowed divergence due to collimation of also about 2 mrad in the previous experiment. The slit patterns have an assumed transmission of 10\% each. The transmitted neutron intensity for the same measurement duration is enhanced by 10$^3$ to give an additional enhancement factor of 33. The overall improvement in sensitivity is a factor of 700.

\subsection{Searches for electric dipole moment of the neutron (EDM)} 
\label{edm}

\begin{figure}[htb]
\begin{center}
\includegraphics[width=.8\textwidth]{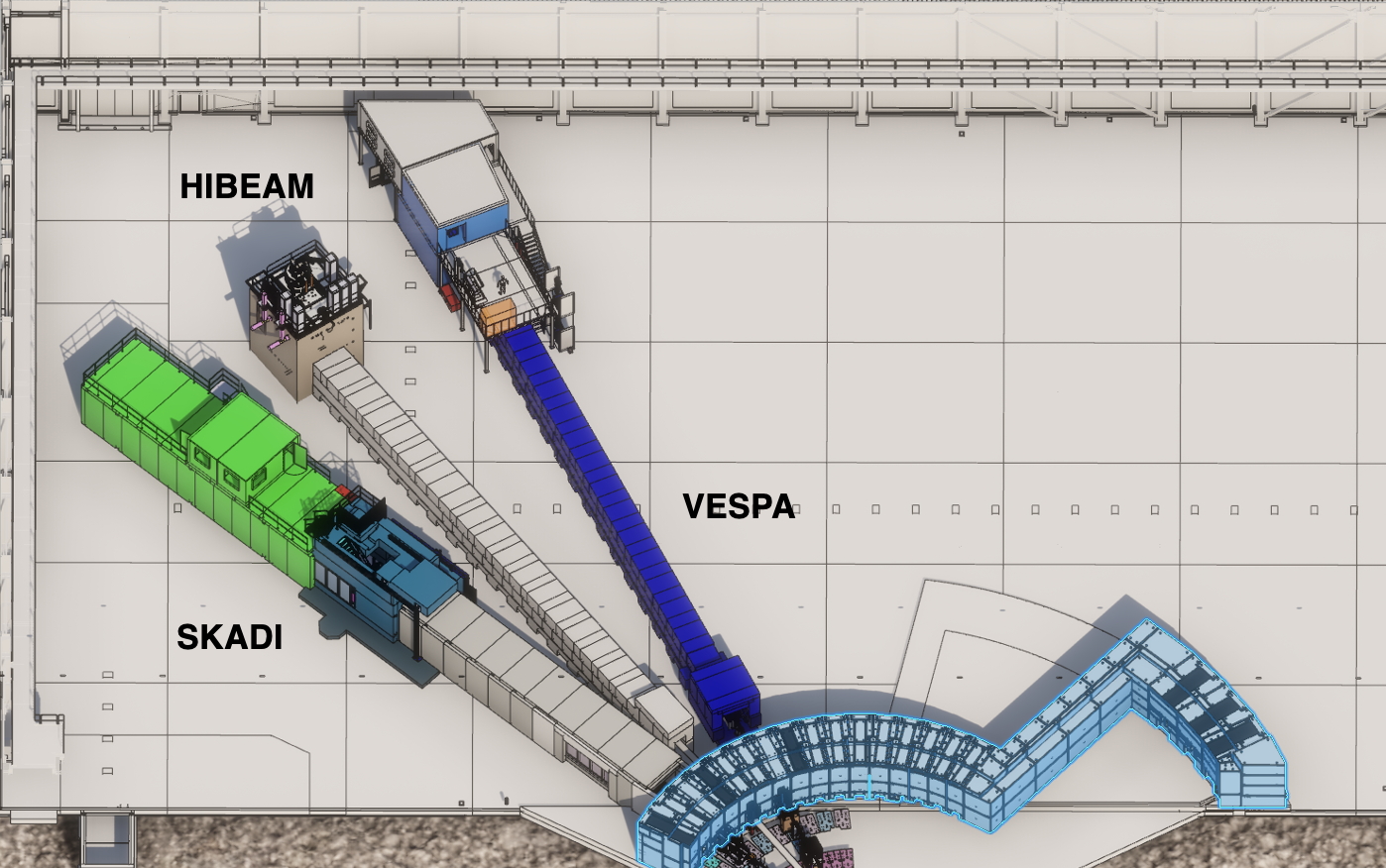}
\caption{The neutron electric dipole moment experiment at the HIBEAM beamline.
} 
\label{fig:edm}
\end{center}
\end{figure}

The search for a nonzero neutron EDM has been a key priority at neutron facilities worldwide for over 60 years \cite{PhysRev.108.120}, owing to its strong potential for uncovering physics beyond the Standard Model. While searches are currently ongoing at various laboratories around the world, some ideas have been proposed in the past at the ESS, including the use of a cold beamline \cite{Abele_2023} or a dedicated ultracold neutron (UCN) source \cite{Santoro:2024lvc}.

While these ideas require further developments, a remarkable opportunity to mount such an experiment at the ESS emerged in late 2023 due to the cancellation of a U.S. experiment to measure the EDM of the neutron: the nEDM@SNS project \cite{Ahmed_2019} at the Spallation Neutron Source (SNS) at ORNL. The experiment was in an advanced design stage and partly constructed. 
Some or all of the equipment constructed for this experiment could, in principle, be used at the ESS. 

The experiment is based on an idea to combine ultracold neutron production in superfluid $^4$He with real-time measurement of the precession frequency using the capture of polarized neutrons on polarized $^3$He \cite{GOLUB19941}. 
At the ESS with a beamline having a higher flux than that at the SNS, there is a possibility to reach toward the 10$^{-29}$  e·cm level for the EDM.

Dedicated design studies are underway to assess the feasibility and performance of the experiment at ESS. Preliminary results indicate that the nEDM@SNS experimental setup can be accommodated in the ESS instrument hall (see Figure \ref{fig:edm}) at the HIBEAM beamline without significant engineering constraints. To conduct nEDM experiments at the HIBEAM beamline, the guide system in the ESS bunker (see Section~\ref{sec:ess}) must be modified to eliminate the direct line of sight to the ESS source. A dedicated study is currently ongoing with a focus on optimizing this guide system; however, preliminary results already suggest a substantial performance improvement over SNS.

\subsection{Other activities} 
\label{oth}

It is important that the HIBEAM concept remains sufficiently flexible to accommodate, with appropriate design and infrastructure adjustments, a variety of activities typically conducted on a neutron beamline, leveraging the unique properties of ESS.

Ongoing future work includes quantifying the potential of HIBEAM for measurements of hadronic parity violation~\cite{Snow:2016zyq,deVries:2020iea}, such as parity-odd gamma asymmetries, as well as investigations into neutron decay~\cite{Abele:2008zz,Falkowski:2020pma,Baessler:2014gia}.

While this paper primarily focuses on the design of HIBEAM as described in Section~\ref{nnsearches},~\ref{sec:sterileres}, ~\ref{ultraaxion} and~\ref{qneutron} it is important to emphasize that this beamline has the potential to serve as a generic particle physics beamline. With appropriate modifications, it could support a wide range of experiments beyond those currently considered, providing a valuable resource for the particle physics community throughout the lifetime of ESS operations.

\section{The HIBEAM beamline}
\label{sec:hibeambeamline}

 \begin{figure}[htb]
	\centering
	\includegraphics[width=.9\textwidth]{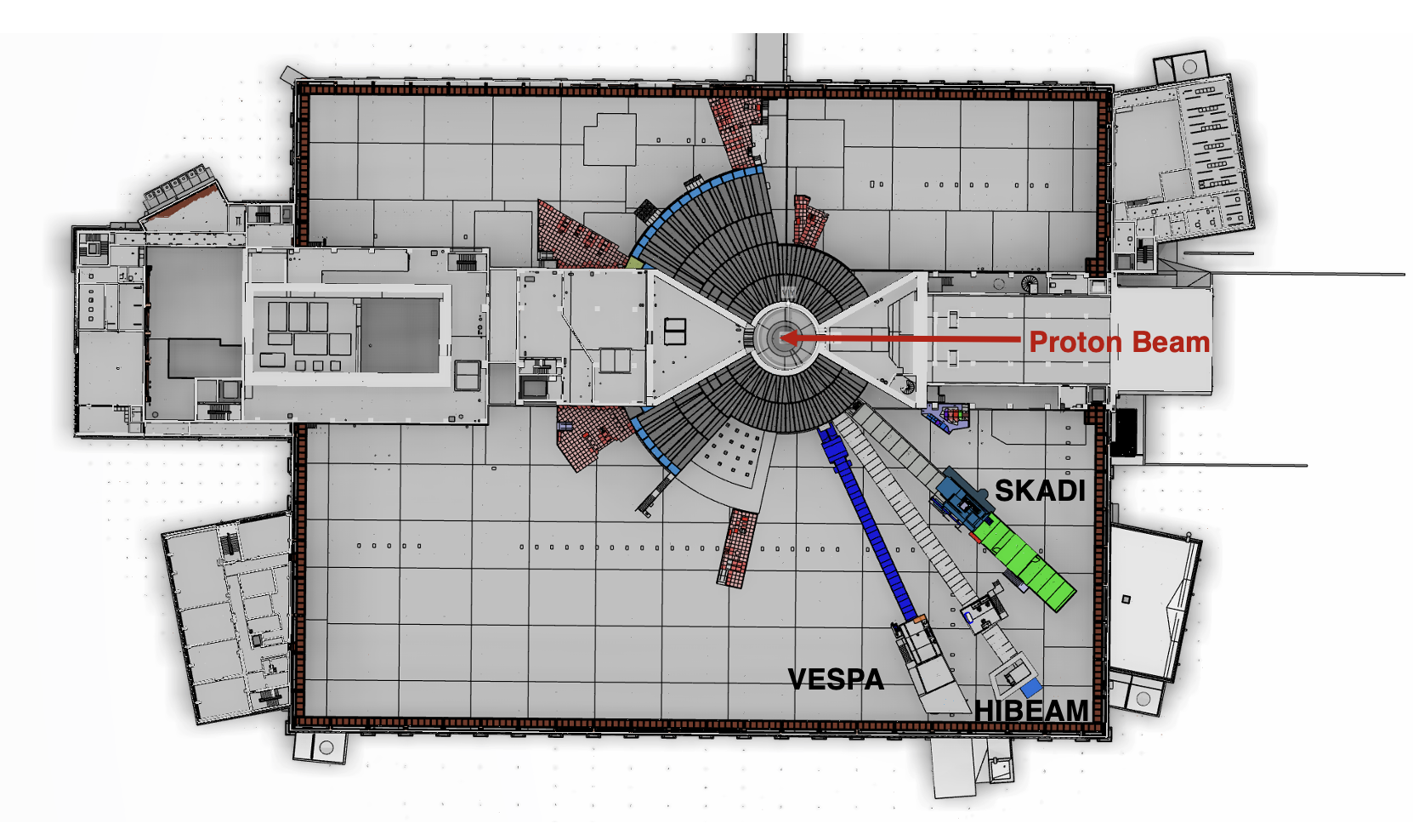}
	\caption{Overview of the ESS instrument hall: The picture shows only the instruments currently under construction in the east sector—VESPA and SKADI—along with the location of the HIBEAM beamline in the E5 position.}
 
	\label{essinstruments}
\end{figure} 

\begin{figure}[tb]
	\centering
	\includegraphics[width=.8\textwidth]{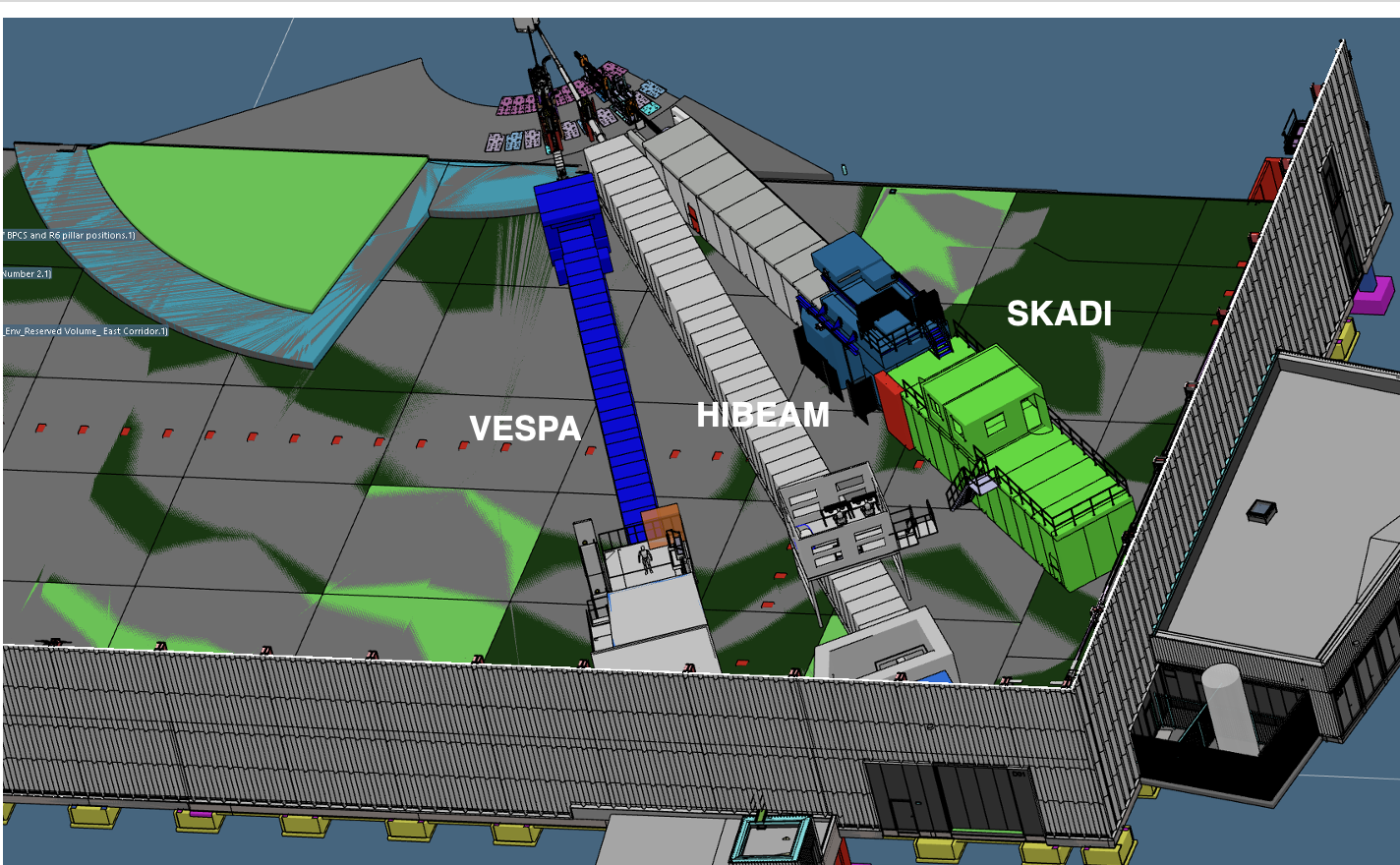}
	\caption{Expanded view of the ESS east sector. In the picture is shown the location of the HIBEAM beamline between VESPA and SKADI.}
	\label{eastsector}
\end{figure} 

To conduct the searches described above, a neutron beamline capable of delivering the highest possible flux of free cold neutrons to the experimental area is essential. Figure~\ref{essinstruments} shows the ESS instrument halls, highlighting the instruments currently under construction in the east sector. Figure~\ref{eastsector} provides a more detailed view of this region. The E5 beam port, located between the VESPA and SKADI instruments, has been identified as a potential site for HIBEAM. The beamline will extend 65 meters from the ESS moderator and will feature a vacuum pipe with a diameter of 40 cm and a thickness of 5 mm. The magnetic control beamline will start at 15 meters from the exit of the bunker wall and will extend for 50 meters\footnote{The magnetic infrastructure could also be extended inside the ESS bunker, starting at 6 meters. However, implementing this approach would be more challenging due to the limited accessibility of the ESS bunker once operations have commenced, as well as the additional requirement of integrating magnetic shielding within the bunker wall.}. The optical systems are described in Section~\ref{sec:optics}, while the magnetic control system for the vacuum pipe is discussed in Section~\ref{sec:magin}. The beamline must also be surrounded by radiation shielding to comply with ESS radiological safety requirements (see Section~\ref{sec:beamline}). Different detectors will be placed in the HIBEAM experimental cave, depending on the specific experiment (see Fig.~\ref{beamlineprinciple}). For the configuration where the final state includes an antineutron, an annihilation target with a radius of 20 cm will be positioned 65 meters from the ESS moderator inside the experimental cave. This target will be surrounded by a TPC and the Wide Angle Shower Apparatus (WASA) Scintillator Electromagnetic Calorimeter~\cite{wasa} (see Section~\ref{sec:annihilationdetector}).


\subsection{Neutron Optics}\label{sec:optics}
\begin{figure}[tb]
  \begin{center}
    \includegraphics[width=0.8\textwidth]{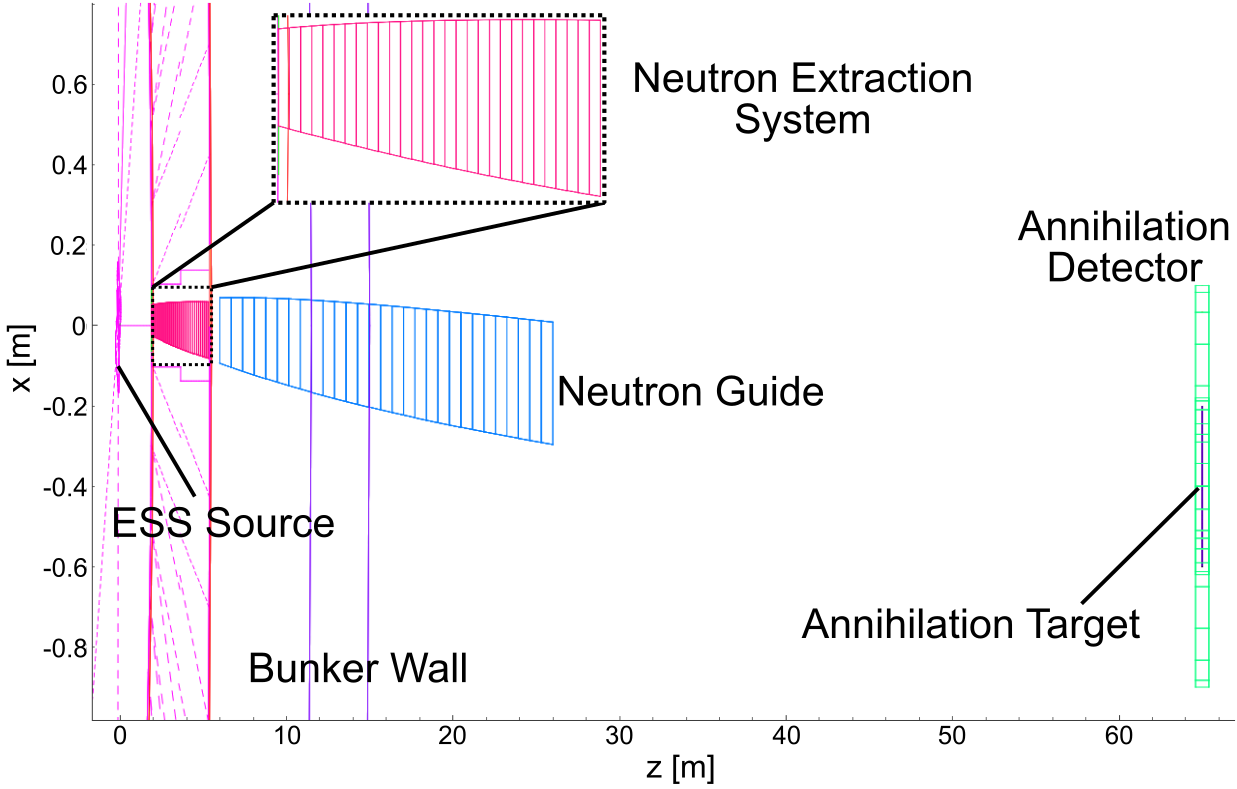}
  \caption{Top view of the optics system for the HIBEAM experiment as modelled in \software{McStas}. Neutrons from the ESS source are focused towards onto the annihilation target using the NES and a second neutron guide. }
  \label{guide}
 \end{center}
\end{figure}

The neutron optics system consists of two components, the Neutron Extraction System (NES) and a second neutron guide. A schematic of the optics is shown in Figure \ref{guide}. The NES is situated within the ESS target monolith, see Figure \ref{target}. It plays a crucial role in transporting neutrons outside the monolith, and it will be positioned between 2~m and 5.5~m from the moderator. Currently, a suitable system for the HIBEAM beamline is under construction, set to be installed in the ESS East sector.

\begin{figure}[tb]
	\centering
	\includegraphics[width=.48\textwidth]{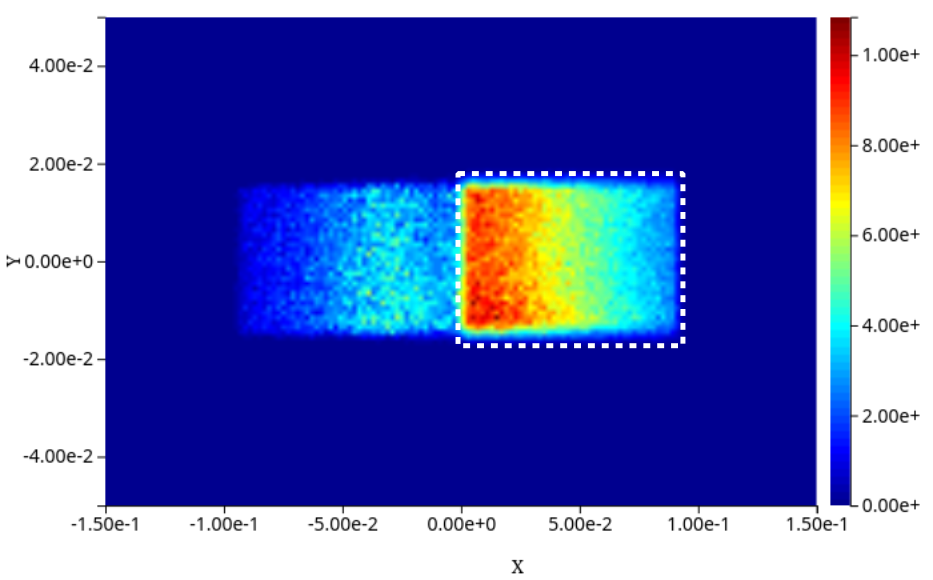}
	\includegraphics[width=.48\textwidth]{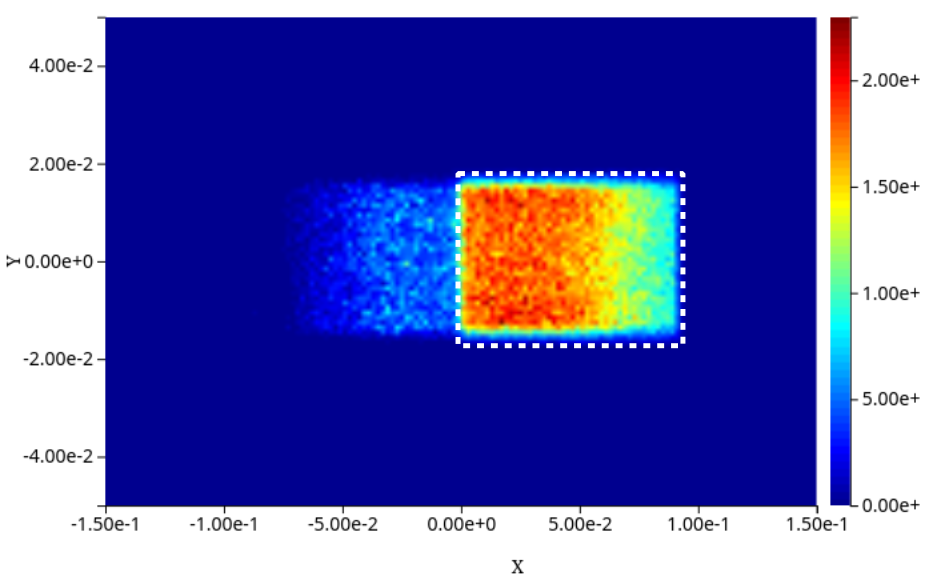}
 \caption{Point of origin in the moderator for neutrons reaching the exit of the NES. The dotted rectangle indicates the area corresponding to the cold wing of the moderator~\cite{Andersen:uh5002}. Left: Symmetric elliptic NES. Right: Optimized asymmetric NES.}
	\label{n_origin}
\end{figure} 

To develop the NES, simulation studies were carried out using the \software{McStas}~\cite{willendrup2004_McStasNew} software package. Its design is presented in the insert of Figure \ref{guide}. It has an entrance opening of 7.8$\times$7.8~cm$^2$ at 2~m from the moderator, and exit dimensions of 14.1$\times$12.9 cm$^2$. The reflective surfaces are coated in a $m=4$ Ni/Ti supermirror. Their shape follows a quadratic equation,
\begin{equation}
    a_{xx} \cdot x^2+a_{yy} \cdot y^2+a_{xy} \cdot x \cdot y + a_{x} \cdot x + a_{y} \cdot y - 1 = 0\ .
    \label{Eq:quad}
\end{equation}

The optimization of the guide shape was carried out using the \software{guide\_bot} software~\cite{Bertelsen2017}. This process resulted in an asymmetric design, which, in combination with the large cross-section of the opening, maximizes acceptance for neutrons originating from the entire closer cold wing of the moderator (see Ref.~\cite{Andersen:uh5002} for further details on the ESS moderator). Figure~\ref{n_origin} illustrates this by comparing the points of origin in the moderator for neutrons reaching the NES exit between the chosen design and a symmetric elliptic guide with the same dimensions.
Using the NES alone, $2.0 \times 10^{11}$~$n$/s with an average wavelength of 3.4~\AA\ can be focussed onto the $r=20$~cm target located in the experimental area assuming 5~MW operating power. This corresponds to a $FOM$ of $8.4 \times 10^{8}~n\cdot s$.
\begin{figure}[tb]
	\centering
	\includegraphics[width=.8\textwidth]{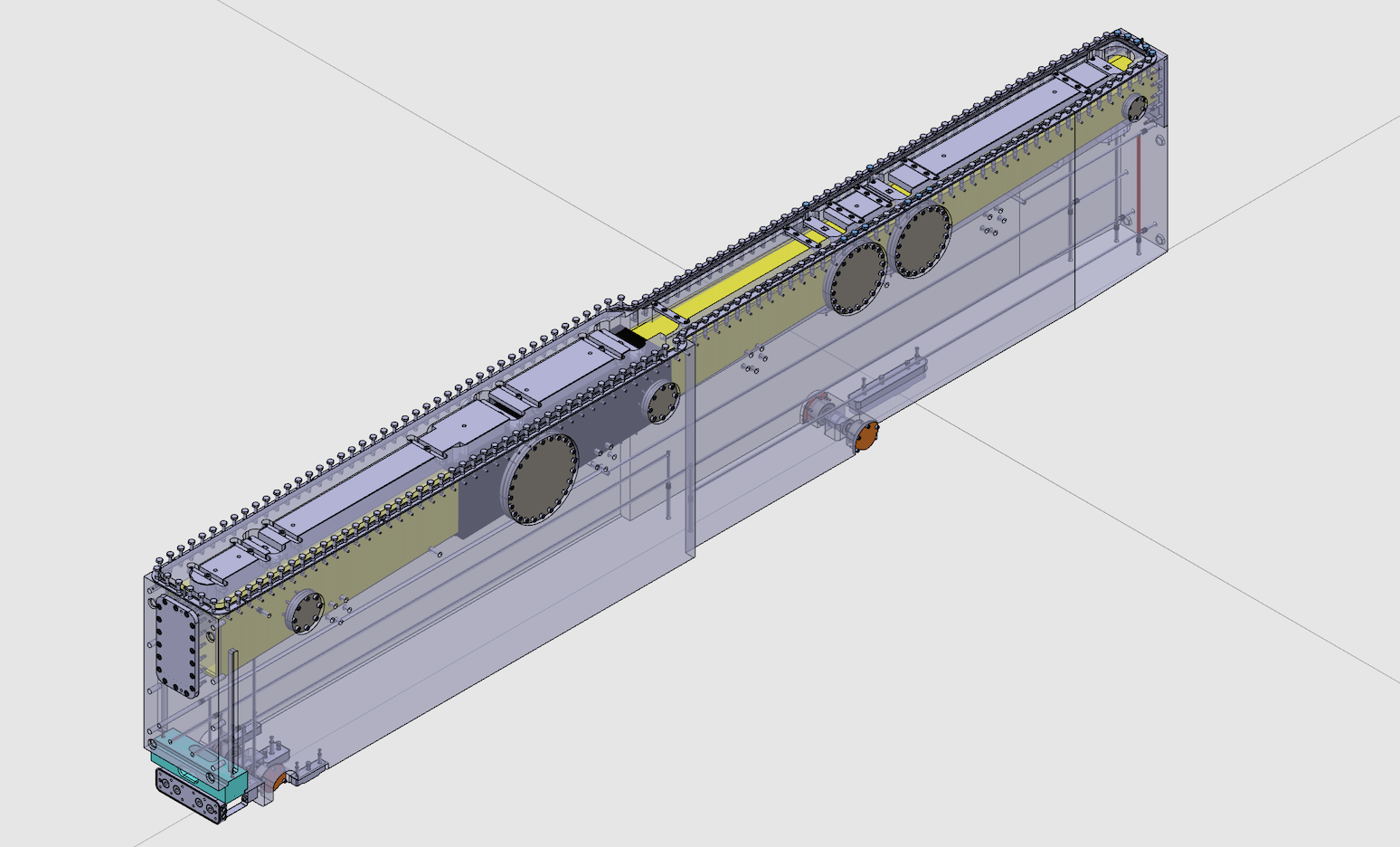}
 \caption{Neutron Extraction System CAD drawing. The NES is long 3.5 and it is installed in the ESS target monolith. The beam is coming from the left. The size at the entrance is 7.8$\times$7.8~cm$^2$ at 2~m from the moderator, and exit dimensions of 14.1$\times$12.9 cm$^2$.}
	\label{monolithinsertcad}
\end{figure} 

This NES design is currently in the engineering phase. Figure~\ref{monolithinsertcad} presents a CAD drawing of the optics and their mechanical support. The system consists of two main components: the Neutron Beam Port Insert (NBPI) and the Neutron Beam Optics Assembly (NBOA), represented by the grey and yellow structures in Figure~\ref{monolithinsertcad}, respectively.

Following the NES, an additional neutron guide is required to improve the focusing of the neutrons onto the annihilation target. The current design of this guide, shown in Figure \ref{guide}, consists of a 20~m-long guide with the shape of the reflectors following Equation~\ref{Eq:quad}. The shape was optimized, as in the previous case,
using \software{McStas} and \software{guide\_bot}, resulting in opening sizes of 16.4$\times$14.4~cm$^2$ at the entrance and 30.5$\times$23.0~cm$^2$ at the exit. With this configuration, it is possible to deliver $1.0 \times 10^{12}$~$n$/s with an average wavelength of $3.8$~\AA\ to the target, corresponding to the $FOM$ for the neutron to antineutron search of $3.35 \times 10^{9}~n\cdot s$ at 5~MW operating power for an annihilation target radius of 20~cm. This size is chosen as it allows the possibility of using the WASA calorimeter as a component of the annihilation detector, as described in Section~\ref{sec:annihilationdetector}.

\begin{figure}[bt!] 
\centering
\includegraphics[width=0.8\textwidth]{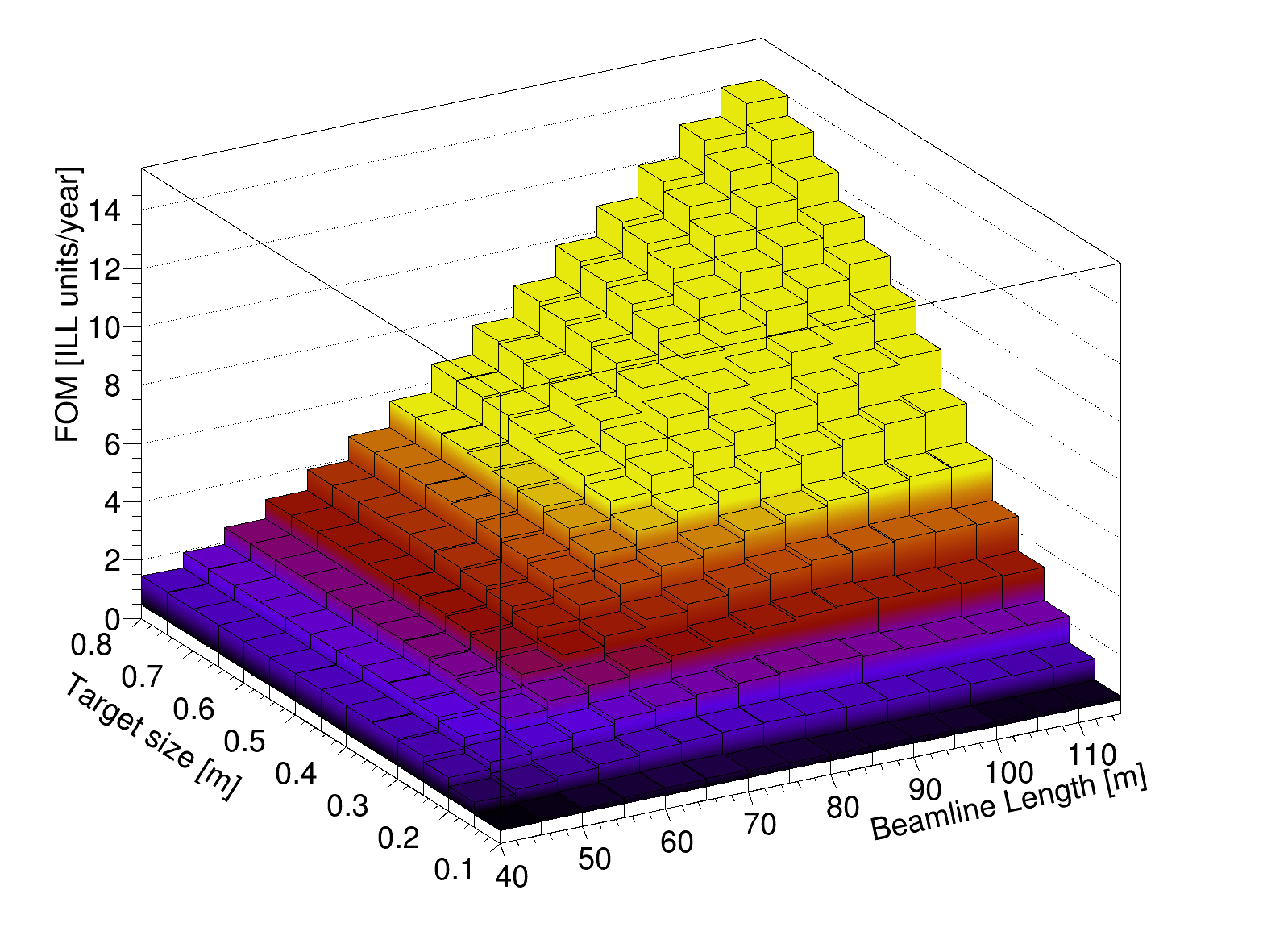}
\caption{$FOM$ (see Section~\ref{nnsearches}) for the neutron to antineutron search as a function of target size radius and the length of the beamline.}
\label{biggerdetector} 
\end{figure}

The $FOM$ depends on both the aforementioned radius and the distance over which the neutrons propagate as quasi-free particles. This is illustrated in Figure~\ref{biggerdetector}. The $FOM$/year expressed here is normalized such that one unit corresponds to that of the earlier ILL experiment. As shown in the figure, depending on the target size and propagation length, an annual $FOM$ of approximately an order of magnitude higher is, in principle, achievable. However, it is important to note that while a length of 75~m is available at the ESS within the existing instrument hall, a longer baseline would require extending the experimental hall or operating the experiment outside the instrument hall. Furthermore, the cost of the annihilation detector is strongly dependent on its size. For a larger-radius annihilation detector than the WASA-based option, a design based on the detector concept developed for NNBAR~\cite{Santoro:2024lvc} is being considered. This design incorporates a scintillator and lead-glass calorimeter along with a TPC.
The $FOM$ values stated above and shown in Figure~\ref{biggerdetector} assume an $m=4$ coating on the reflecting surfaces of the second optic system. However, using $m=3$ or even $m=2$ results in only a minor reduction of the $FOM$ ($<$5~\%).

Lastly, it is important to mention that the second optics system shown in Fig.~\ref{guide} will be used for neutron oscillation searches, ALP searches, and the neutron nonzero charge experiment. However, for future experiments such as the search for the neutron EDM, hadronic parity violation studies, or neutron lifetime measurements, this optics system must be replaced with one that eliminates the direct line of sight from the moderator. While the optimization of such a system is still ongoing and will be the subject of future work, preliminary studies indicate that the performance of an optimized bender eliminating the line of sight is comparable to that of a previous beamline design optimized for similar measurements~\cite{torsten}.

\subsection{Magnetic Infrastructure}\label{sec:magin}

As described above, after passing through the second optics system, the neutrons reach the magnetic control area. For the neutron-to-antineutron search, the magnetic field in this region must be kept below 5~nT to ensure free-flight conditions. This requirement presents a significant technological challenge. However, prototyping for an arbitrarily extendable, detachable low-frequency shield has been explored previously \cite{10.1063/1.5141340}, and ongoing efforts are focused on developing a more sophisticated prototype system (see Section~\ref{sec:magprot}).

Major challenges here are (i) extending the low magnetic field region to 50 m from the typically few meters length of most existing magnetic shields; and (ii) shielding in a constrained space and high radiation environment.

The free condition is fulfilled if the average magnetic field is below $<$ 5~nT. With a minimum particle velocity of 400 ms$^{-1}$, the flight time in the low field region is 0.1625 s, corresponding to a characteristic frequency of 6.2 Hz. It is, therefore, necessary that as well as limiting static magnetic fields, any varying magnetic fields with frequencies up to $\sim$10 Hz must also have an amplitude below 5~nT. To achieve a field-free (or quasi-free) region, a combination of magnetic field line redirection and absorption of magnetic energy through eddy currents is required. Additionally, magnetic shielding is necessary to minimize the influence of the magnetic fields from the detector and spallation target. A larger scale shield with similar specifications as for HIBEAM was designed for NNBAR~\cite{Santoro:2024lvc} using \software{COMSOL}~\cite{comsol}.   

\begin{figure}[tb]
 \begin{center}
    \includegraphics[width=0.75\textwidth]{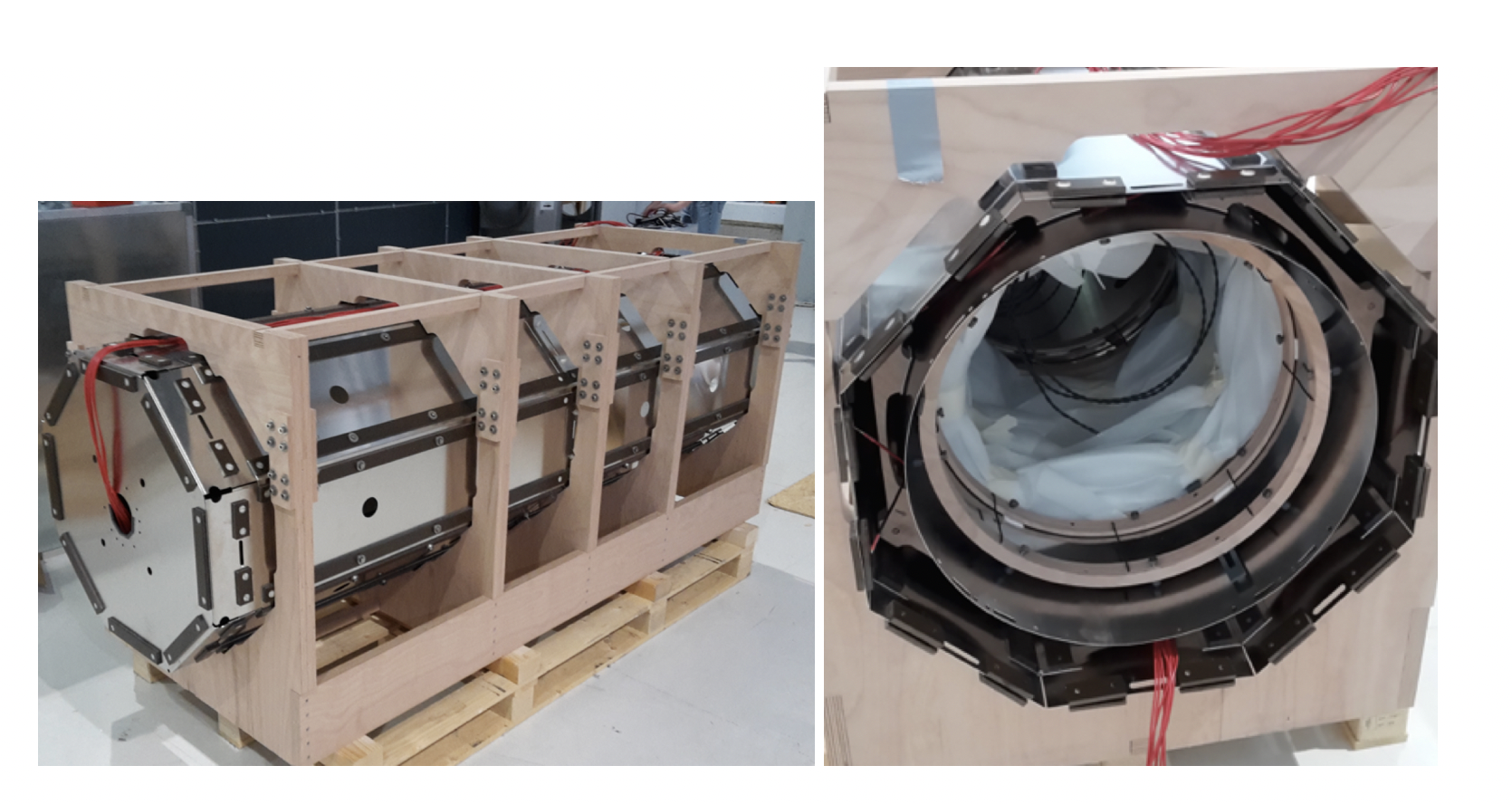}
 \end{center}
  \caption{A schematic view of vacuum chamber, magnetic shield and magnetic field coils as needed for the experiment. Inset left: Vacuum chamber and modified end section of magnetic shield for the detector region; Inset right: Detailed layout of a chamber and magnetic shield section. The magnetic shield thickness is increased in the middle region of the shield. Note the ring-shaped shield sections that “catch” magnetic flux at the ends of the shield, at connections or access holes.}
  \label{fig:magneticsshielding}
\end{figure}

Figure \ref{fig:magneticsshielding} shows the concept, which includes a vacuum chamber made from aluminum, surrounded by a two-layered magnetic shield made from mu-metal.

The aluminum vacuum chamber will have a thickness of 5 mm to withstand vacuum pressure and a diameter of 0.4 m, with high material purity to avoid neutron activation. Circling currents in the material shield external magnetic fields above 1 Hz. Access for vacuum equipment and instrumentation will be provided from the side, with 100 mm ID aluminum tubes placed periodically at 2 m spacing, which connect after 40 cm distance to pumps, placed outside the magnetic shield. The branch tubes act as waveguides to prevent low-frequency distortions from entering the inside volume, while providing sufficient vacuum conductivity to obtain $<$~10$^{-6}$~mbar inside.

The mu-metal shields are composed of an octagonal assembly in two nested shells with an inner diameter of 0.45 m and 0.75 m, from 4 individual bent segments around the circumference of maximum 750 mm width and 300 mm length in axial direction. The segments overlap by 50 mm and can be assembled after the vacuum chamber is installed to provide flexible access for instrumentation of the beam line. 

To achieve a residual field of $<$ 5 nT within the entire volume of the vacuum chamber, the remanent magnetization of the passive magnetic shield must be reduced through magnetic equilibration. With state of the art equilibration \cite{9075432} by a sinusoidal current with linearly decreasing envelope over 20 s into a set of toroidal coils wound around the inner octagonal shield (also in independent sections of the shield), this goal is feasible (as demonstrated in \cite{10.1063/1.5141340}). 
To achieve optimal performance along the beamline, the increased diameter of the shield ensures that remanent fields near the shield surface do not exceed 5 nT inside the vacuum chamber. The shield will be mounted at 1.5 m intervals using screws on detachable plastic frames, which mechanically and electrically isolate the shields from the vacuum chamber and from each other. Additionally, the shields will be thicker in the middle third of their length to compensate for the increased flux from the Earth's magnetic field. However, in the free flight region, the magnetic field requirement is only $<$ 5~nT, which is achieved by a separated mu-metal ring-section placed to catch residual flux from the detector, while not guiding the flux into the shielded region. Also, for the vacuum chamber, additional constraints apply to the detection region with the annihilation target inside the vacuum chamber and the detector placed outside. Due to the geometry of the annihilation target and the stringent background requirements, the quality of the reconstruction of the reaction particles must be of very high quality.  In turn, the material thickness of the chamber must be small to avoid angular deflection $\theta_0$ of the annihilation products passing the material due to Moliere scattering. For aluminum with 5 mm wall thickness, this effect is small enough to be accommodated with the reconstruction precision required by the 100 $\mu$m thick annihilation target, taken into account in the design of the vacuum chamber. Here, the chamber wall is milled down to the required thicknesses and deformed due to the vacuum pressure, with only thick strips of material left to match the structural stability requirements. The loss in shielding efficiency for 1-100 Hz due to the reduced material thickness does not extend more than 0.5 m into the shield. With the specific requirements for this setup, the experiment is also a future technology demonstrator in its magnetic aspects: the ring-geometry for field reduction was specifically developed for the HIBEAM instrument, which enables the modularization of magnetic shields, easy access through shields for pumping and sensing, as well as minimization of large stray fields at the end of the shields. 
Thickness variations of the shield layers and the segmentation of the magnetic equilibration procedure were also specifically investigated and developed in preparation for this project.  Magnetic equilibration is also done to ensure stability of applied magnetic fields inside the shield.

 In sterile neutron searches, magnetic coil systems are used to generate both longitudinal and transverse magnetic fields. Transverse fields up to 300~$\mu$T are generated using 2 sets of so-called cosine-theta coils with 90$^{\circ}$ rotated alignment with a homogeneity of $\Delta B / B < 10^{-3}$; the longitudinal field is generated with a loosely wound solenoid axially with the shield and vacuum chamber. Here, the homogeneity is improved by the shield, which forms a configuration known as “magic box” when used with lids and short aspect ratio, resulting also in close to $10^{-3}$ relative homogeneity. Magnetic characterization of the shield is done using fluxgate magnetometers, which are moved through the shield on a trolley. Magnetic equilibration resets the field distribution and amplitude to typically within 1~nT after any arbitrary treatment in the shield, extrapolating from experience~\cite{10.1063/1.5141340}. Online magnetic characterization can optionally be conducted using polarized neutrons, similar to the method in~\cite{SCHMIDT1992569}, where polarized neutrons and a spin-echo technique were employed. In this setup, this approach could be further advanced by deploying a small $^3$He cell or removable polarizing optical components at the entrance of the shielded section to polarize neutrons at different positions along the beam. A 0.5 m-long RF section, operating at a few hundred Hz at 10~$\mu$T for $\pi/2$ flipping, placed at the beginning of the beamline and a second one at the end, could enable spatial online mapping while also being used for performing the ultralight axion dark matter searches described in Section~\ref{ultraaxion} and Ref.~\cite{PhysRevLett.133.181001}.

\subsubsection{Prototype of the magnetic infrastructure}\label{sec:magprot}

As discussed above, achieving a low-field region requires optimizing the magnetic shielding structure, including increasing the shield thickness in the middle region to compensate for residual flux. To further develop this concept, a 10-meter prototype of the magnetic infrastructure is currently being designed and later will be build and tested. The geometry of this prototype has been simulated to ensure that the shielding provides an approximately uniform field distribution inside, requiring a symmetrical design while maintaining manufacturability and ease of assembly.

\begin{figure} \centering \begin{subfigure}{.7\textwidth} \centering \includegraphics[width=0.9\textwidth]{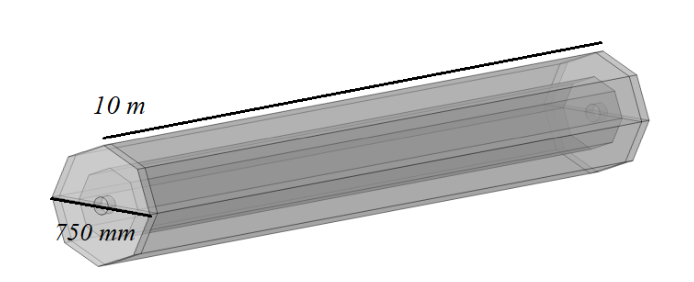} \caption{Overview of the shield geometry.} \label{fig:sub1} \end{subfigure} \vspace{0.2cm} \begin{subfigure}{.7\textwidth} \centering \includegraphics[width=0.5\textwidth]{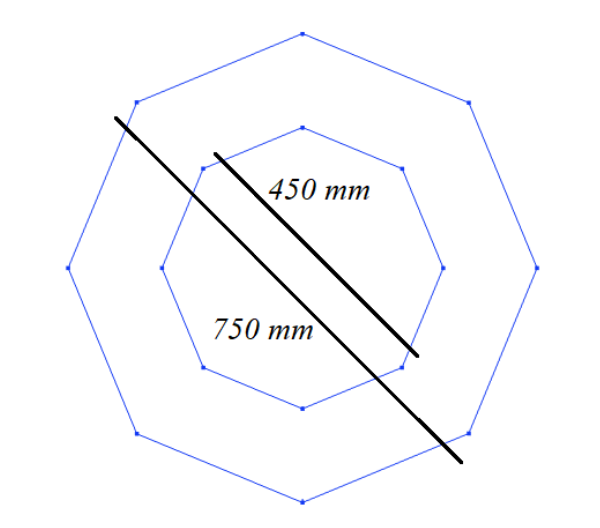} \caption{Transversal cut showing the two nested shielding layers.} \label{fig:sub2} \end{subfigure} \begin{subfigure}{.7\textwidth} \centering \includegraphics[width=1.0\textwidth]{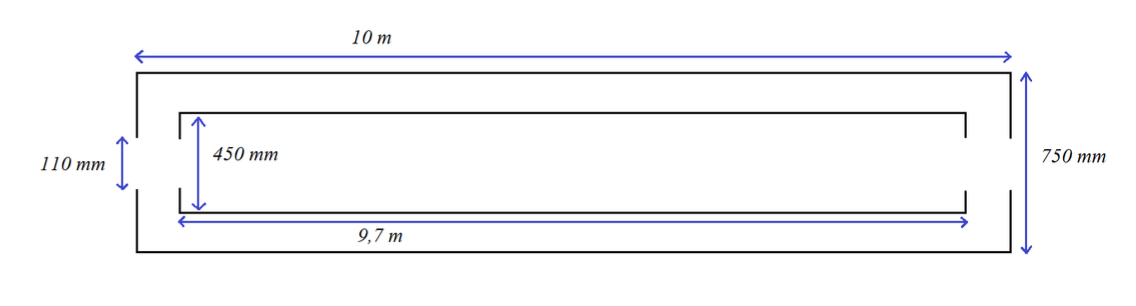} \caption{Schematic blueprint of the nested shielding configuration.} \label{fig:sub3} \end{subfigure} \caption{Visualization of the 10-meter shielding prototype.} \label{fig:magprot} \end{figure}

This prototype is being developed with support from the Swedish Foundation for Research Strategy. Its reference geometry is based on previous designs~\cite{10.1063/1.5141340}, where passive shielding consists of two nested octagonal tubes measuring 9.7 meters and 10 meters in length. The inner and outer tubes have circumference diameters of 450 mm and 750 mm, respectively, as shown in Fig.~\ref{fig:magprot}.

Simulations conducted with a uniform magnetic shield thickness revealed a local magnetic field maximum at the center of the tube. This observation confirmed the need for a design modification: to achieve a flat magnetic field profile, the mu-metal thickness should increase progressively towards the center of the shield.

\begin{figure} \centering \includegraphics[width=0.9\textwidth]{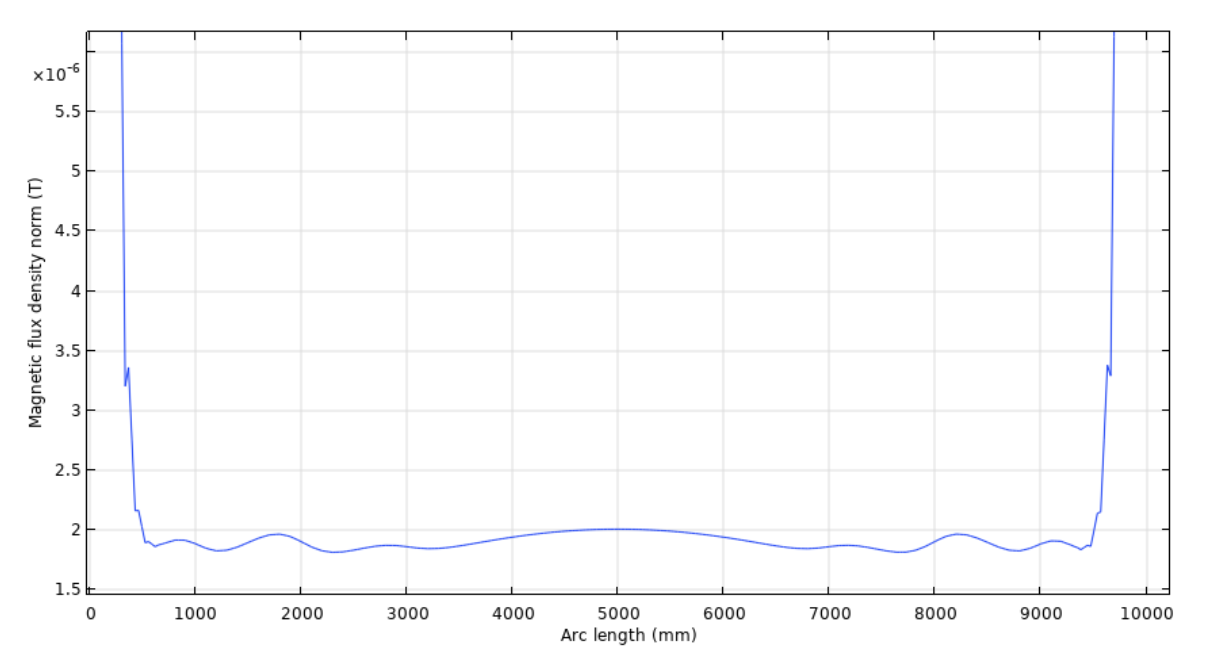} \caption{Magnetic flux density (T) measured along the central axis of the shield in a model with progressively increasing mu-metal thickness.} \label{fig:magprotres} \end{figure}

Figure~\ref{fig:magprotres} illustrates the effect of implementing this approach. The shield is divided into sections of varying thickness, increasing toward the middle, effectively reducing field variations and creating a more homogeneous shielding environment. While the current simulations focus solely on passive shielding, further studies will integrate active magnetic coils to perform detailed field scans for sterile neutron searches and potential applications in ultralight axion dark matter detection.

\subsection{Beamline simulation} \label{sec:beamline}
To ensure that HIBEAM meets all relevant radiological requirements, beamline simulations have been performed. The ESS requirement is that the dose rate on the shielding surface may not exceed \SI{3}{\micro\sievert/\hour}. With a safety factor of 2 being applied for all Monte Carlo simulations, an ultimate threshold of \SI{1.5}{\micro\sievert/\hour} has guided the design of the shielding.

\subsubsection{Beamline geometry} \label{beamlinegeom}

\begin{figure}[tb] 
\centering
\includegraphics[width=\textwidth]{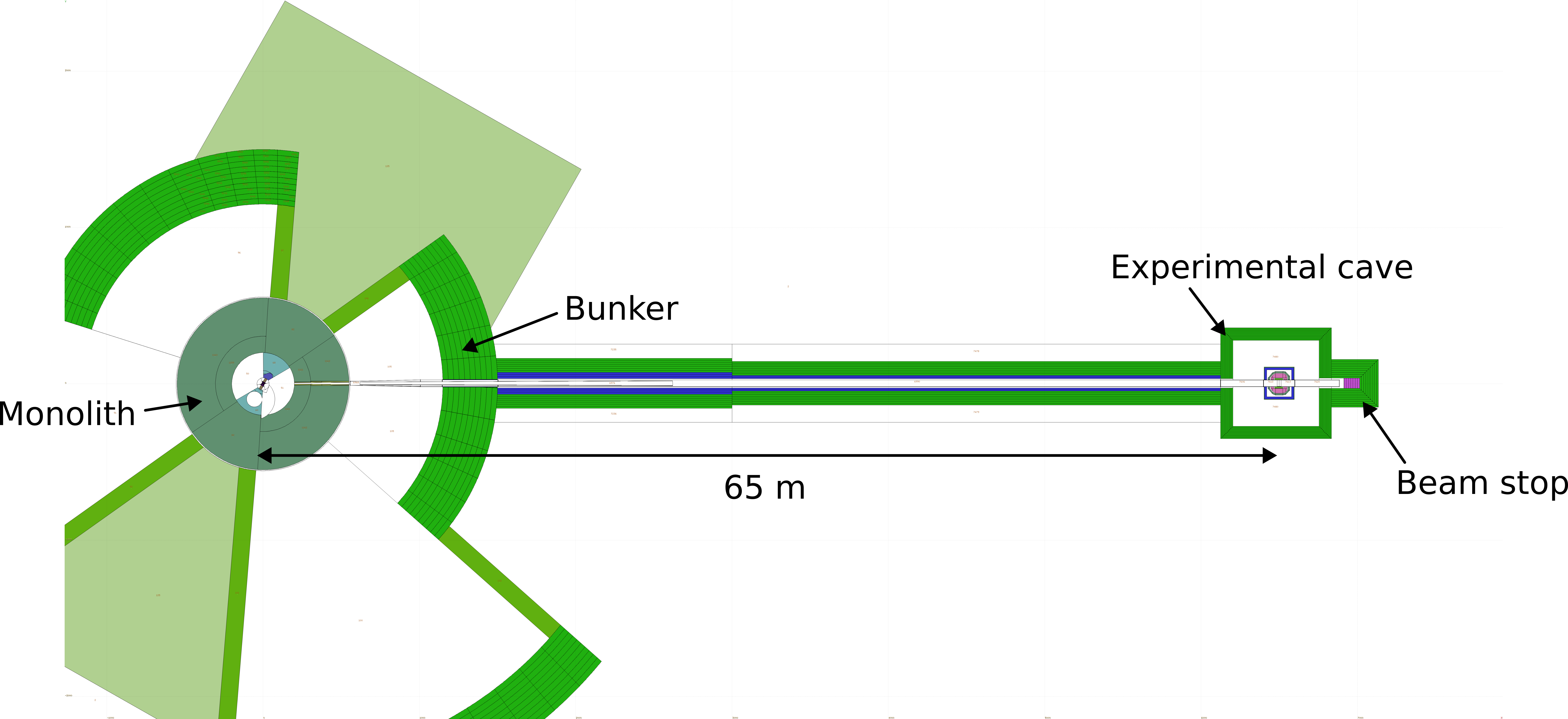}
\caption{Overview of the geometry developed for HIBEAM beamline simulations.}
\label{beamlineoverview} 
\end{figure}

\begin{figure}[tb] 
\centering
\includegraphics[width=0.8\textwidth]{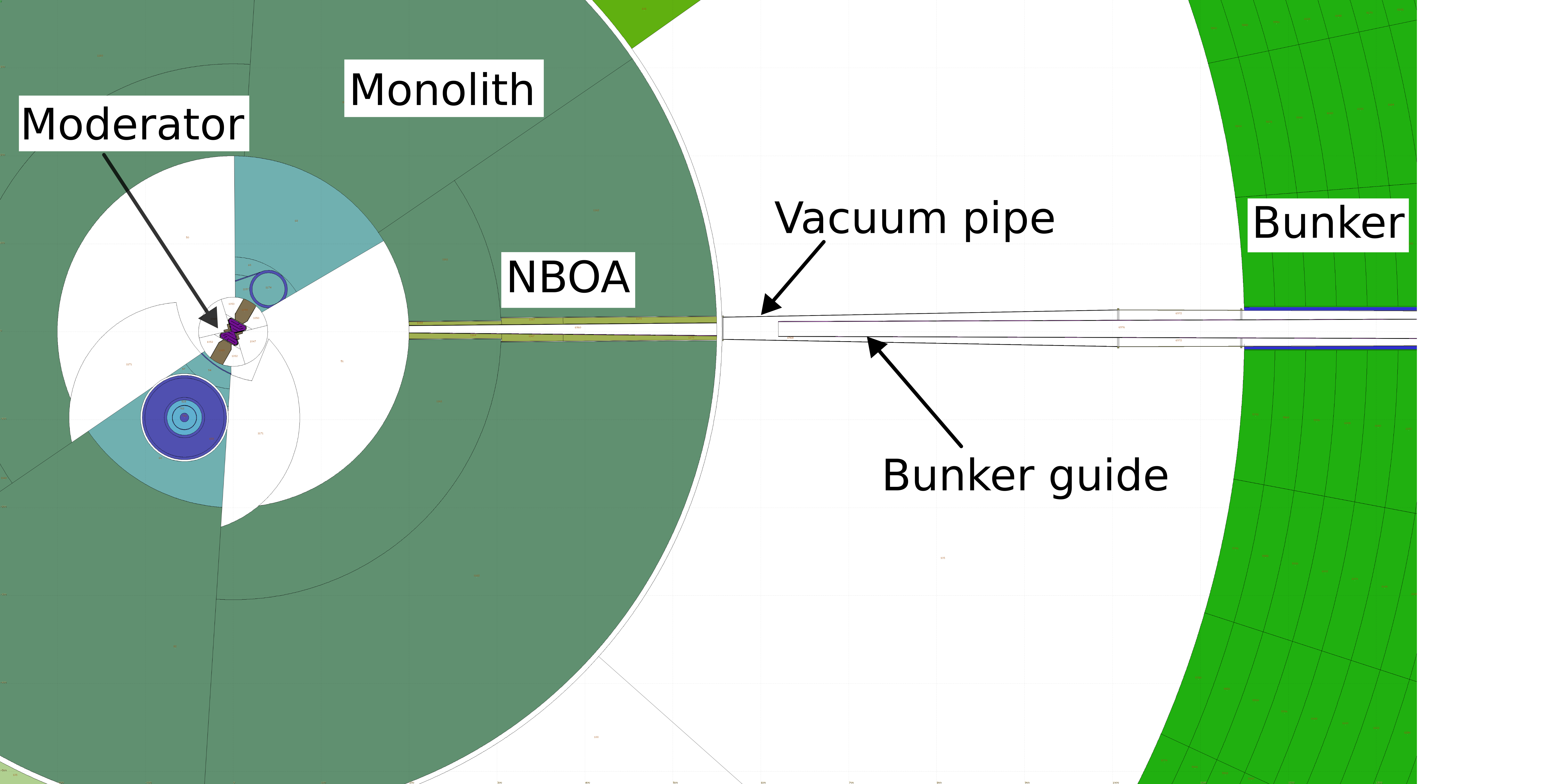}
\caption{The image provides a close-up view of the beamline within the target-bunker area, highlighting the moderator, the NBOA, and the guide inside the bunker. Details of the simulations are provided in the text.}
\label{beamlinebunker} 
\end{figure} 

The beamline geometry was constructed using \software{CombLayer}~\cite{CombLayer}, a C++ geometry constructor designed for handling intricate and highly parametric models. Simulations were performed using \texttt{PHITS} \cite{PHITSref} (see Section~\ref{beamsimcode} for details on the simulation setup). Figure \ref{beamlineoverview} provides an overview of the model, while Figure \ref{beamlinebunker} focuses on the first part of the beamline in the target-bunker area. The simulated geometry includes a detailed model of the ESS moderator, the surrounding monolith and bunker (see Section~\ref{sec:ess}), and the optics system described in Section~\ref{sec:optics}.

\begin{figure}[bt!] 
\centering
\includegraphics[width=0.7\textwidth]{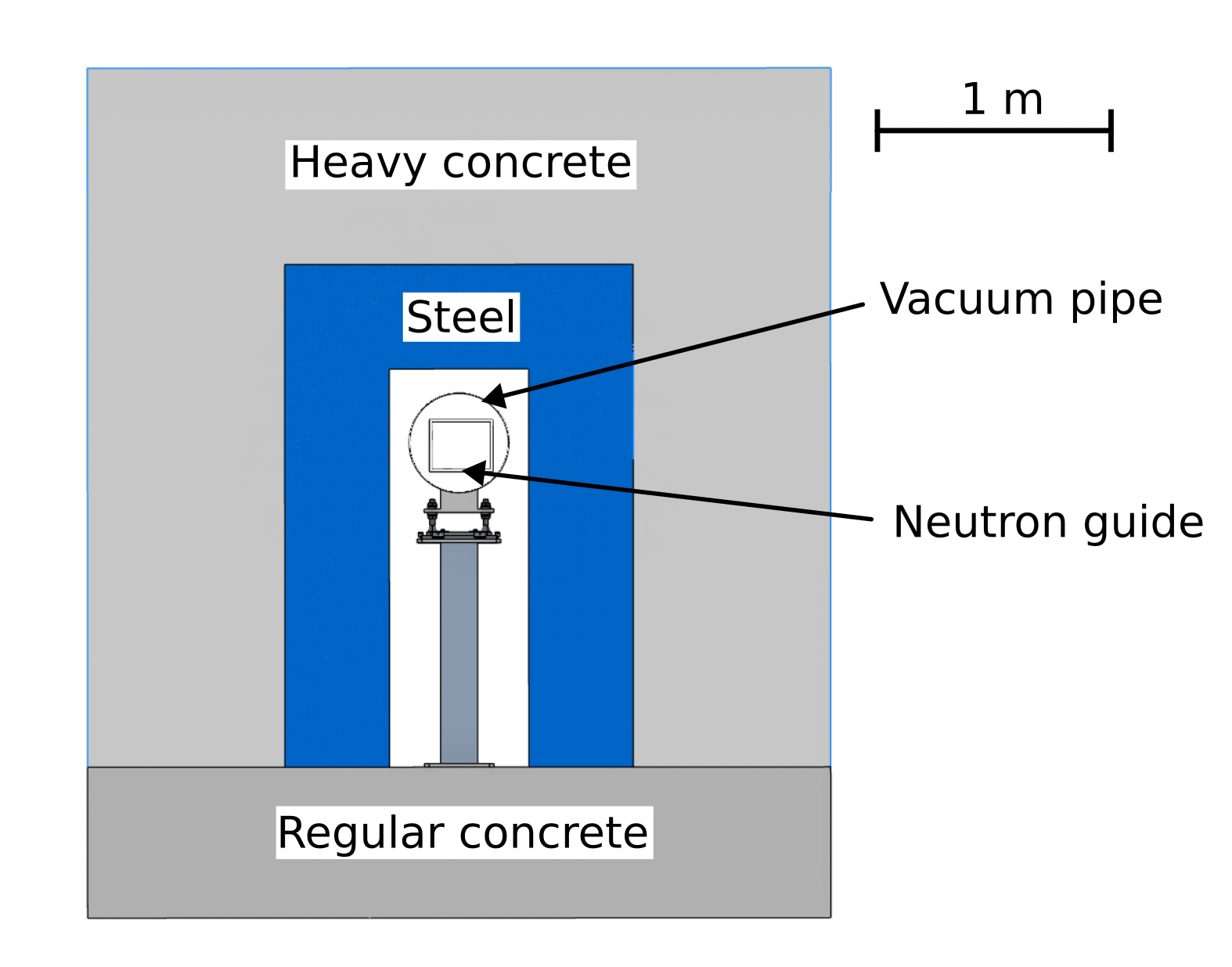}
\caption{Cross-section of the beamline shielding outside the bunker, showing the usage of steel and heavy concrete around the vacuum pipe.}
\label{crossec} 
\end{figure}

Outside the bunker, the beamline is enclosed in dual-layer shielding consisting of steel and heavy concrete on the sides and top, while the regular concrete facility floor lies below, as shown in Figure \ref{crossec}. In the first 12 meters outside the bunker, the shielding consists of 45 cm of steel and 85 cm of heavy concrete. Beyond this section, extending up to the experimental cave, the shielding is reduced to 30 cm of steel and 80 cm of heavy concrete.

\begin{figure}[bt!] 
\centering
\includegraphics[width=0.8\textwidth]{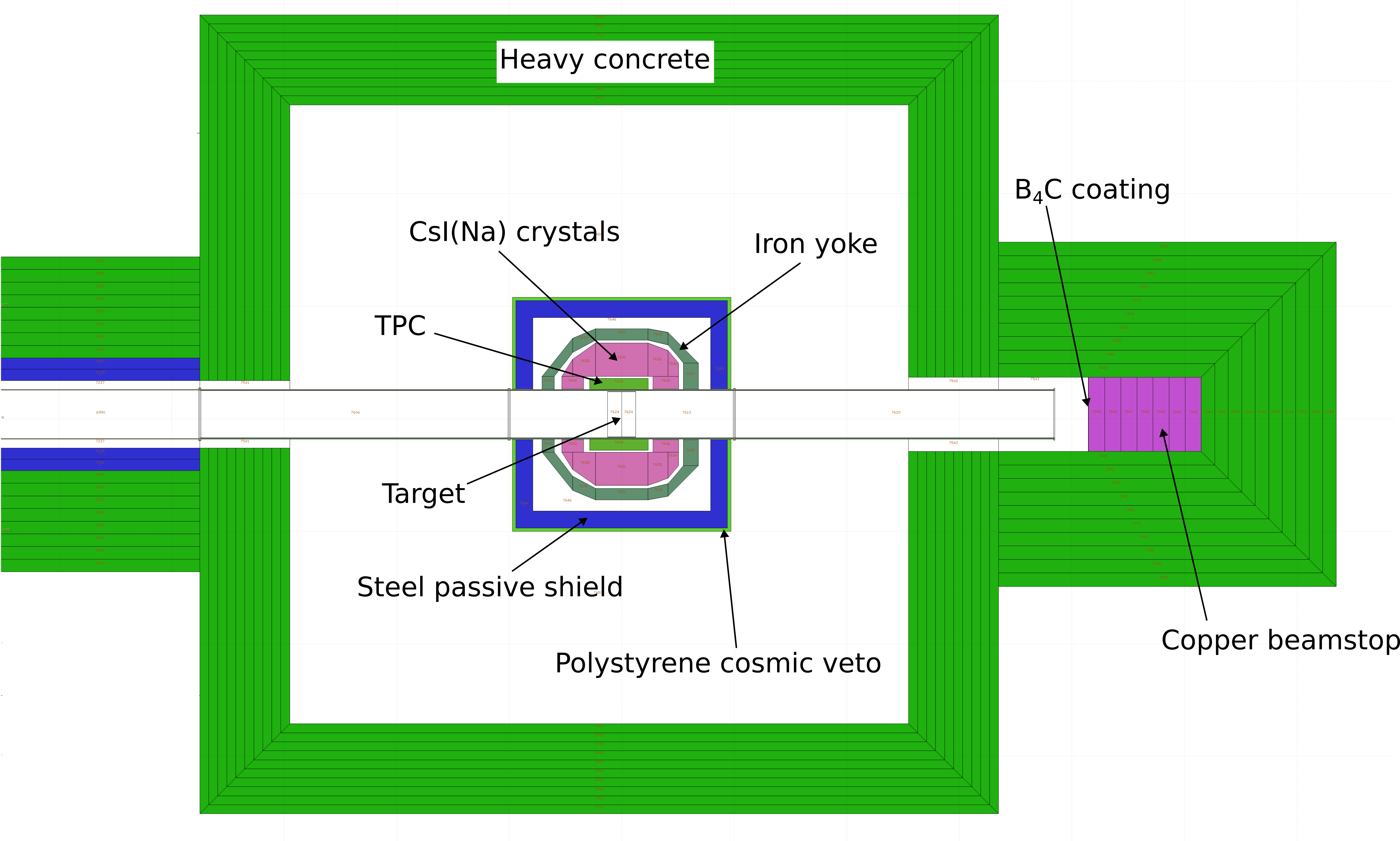}
\caption{View of the experimental cave as modeled for the beamline simulations.}
\label{expcave} 
\end{figure}

The experimental cave, shown in Figure \ref{expcave}, includes a simplified model of the annihilation detector described in Section~\ref{sec:annihilationdetector}. The cave walls are constructed from 80 cm of heavy concrete. Positioned behind the detector is a beam stop consisting of 1 meter of copper with a 2.5 mm coating of \ce{B_4C}. The heavy concrete wall behind the beam stop is 1.1 meters thick, while the side walls are 1.2 meters thick.

\subsubsection{Neutron source construction and variance reduction}
\begin{figure}[bt!]
    \centering
    \subfloat{{\includegraphics[width=0.5\linewidth]{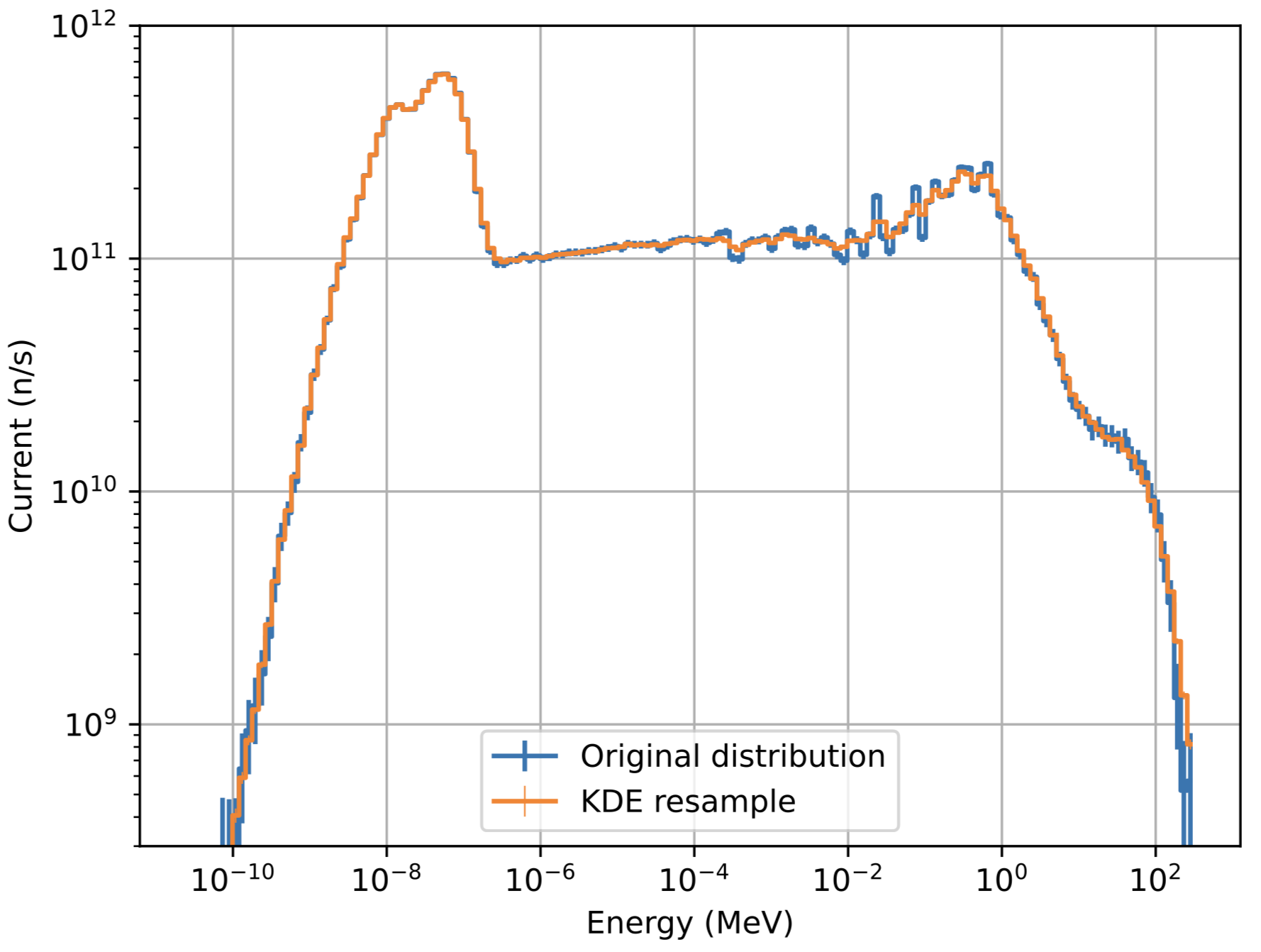} }}%
    \subfloat{{\includegraphics[width=0.5\linewidth]{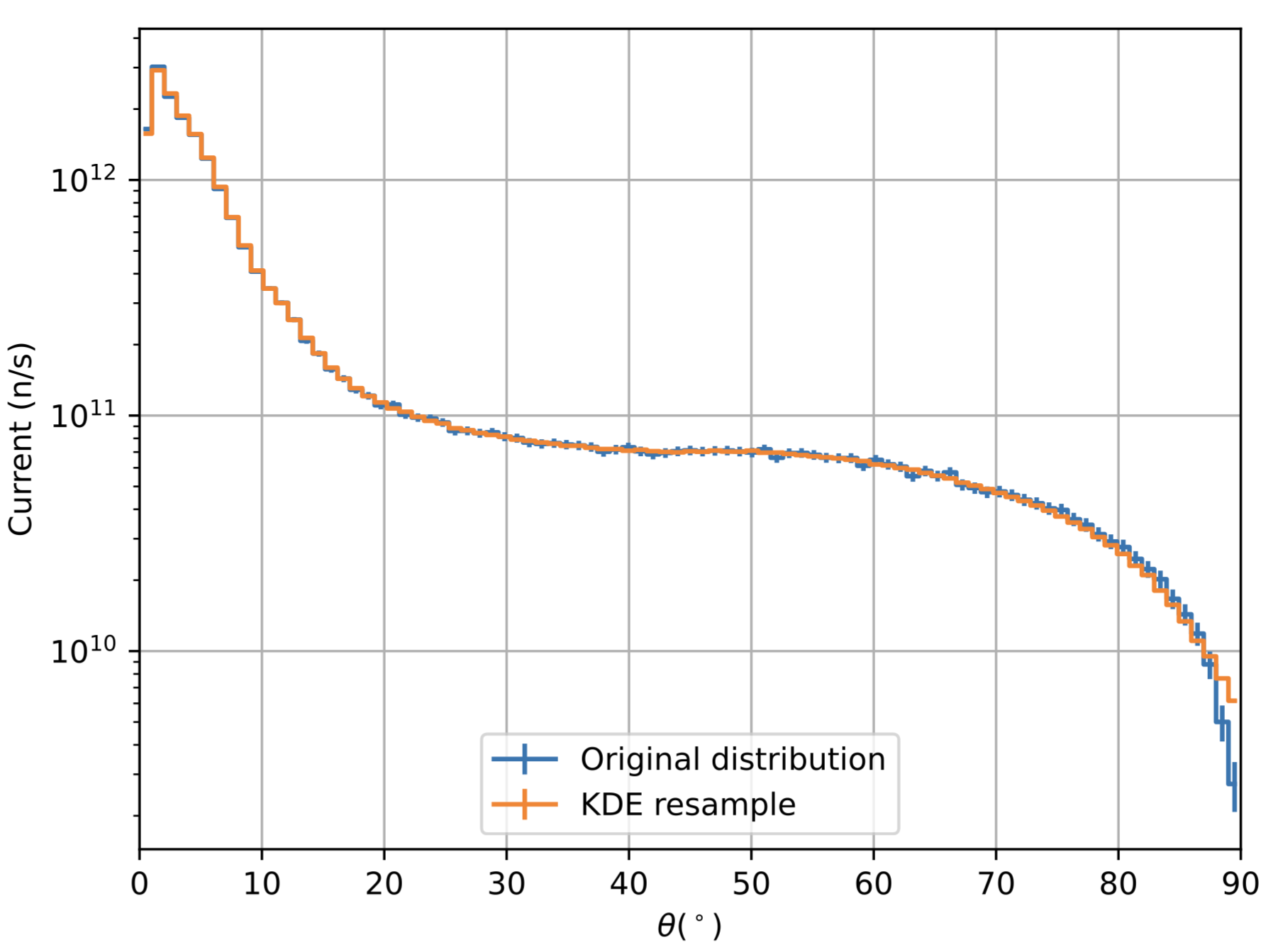} }}%
    \caption{Panel (a) shows the energy spectrum of the neutrons entering the HIBEAM beamport in blue, and the resampled and normalised distribution obtained using multivariate KDE in yellow. Panel (b) shows the corresponding distribution in polar angle with respect to the beamline axis.}
    \label{spec}
\end{figure}

Initial simulations were performed using a source of monoenergetic (2 GeV) protons incident on the ESS tungsten targets. 
The neutrons entering the HIBEAM beam port were recorded in an \software{MCPL} (Monte Carlo Particle Lists) file \cite{MCPLref}. To generate an unlimited number of neutrons with reasonable usage of computational resources, multivariate kernel density estimation (KDE) was performed using \software{KDSource}~\cite{KDSource}. Gaussian kernel functions were used with the optimised variables being the lethargy as well as the positions and directions of the neutrons in Cartesian coordinates. The energy and polar angle distributions (with respect to the beamline axis) at the entrance of the NES of the original and resampled distributions are compared in Figure \ref{spec}. Satisfactory agreement is noted, and this was further established by flux comparisons in other regions of the beamline between simulations from protons and simulations using KDE resampling.

Further variance reduction was achieved using cell-based geometry splitting. The shielding was divided into 10 layers, with the cell importance increasing by a factor of 2 per layer, resulting in an importance factor of 1024 in the void region outside the beamline.

\subsubsection{Beamline simulation specification and results}
\label{beamsimcode}
\begin{figure}[bt!]
    \centering
    \includegraphics[width=1.2\textwidth]{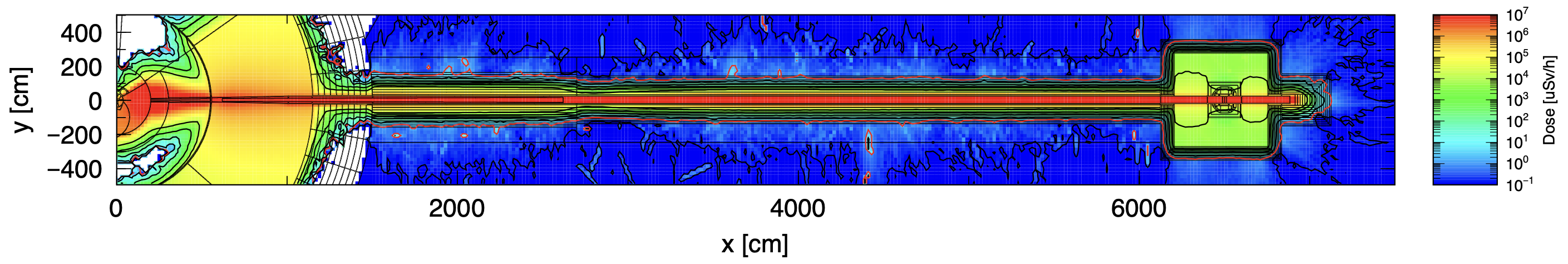}
    \caption{Dose map showing the total dose rate contributions from neutrons and photons, calculated using 1 billion neutrons generated with multivariate kernel density estimation. The red contour line represents the dose rate limit of \SI{1.5}{\micro\sievert\per\hour}. Note that the color scale on the right also includes red, corresponding to higher dose values.}
    \label{beamlineres}
\end{figure}

Simulations were performed using \texttt{PHITS} version 3.33. Neutron interactions above 20 MeV are modelled using the intranuclear cascade model INCL-4.6 \cite{INCL46} combined with the Generalized Evaporation Model~\cite{GEM}, while low-energy neutron interactions and photon interactions are treated using the nuclear data library ENDF/B-VIII.0~\cite{ENDF}. Fluxes of neutrons and photons were tallied in cells of volume $20 \times 20 \times 20 \, \SI{}{cm^3}$ and converted to effective dose rates using the whole-body conversion coefficients from ICRP 116 \cite{ICRP}, with the worst-case irradiation configuration being used for each energy bin. The dose contributions from neutrons and photons are summed. An example of a resulting dose map, is shown in Figure \ref{beamlineres}. The red contour line indicates the dose rate limit of \SI{1.5}{\micro\sievert\per\hour}. The present design satisfies the ESS dose rate requirements along the entire beamline. 

\section{Detectors}
\label{sec:det}

To carry out the HIBEAM program, both neutron and antineutron detectors are required, as shown in Fig.~\ref{beamlineprinciple}. For the neutron detector system, depending on the experiment, different requirements apply: a detector capable of handling very high rates is necessary for the disappearance mode, the axion-like particle search, and the search for nonzero electric charge of the neutron experiment. Two possible solutions are described in Sections~\ref{sec:neutrondet1} and~\ref{sec:neutrondet2}. For the regeneration mode, a standard neutron detector with low background will suffice, as this is a widely used technology will not be discussed further in this work. The antineutron detector is detailed in Section \ref{sec:annihilationdetector}.

\subsection{Neutron detector for the disappearance mode: the current mode option}\label{sec:neutrondet1}

The very high instantaneous neutron rates in the neutron detectors required for different experimental configurations in this work are far too high to count individual neutron pulses (see Section \ref{sec:neutrondet2} for proposed future studies on alternative detection methods). The most practical solution is to use current-mode detection. Since one cannot apply the usual types of signal discrimination in this case, one must be careful to ensure that the extra noise in the measured current from background processes and from fluctuations in the number of current-generating quanta from each individual neutron capture in the detector are both small compared to $\sqrt{N}$ where $N$ is the number of neutrons. The burst of fast neutrons from the ESS target/moderator system is gone by the time the slow neutrons of interest arrive at the detector, so the most serious background for neutron detectors in the direct beam comes from gammas. It is therefore important for the detector to be able to withstand fast neutron bombardment with no ill effects and be insensitive to gammas. It should also incorporate a method to subtract the gamma-induced component of the neutron detector signal without using the usual method of pulse shape discrimination.

\begin{figure}[ht]
 \begin{center}
    \includegraphics[width=0.75\textwidth]{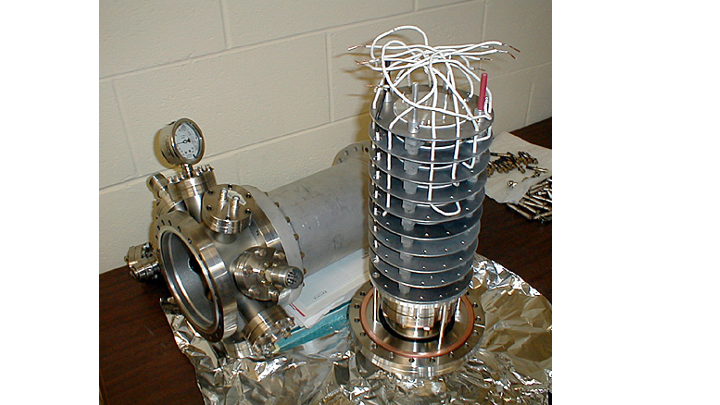}
 \end{center}
  \caption{Disassembled current-mode $^{3}$He ion chamber used in the n-$^{4}$He parity violation experiment~\cite{PhysRevC.100.015204}, showing the ultra-high vacuum (UHV) chamber and ion chamber plates.}
  \label{fig:ionchamber}
\end{figure}

An ion chamber filled with $^{3}$He gas, utilizing the $n+^{3}$He $\to ^{3}$H $+p+0.78$ MeV reaction, serves as a gamma-insensitive and radiation-damage-resistant detector for neutrons. It has a high absorption cross-section for slow neutrons and releases enough energy per neutron capture to produce a strong ionization signal. As a result, the detector's noise remains low and is primarily limited by the intrinsic statistical fluctuation of the detected neutron count, which follows the $\sqrt{N}$ scaling. The small fraction of the signal induced by background gammas in the beam in such a $^{3}$He ion chamber can be subtracted by placing an identical $^{4}$He ion chamber just in front of the $^{3}$He ion chamber. To an excellent approximation, the induced signal in the $^{4}$He ion chamber comes only from the gammas in the beam, and a $^{3}$He ion chamber with the same dimensions and gas pressure directly behind the $^{4}$He ion chamber will possess a nearly identical response to gammas. Therefore, one can subtract these two signals to leave only that from the neutrons. Szymanski et. al~\cite{Szymanski:1994nima} successfully demonstrated this concept at the LANSCE (Los Alamos Neutron Science Center) pulsed spallation neutron source. Possible backgrounds from neutron activation of ion chamber components by high energy neutrons and gammas can be quantified at ESS using the signal from the fraction of the time-of-flight frame outside the slow neutron arrival time and also by taking data during dropped pulses from the ESS proton accelerator. Fast neutrons in the beam appear too early in the time-of-flight frame to be confused with the slow neutrons of interest for this work.

Addition of other species to the gas can reduce the range of the $^{3}$H and $p$ to a few mm for easily-achievable gas pressures. A helium-argon gas mixture in the segmented ion chamber was developed for polarized slow neutron spin rotation measurements of parity violation~\cite{PhysRevC.100.015204}. Figure~\ref{fig:ionchamber} shows a picture of the detector used in this measurement. The detector ion collection plates can be segmented transversely to the beam to enable imaging and to reduce common-mode noise caused by pulse-to-pulse fluctuations in the beam phase space, which are not captured by the low-efficiency upstream monitor. The ion chamber detector can also be segmented along the beam direction to take advantage of the fact that the slower neutrons are more likely to be absorbed at the entrance to the chamber due to the $1/v$ behavior of the neutron absorption cross section. Ionization signals from different depths into the detector therefore give information on the neutron energy spectrum. The comparison of this data with the neutron time-of-flight information enables a check of boundary effects in the ion chamber response from ions that are lost to the collection plates by absorption in materials.  
The required precision and stability of the total charge measurement from the absorbed neutron beam should be ${10}^{-7\ }$ for this work~\cite{Addazi:2020nlz}. This level of precision applies to both short-term stability (within a single measurement cycle) and long-term stability (over multiple experimental runs). A $^{3}$He-based neutron ion chamber of design similar to that described above was used in a recent neutron-$^{3}$He parity violation experiment at SNS which reached ${10}^{-8\ }$ precision~\cite{n3He:2020zwd}. This success bodes well for our ability to meet this specification. The stability of the efficiency of such an ion chamber can be quite high as it depends only on chamber geometry and stability of ion collection fields, constant gas composition and density, and electronics properties. The current mode signals are large enough that no electronics gain stage needs to be employed: one needs only a current-to-voltage converter.  

A detector with variable efficiency can be achieved by adjusting the $^{3}$He gas pressure. 
Commercially available clean metal bellows compressors allow for the safe handling of tritium gas,  as a nontrivial amount is generated in the $^{3}$He ion chamber. For HIBEAM, the expected tritium production due to neutron flux is on the order of tens to a few hundred mCi per year, depending on gas pressure and chamber volume. This amount is manageable using standard tritium safety procedures, and the gas can be safely removed at the end of the experiment using hydrogen getters, as demonstrated in~\cite{n3He:2020zwd}.

\subsection{Neutron detector for the disappearance mode: the single neutron counting mode}
\label{sec:neutrondet2}
As described above, the single-neutron counting mode presents a significant challenge, and current technology is not yet capable of achieving it. However, it offers substantial advantages for the HIBEAM disappearance experiment by enhancing background rejection, statistical sensitivity, and systematic consistency checks. For instance, the ability to perform precise time-of-flight (ToF) measurements enables the separation of different velocity components of neutrons, allowing genuine disappearance signals to be distinguished from spurious effects caused by beam fluctuations. Additionally, the segmentation of the detector along its length facilitates multiple independent neutron measurements to be combined, reducing uncertainties beyond the Poisson limit and significantly improving the statistical significance of the experiment. The longitudinal imaging capability further enables the verification of expected beam intensity and spectral variations, allowing for consistency checks in signal identification and background suppression. Additionally, the single-neutron counting mode optimizes detector performance by fully utilizing the neutron spectrum while ensuring that the readout electronics can handle high instantaneous count rates without saturation. Despite the technical challenges associated with its implementation, these advantages make it a highly promising approach. Consequently, while current-mode detection is the established baseline for the disappearance detector, the HIBEAM collaboration is actively exploring strategies to implement single-neutron counting, as described below.

\subsubsection{Detector requirements}

The main requirements for such a detector are in terms of its efficiency, which must be sufficiently stable to ensure efficiency variations are not mistaken for neutron disappearances. Given Equation~\ref{eq:free}, this means the detector efficiency should be stable to one part in $10^{7}$. Such an extreme value suggests adopting a detector with efficiency very close to $1$. This is particularly challenging given the high neutron flux: if operated in single neutron counting mode, the detector is effectively required to handle rates around $10^{11}~\mathrm{s}^{-1}$.

Another issue is that the ESS itself is not stable to within $10^{-7}$. This necessitates normalizing measurements to each individual neutron pulse, which requires not only the design of the disappearance detector itself but also the inclusion of a beam monitor located at the beginning of the magnetic control infrastructure. Additionally, the stability of the detector must be carefully monitored over long operational periods to mitigate systematic errors and environmental influences, such as temperature fluctuations or radiation-induced degradation of detector components. 

This stability requirement is the same as the requirement for the total charge measurement from the absorbed neutron beam, which must also be maintained at $10^{-7}$ precision to ensure consistency across all detection methods.



\begin{figure}[ht]
 \begin{center}
    \includegraphics[width=0.45\textwidth]{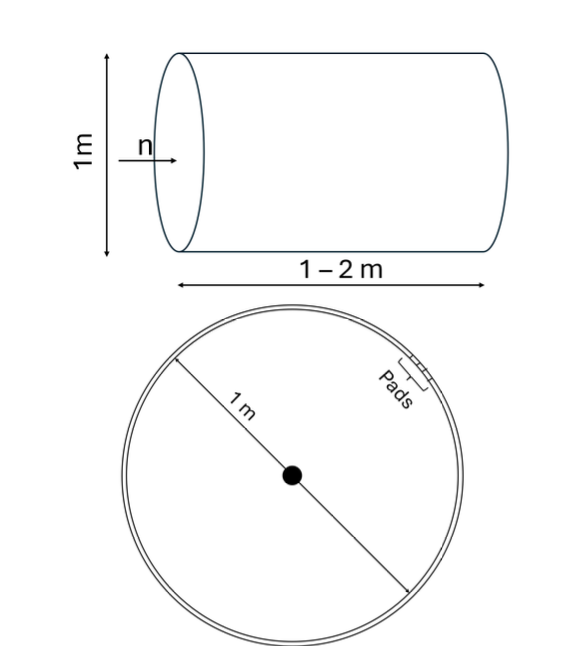}
 \end{center}
  \caption{Overview of the proposed concept for the neutron disappearance experiment using single-neutron counting.}
  \label{fig:nedet}
\end{figure}

\subsubsection{Detector concept}

The preliminary design for the detectors used in the disappearance measurement consists of ${}^{3}$He-based gas detectors with a cylindrical geometry, operating in drift mode (see Fig.~\ref{fig:nedet}). Given the large footprint of the HIBEAM neutron beam, the cylinder diameter is set to 1 m. The neutron beam, propagating parallel to the detector axis, interacts with the gas mixture (e.g., containing \ce{^3He}), leading to the production of charged particles via the reaction $n+^{3}$He $\to ^{3}$H $+p+0.78$  as described previously. These charged particles subsequently generate electrons through collisions with the gas molecules.  A central cathode and an external anode establish a radial drift field for these electrons, inducing a signal at the external anode. The detectors operate in single-neutron-counting mode with a segmented (padded) readout and employ a Gas Electron Multiplier (GEM) for signal amplification. Notably, the signal is carried by electrons rather than ions, distinguishing this design from some previous proposals.

\begin{figure}
\centering
\begin{subfigure}{.7\textwidth}
  \centering
  \includegraphics[width=0.85\textwidth]{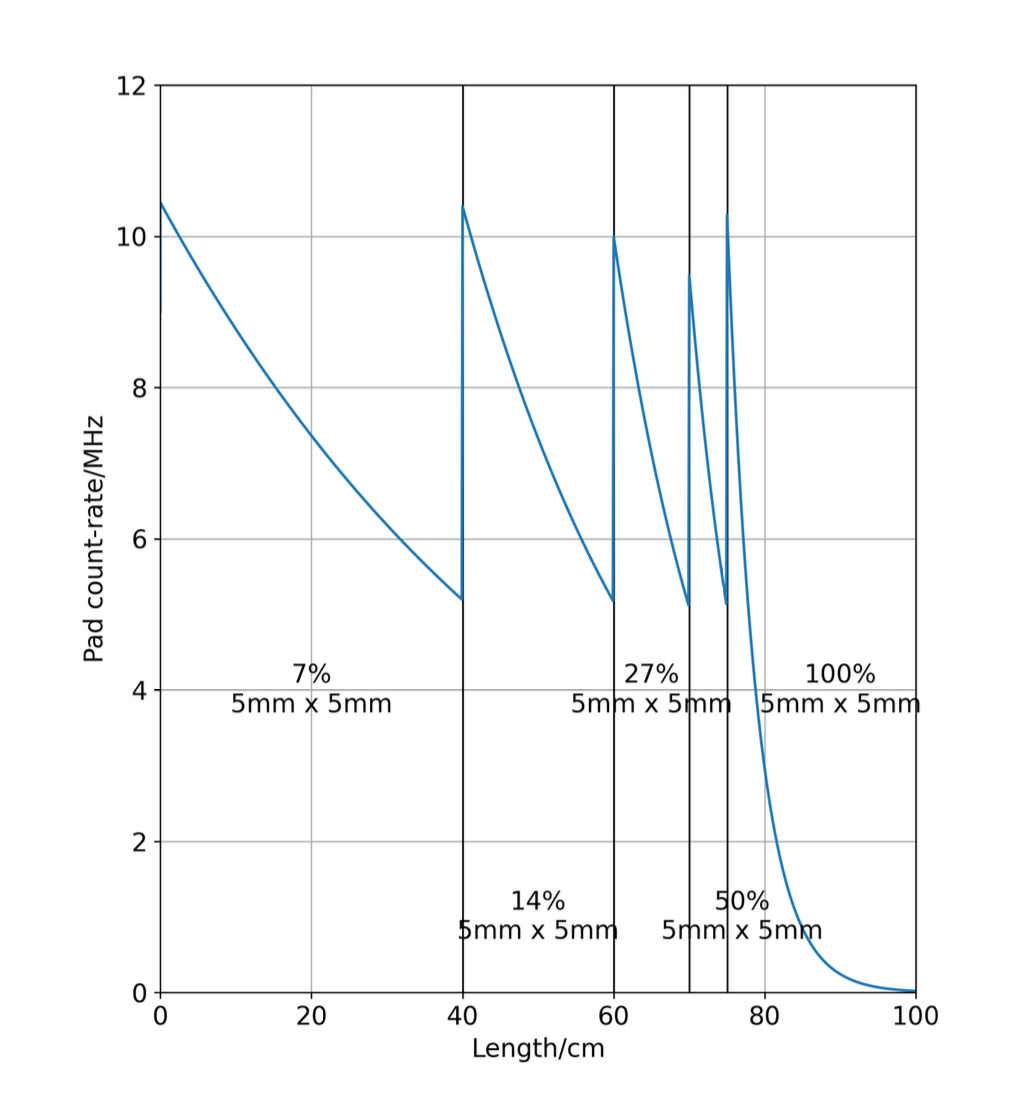}
  \caption{Detector 1m-long with 5 mm $\times$ 5mm pads along the whole detector.}
  \label{fig:sub1}
\end{subfigure}
\vspace{0.2cm}
\begin{subfigure}{.7\textwidth}
  \centering
  \includegraphics[width=0.9\textwidth]{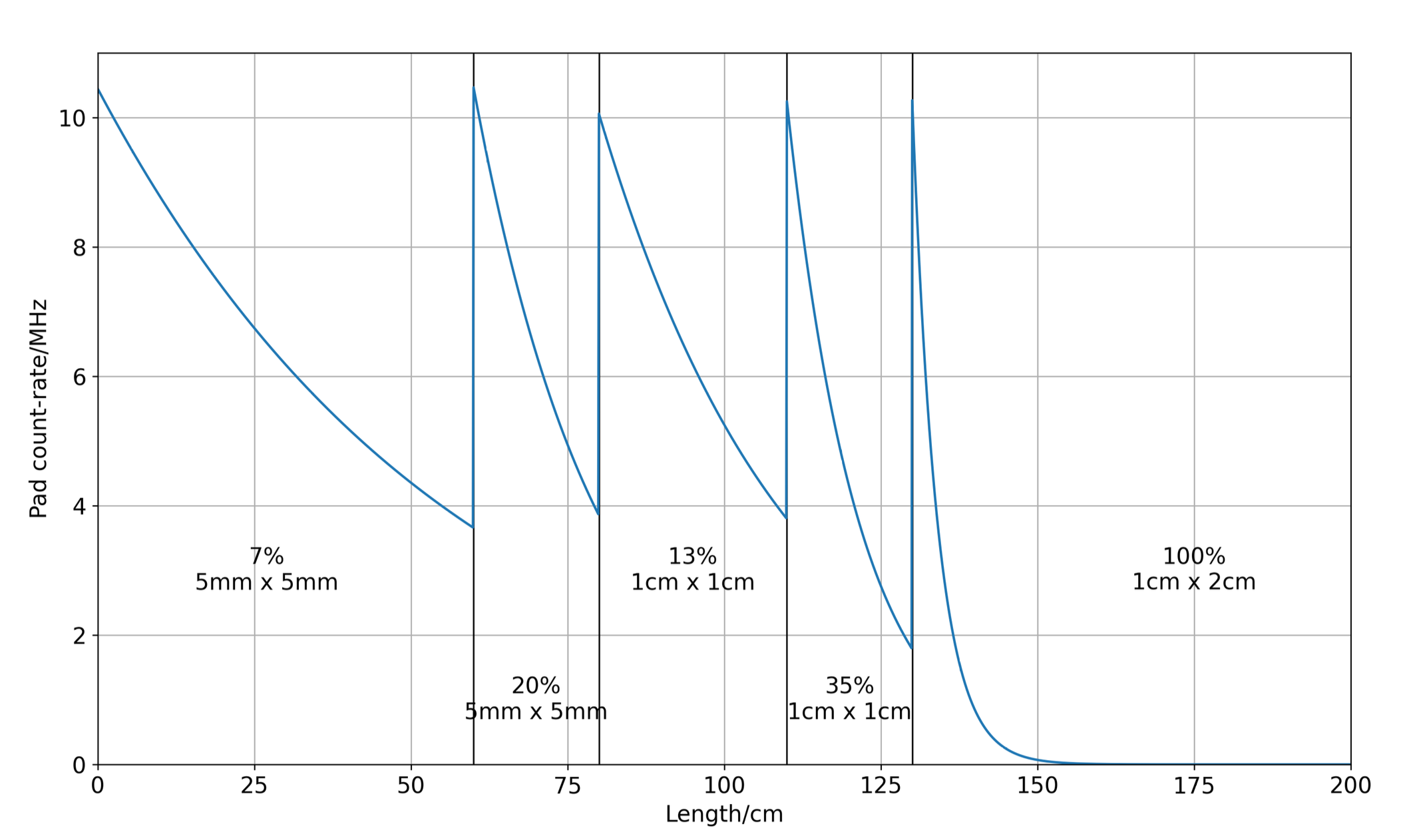}
  \caption{Detector 2m-long with pads of variable size.}
  \label{fig:sub2}
\end{subfigure}
\caption{Possible designs of the readout pads. The plots indicate \ce{^3He}  concentration and pad size. The gas pressure is always assumed to be 1 bar.
}
\label{fig:neutrondetectorab}
\end{figure}

The electrode design is yet to be finalized. In particular, the standard approach used in cylindrical gas detectors—a single central wire as the cathode—is not feasible. The high local field required to maintain a sufficient drift field across the entire detector volume would lead to avalanche multiplication at the wire. Alternative configurations are under consideration, including multi-wire designs and a large off-axis cathode.

Apart from the necessity of ${}^{3}$He, the optimal gas mixture remains to be determined. It must ensure efficient ion transport to minimize space-charge effects while also providing good electron transport for a fast response. Several candidates are being evaluated, ranging from H$_{2}$, which offers excellent ion mobility, to CF$_{4}$, which provides favorable electronic properties. Compatibility constraints must also be considered, particularly to prevent the formation of chemically aggressive species such as HF and radicals.

Due to the high counting rate of the HIBEAM beamline, maintaining a uniform \ce{^3He} concentration throughout the detector would pose significant challenges. It would either result in prohibitively high count rates at the first readout pads—given any realistic pad size—or require an unreasonably long detector. To address this, longitudinal segmentation is necessary.

Assuming a maximum pad count rate of 10 MHz, slightly above the current state-of-the-art GEM electronics, feasible detector designs can be realized. Figure~\ref{fig:neutrondetectorab} presents two such designs: one featuring a 1 m-long detector with small pads distributed along its entire length, and another employing a 2 m-long configuration with decreasing pad size along the detector. Both designs utilize approximately 10$^{5}$ readout channels and a substantial amount of \ce{^3He}.

\subsubsection{Future Work and Open Questions}

The primary challenge of the design proposed in the previous section is its cost. While technically feasible, the large number of readout channels and the substantial amount of ${}^{3}$He required would result in an expense of several million euros.   If this cost proves prohibitive, an alternative approach could involve an initial current-mode section with a short length and high ${}^{3}$He concentration. This would rapidly attenuate the beam intensity, reducing the overall detector length and enabling the use of larger pads in subsequent sections. This compromise would mitigate costs while still preserving the advantages of single-neutron counting.

There are still uncertainties regarding the impact of scattering from the detector itself. Qualitative analysis does not yield a definitive conclusion, making it necessary to investigate this effect through simulations. If scattering proves to be a significant issue, the detector's geometry may be adjusted by adding lateral protrusions to mitigate its influence. Future research will focus on selecting an optimal gas mixture and refining the electrode design. Since both factors are crucial for charge transport and diffusion, computationally intensive Monte Carlo simulations will be conducted to ensure optimal performance.

\subsection{The annihilation detector}
\label{sec:annihilationdetector}

\begin{figure}[h]
  \begin{center}
    \includegraphics[width=0.85\textwidth]{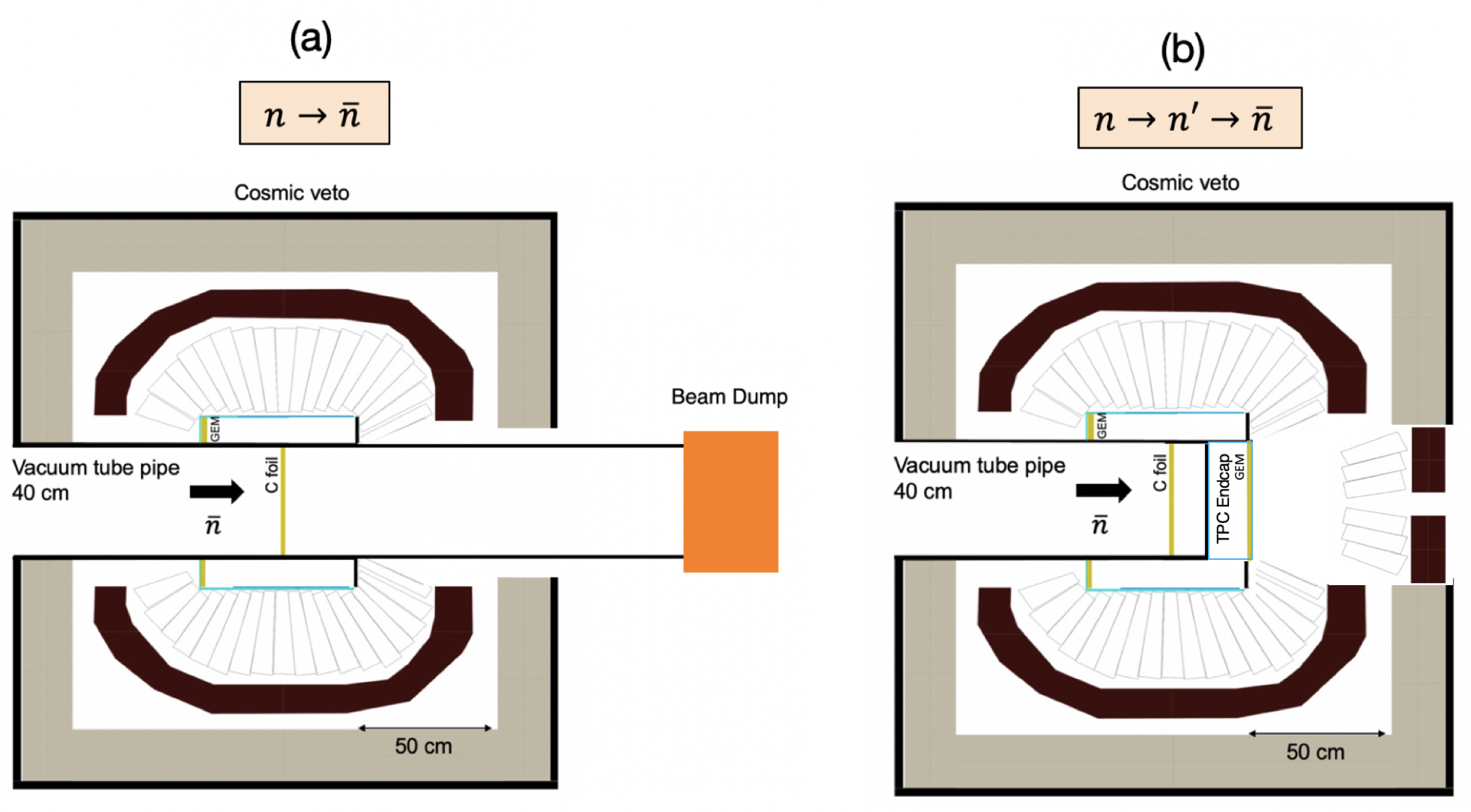}
  \end{center}
  \caption{The annihilation detector for free neutrons converting to antineutrons (left) and for the conversion process via sterile neutrons (right). The figure shows the annihilation foil, the time projection chamber, the WASA crystal calorimeter, the absorber between the calorimeter and the cosmic veto and the cosmic veto itself. }
  \label{fig4}
\end{figure}

The signature of the neutron-antineutron transition is via the annihilation of an antineutron on a carbon target \footnote{Carbon is chosen as the preferred material due to its low neutron absorption cross-section, which minimizes neutron capture and the subsequent production of gamma radiation. Furthermore, the annihilation cross-section for antineutrons in carbon is extremely large (kilobarns) compared to typical neutron capture cross-sections (millibarns), meaning that even a very thin carbon foil provides a high probability for antineutron annihilation, making it an efficient and practical choice for such experiments. Additionally, carbon is structurally stable, radiation-hard, and widely used in neutron beam experiments.} surrounded by an annihilation detector (see Figure~\ref{fig4}). The antineutron-nucleon annihilation signature is the classic pionic  star~\cite{Bressani:2003pv} i.e. around five pions isotropically produced.    

The details of the final state can be studied with a model of $\bar{n} + C$ annihilation~\cite{ann1,ann2,ann3}. The model considers $\sim100$ independent annihilation branching channels. Decays of heavy resonances are also included. The simulations were validated against available $\bar{p} + p$ and $\bar{p} + {}^{12}C$ datasets.  The simulation includes intranuclear cascade and particle transport through the nuclear medium. 

\begin{figure}[ht]
  \begin{center}
    \includegraphics[width=0.8\textwidth]{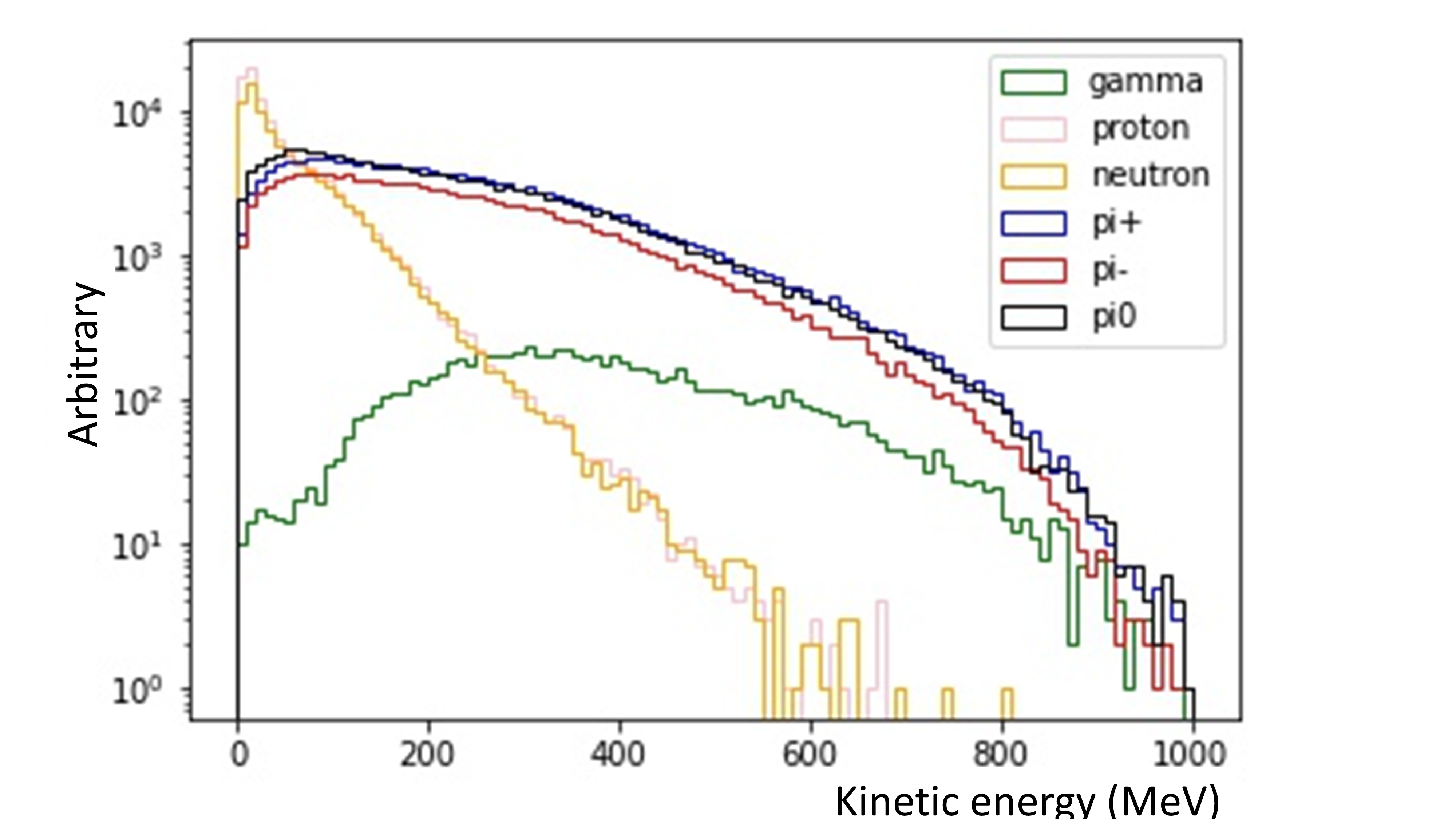}
  \end{center}
  \caption{Simulated kinetic energy spectra for different particle types in $\bar{n} + C$ annihilations. }
  \label{fig5}
\end{figure}

Figure~\ref{fig5} shows the kinetic energy distributions of particles produced in antineutron-nucleon annihilation in carbon. The invariant mass extends up to around twice the neutron mass but is, in practice, lower owing to nuclear scattering effects~\cite{ann1,ann2,ann3}.

Given the more advanced techniques, described in this section, to be used for HIBEAM compared to the ILL experiment of thirty years ago, an efficiency that at least matches the earlier result (i.e.,  around 50\% with full background suppression) can be expected. Furthermore, as shown for the NNBAR CDR~\cite{Santoro:2024lvc}, the use of particle identification, higher precision 3D tracking and a dedicated electromagnetic calorimeter, together with more advanced analysis techniques, such as the use of ML, can lead to higher efficiencies.

\subsubsection{Annihilation detector using the WASA calorimeter}\label{det:wasa}

In order to detect such a signal a dedicated annihilation detector comprising tracking and calorimetry is needed.
A schematic of the annihilation detector that will make use of the 
WASA experiments \cite{wasa} calorimeter can be seen in Figure~\ref{fig4}. Incoming antineutrons traveling from the left in the vacuum tube annihilate in a thin carbon foil. Energies of annihilation products are measured in the CsI(Na) calorimeter previously used in the WASA experiments \cite{wasa} at CELSIUS in Uppsala, at COSY in Julich and presently at FAIR, Darmstadt. The WASA calorimeter is very compact, can be transported and will be available for the HIBEAM project after 2027. Charged particle track directions for event topology are measured in 3D by a cylindrical TPC. Particle identification is done with a combination of TPC and calorimeter. 

Since the only energetic background capable of mimicking an annihilation signal is cosmic rays, an active shield of scintillator bars will surround the experimental setup. Events coincident with a charged particle signal in the cosmic veto will be rejected. To prevent self-inhibition, an absorber thick enough to stop all charged pions from the annihilation is required between the calorimeter and the scintillators.

The annihilation detector will collect as much evidence as possible that a  $n-\bar{n}$ annihilation has occurred.  The task is to detect and identify the final state particles, verify the energy and momentum balance to be compatible with $n-\bar{n}$ annihilating at rest and to verify the event topology with a common vertex in the annihilation foil for all particles. Annihilation in a nucleus opens additional ways  to dissipate energy and momentum by re-scattering in the nucleus.
 Figure~\ref{fig5} shows the simulated kinetic energy distributions of particles emitted when annihilation takes place in a carbon nucleus. Consequently the requirements on energy and momentum conservation cannot be as strict in the $\bar{n} + C$ annihilation as in the free $n-\bar{n}$ case, as long as not all particles can be detected (neutrons and nuclear remnant are missing) but it will anyway be in an energy regime where no natural sources can contribute other than those of cosmic origin.
Around 91.5\%  of the annihilations give at least one  $\pi^{0}$. A good electromagnetic calorimeter which can trigger on  $\pi^{0}$  is thus essential. Also at least 2 charged pions are present in 98.5\% of the cases and in addition protons may be emitted from the nucleus. So charged particle tracking can reconstruct the event topology with a common vertex in the foil. Both tracking and calorimetry need to have as large geometrical coverage as possible. 

Figure~\ref{fig4} shows the detector for the mode for the~\ntoantin~and  ~\nbarviasterile~searches. The brown frame is a solid iron construction in which the CsI(Na) modules are mechanically fixed by their light guides to Photo Multiplier (PM) tubes on the outside of the iron. 
The two search modes~\ntoantin~and~\nbarviasterile~are quite different. 
The traditional search for~\ntoantin~transformation based on degenerate states in the absence of external electromagnetic and strong (nuclear) fields will have a very low probability and must be searched for by observing a maximal number neutrons over as long flight time as possible. Instead, for the~\nbarviasterile~mode by scanning systematically over weak $B$-fields a resonant behavior will enhance the production of $n'$ which in turn transforms to $\bar{n}$. If the process exists, it should be governed by drastically higher probabilities at the proper resonance setting for the magnetic fields. One expects rates of $\bar{n}$ annihilation events per hour compared to events one per month or even year in the traditional case.



An important difference between the two cases is that in the standard~\ntoantin~search the neutron beam will pass through the annihilation target and therefore neutron-induced nuclear reactions in the annihilation foil, a source of beam-induced pile-up background,  will be present. 
For the~\nbarviasterile~case, this source of background will be absent.  It will also be possible to detect annihilation products in the very forward direction. This gives larger solid angle coverage so the annihilation can be more completely characterized.
\subsubsection{The beam tube}
The beam tube will be made as thin as possible since any charged particles stopping in it will remain undetected. With 40~cm diameter the cylinder wall is estimated to need to be 5~mm aluminum to withstand atmospheric pressure. The detection threshold for protons will then be a few tens of MeV. As seen in the kinetic energy spectra (Figure~\ref{fig5}) the threshold will only cut off the lowest 2-3 bins. For charged pions the loss is even smaller and essentially marginal, while for protons the effect is a little larger but still not a problem.  Another effect of the vacuum tube wall is multiple scattering. A measured direction outside the wall will not point back correctly to the vertex due to this. The impact on the vertex accuracy will be reduced by the thin wall and short pointing distance to the foil. The conclusion is that no tracking needs to be implemented inside the tube. 

\subsubsection{The Time Projection Chamber  }

The cylindrical TPC is compact with an outer diameter of 60~cm and inner 40~cm. The maximal drift length is about 35~cm. The short length means that the absolute voltage on the negative terminal of the drift field (typically 200V/cm) will be 10kV at most, which is rather trivial to handle. One uncertainty for the TPC field
cage and drift is that the track has to be registered close to the edges of the field. It will be an important R\&D
task to optimize the useful track length if necessary by corrections of static distortions at the field edges so that
the whole TPC thickness is usable. The avalanche detector will be a GEM stack with 4 layers. GEM readout is
superior to wire chamber readout for this application since the track image on the pad plane will retain the
small extension of the electron cloud after the short drift. This will give the best position resolution. With
GEMs one will also be able to run the TPC without a gating grid and record the full track history. Also the missing 8.5\% of annihilations without $\pi^{0}$ i.e. charged particles only, would be recorded and matched to particles 
in the calorimeter. Continuous readout results in considerable data volumes from the TPC.  Another possibility with moderate loss of annihilation events is to run in triggered mode using the excellent performance of the calorimeter. The TPC front-end chip, SAMPA (described below), allows for a 10-microsecond trigger latency. Due to the short drift, the electron cloud at the readout plane will be only a few millimeters wide. To ensure charge sharing between neighboring pads without requiring narrow, millimeter-wide pads, zigzag-shaped readout pads can be used, with each pad covering approximately  0.5 cm$^{2}$ on the pad plane. 
The cylindrical part of the TPC with 1600 cm$^2$ readout area will then have 3200 pads/electronic channels. For the endcap TPC the drift length will be about 15 cm. On the area 1200 cm$^2$ it will house 2400 pads/electronic channels. A cylindrical TPC drift vessel with a diameter of 60 cm should be rather straightforward to construct with outer walls of a laminated construction of   Kapton PCB + Honeycomb structures. The inner wall can be integrated with the beam tube itself. From a construction standpoint, the endcap TPC is expected to be mechanically simpler than the cylindrical section, given its smaller size and less complex geometry. Each track will typically occupy 30 pads, 10 along the track and 3 perpendicular to it. A typical annihilation event has typically 5 tracks. A channel can then handle multiple hits separated by a few hundred nanosecond.

The planned method for running the GEM TPC resembles the way the largest TPC in the world is operated: the ALICE TPC at LHC after its upgrade~\cite{alice}. The front-end chip for ALICE TPC (named SAMPA) has 32 channels per chip and contains a preamplifier, shaper, and pipeline memory for each channel. Packaged in Ball Grid Array (BGA) package a chip is 15$\times$ 15mm$^2$. 
With 32 channels per SAMPA, each chip serves an area of 16cm$^2$, so there is plenty of area available to make a quite dense 
construction of the front end board fitting the limited space available. Cooling will of course be a challenge; the arrangement 
with the cylindrical and endcap TPC being read out in opposite directions makes cooling easier.  

If operated in triggered mode a powerful pulse recognition and zero-suppression takes place already in the SAMPA
chip. The SAMPA digitizes continuously and upon receipt of a trigger the result of the previous 192 clock cycles is available for 
readout. At 20Mhz sampling this means a trigger latency of about 10 microseconds is allowed. The data volumes to read out in 
triggered mode with zero suppression will be of the order 1kbit per track. In continuous mode with no zero-suppression each SAMPA will produce 6.4Gbit/s at 20 MHz sampling.
Drawing from the experience with the International Linear Collider TPC (LCTPC) prototype~\cite{Shoji:2018kaa}, which shares key characteristics with HIBEAM’s TPC, it is possible to anticipate aspects of its expected performance. Compared to the HIBEAM experiment, the LCTPC test configuration features a track length of 17 cm, whereas HIBEAM is expected to have a shorter track length of approximately 10 cm. Consequently, the dE/dx resolution is expected to be lower due to the reduced track length. However, in other respects, the dE/dx results from the LCTPC beam tests are expected to be comparable to those of HIBEAM, as both setups exhibit similar charge spreading over pads. The ionization electron density per unit track length is another critical parameter, but since both HIBEAM and LCTPC primarily use argon as the main gas component, the difference in this aspect is minimal. HIBEAM’s pad dimensions remain subject to optimization through prototype studies, with the goal of achieving a dE/dx resolution within the 15-20\% range. Simulations indicate that this level of resolution is sufficient to achieve a clear dE/dx vs.~E separation between pions and protons in the kinetic energy regions of interest.

In terms of spatial resolution, it is worth noting that position resolution is not a very relevant metric for a TPC in HIBEAM. What is critical is the pointing accuracy when fitting and extrapolating a straight line to the multiple points along the track. For HIBEAM's TPC, this accuracy is limited by multiple scattering in the aluminum of the vacuum vessel tube. HIBEAM cannot use a magnetic field \footnote{The beamline must remain magnetically shielded to preserve the quasi-free condition required for neutron oscillation searches. Introducing a magnetic field on the order of a Tesla in the detector area would compromise this shielding, disrupting the necessary field-free environment in the beamline and suppressing the neutron oscillation process.}, which would reduce the diffusion spread of the drifting electron cloud and thus improve position resolution. However, drift lengths are short, which compensates for this to a large extent. The optimization of pad size and shape will be done to achieve the desired pointing resolution, including multiple scattering rather than striving for the best possible TPC-internal resolution. The role of tracking is also to match tracks to other detector systems. They are placed very close, and the pointing resolution outwards will not be a limiting factor.

\subsubsection{The Calorimeter}

The WASA calorimeter consists of approximately 1000 CsI(Na) modules of slightly varying depth~\cite{wasa}. Each module is typically 16 radiation lengths deep and 0.8 nuclear absorption lengths. As a scintillation detector, it measures the energy deposited by charged particles, and due to the high atomic numbers of cesium and iodine, it serves as an efficient, fully absorbing electromagnetic calorimeter for the expected photon energies (see Figure~\ref{fig5}). For charged hadrons, the energy resolution is also very good; however, the likelihood of nuclear reactions occurring within the material increases for particles with long ranges in CsI. If no nuclear interaction occurs, the CsI thickness allows the calorimeter to fully stop protons up to 400 MeV and charged pions up to 190 MeV. Consequently, all expected protons will be fully absorbed, while a significant fraction of charged pions will traverse the calorimeter without stopping.

The challenge of measuring charged hadron energies by stopping them in this energy range is a fundamental limitation of the interaction process. The only alternative would be to measure their momentum by tracking them in a magnetic field. However, a magnetic field cannot be used in this experiment. Instead, event kinematics will be used to constrain the energy of pions that do not undergo hadronic interactions.
The WASA calorimeter geometry was originally designed for detecting particles originating from a single point at the detector center. However, in this experiment, antineutron annihilation vertices will be distributed across the transverse area of the annihilation foil, leading to variations in incident angles that could affect resolution. For charged hadrons, the TPC will provide angular information, allowing for corrections. For photons, the energy measurement is expected to be less affected, as the total detected energy will be the sum of deposits in multiple adjacent modules. For reconstruction of the $\pi^{0}$ mass the opening angle between the two photons will be sufficiently large to ensure their energy deposits remain separable. The WASA calorimeter will be calibrated using cosmic rays before commissioning, as done in previous experiments. Each crystal is equipped with a fiber-based light pulser system, operating at ~5 Hz, to monitor stability and detect potential gain shifts due to HV fluctuations or other effects. Since radioactive sources are impractical once the crystals are mounted, the light pulser system will serve as the primary tool for continuous online calibration and stability monitoring during operation.
\par
To evaluate the performance of WASA for $\pi^0$ reconstruction, a fast detector simulation was developed. The parameterization for energy deposits is based on the detector's intrinsic energy resolution for photon pairs, given by~\cite{wasa}:
\begin{equation} 
\frac{\sigma_E}{E} = \frac{0.05}{\sqrt{E [\text{GeV}]}} .
\end{equation}
The parametrization was implemented in \texttt{Geant4}~\cite{Geant41, Geant42, Geant43}  using its specialized tools for fast simulation. A simplified version of the Simulation and Event Reconstruction (SEC) was used, where the actual calorimeter geometry was modeled as a hollow cylinder with an inner radius of 32.5 cm, an outer radius of 63.5 cm, and a length of 109 cm. When an incoming particle reaches the calorimeter, the deposited energy is determined through a Gaussian smearing of the particle's kinetic energy, with the Gaussian width given by the detector resolution. No smearing was applied to the energy deposited position. The code was validated by simulating decays of $\eta$ mesons into photons and neutral pions, and comparing the simulated response with real data recorded by the WASA calorimeter~\cite{wasa}. Good agreement between the fast simulation and real data was found.
\par
\begin{figure}[h]
  \begin{center}
    \includegraphics[width=0.80\textwidth]{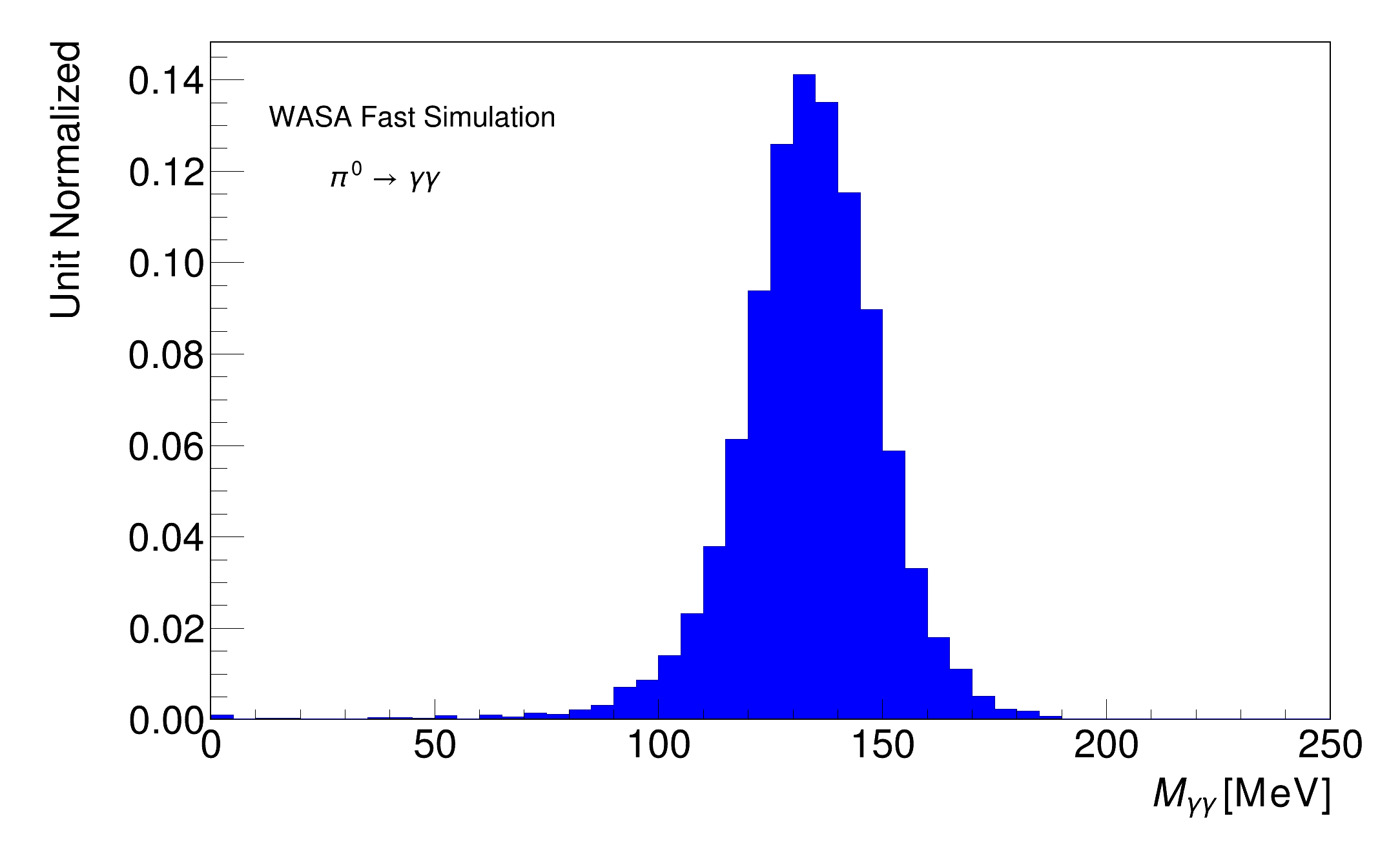}
  \end{center}
  \caption{$\pi^0$ mass reconstruction from a fast simulation of the WASA Scintillator Electromagnetic Calorimeter.}
  \label{fig_fast}
\end{figure}

Figure~\ref{fig_fast} shows the $\pi^{0}$ mass reconstruction from the fast simulation of a pair of photons detected by the calorimeter, originating from the decay of neutral pions with kinetic energies of 200 MeV. The primary $\pi^0$'s were shooting in random directions from a disc of 25 cm radius.

\subsubsection{Cosmic ray veto} 
Cosmic rays are the only background with sufficient energy to mimic an annihilation signal. To suppress these events, the experimental setup will be enclosed by an active scintillator-based veto system (see Figure~\ref{fig4}). Any event coinciding with a charged particle signal in the veto scintillators will be rejected. To avoid self-inhibition, a sufficiently thick absorber will be placed between the calorimeter and the scintillators to ensure that all charged pions from the annihilation are stopped before reaching the veto. Neutral cosmic events, which do not leave a direct signal in the veto, will be identified by the absence of a reconstructed vertex in the annihilation foil. The false annihilation event rate due to cosmic rays and other natural backgrounds will be studied by shutting off the neutron flux at the neutron port and through dedicated Monte Carlo simulations. The scintillator slats will be 3 cm thick, ensuring that minimum ionizing particles deposit approximately 6 MeV, which is well above the typical energy deposit from Compton electrons due to natural gamma radiation. Light readout from the scintillators will be achieved using wavelength-shifting fibers, with silicon photomultipliers (SiPMs) for photon detection.

\subsubsection{Particle Identification } 
The identification of charged and neutral pions is strong evidence that an annihilation event has taken place. Neutral pions are identified by their invariant mass, calculated from the measured energies of the two final gammas and the opening angle between them. Determining the opening angle requires knowledge of the point of origin, which can be estimated through projection of charged tracks in the TPC. Approximately 98.5\% of annihilation events will have at least 2 charged pions allowing a vertex to be defined. Charged hadrons, essentially protons and pions in this case, will be identified using dE/dx measurements in the TPC correlated with the total energy measured in the calorimeter. 
The mass difference between pion and proton is large, so the energy loss resolution does not have to be strict. The expected physics also helps because the proton spectrum is negligible beyond 200~MeV, so the protons are far from minimum ionizing, while the bulk of charged pions will be minimum ionizing. In addition negative pions will be stopped in the material and captured by nuclei and give rise to an additional rather arbitrary energy signal by nuclear fragments. Stopping positive pions will decay to a muon and neutrino. The muon will add 4.2 MeV to the energy signal as the time scale of the decay is 26ns. Subsequent decay of the muon is on a timescale of microseconds. The energy released by the muon decay will remain unmeasured or at least separable from the prompt event. Since the rest masses of pions account for a sizable fraction of energy conservation balance in the annihilation process, it is important to identify the pions and important to know whether the rest mass is part of the visible energy (as it is for the neutral pions) or not.  

\subsubsection{Trigger and data acquisition system} 
Since the current WASA data acquisition system was commissioned in 2006, its architecture is outdated and components now lack continuous support. Therefore we will implement a new WASA Data aquisition system (DAQ) using on a digitization layer containing self-triggering analog-to-digital converters (ADCs) that continuously digitize all interesting signals. The ADCs will use a design developed at University of Uppsala (UU) for the
PANDA experiment at FAIR  \cite{pawel}. Integrating the cosmic veto and TPC with the WASA DAQ and event building will
be accomplished by FPGA-based Data Concentrators, based on corresponding modules designed at UU for the BESIII CGEM readout 
\cite{Amoroso_2021}. Data will be sent over optical links using
standard Gigabit Ethernet physical link protocol.
The WASA calorimeter permits several fast triggering possibilities, based on threshold on measured energy in a module or cluster of adjacent modules. Even the analog sum of all energies in the calorimeter can be used for threshold discrimination. There are consequently many ways to construct a trigger for $\pi^{0}$. Charged hadrons with sufficient energy may also trigger but there is no harm in that if it is not caused by cosmic particles. So these shall be rejected as described above. A $\pi^{0}$ based trigger should find  91.5\% of the annihilation events while 8.5 \% have no $\pi^{0}$.
Matching TPC tracks with their respective hits in the calorimeter  (in case of a trigger) is performed by using the measured trigger pulse timing in the CsI to set the appropriate time zero for the drift time measurement in the TPC. This is of course only applicable for particle trajectories within the geometrical coverage of the detector, which is less than 100\% due to the entrance opening. Seen from the center of the foil the opening has the half opening angle of about 35$^{\circ}$ ($\sim$1.1sr). The solid angle coverage is thus up to  85\%. 

\subsubsection{Signal-like backgrounds}\label{sec:gain}

The expected final state configuration can be produced by neutrons or other fast beam contaminants with momenta of $\mathcal{O}$($100$)~MeV interacting with the carbon foil and beam-related infrastructure. However, the arrival time of such fast neutrons at the foil is synchronized with the linac proton pulse arrival time on the ESS tungsten target. For a 14Hz repetition rate and 2.86ms pulse width, a time window excluding around 5\% of operating time suppresses fast neutrons. Additionally, the ESS linac structure is designed to minimize proton leakage between pulses, reducing the potential for out-of-time fast neutrons to contribute to the background. HIBEAM will further monitor beam-related backgrounds by running in configurations without B-field suppression, providing an empirical method to assess and mitigate any unexpected contributions.


The major background is expected to arise from cosmic rays, as was the case in the ILL experiment~\cite{Baldo-Ceolin:1994hzw}. As demonstrated at ILL, complete suppression of this background is achievable. In addition to employing an active cosmic veto, background rejection can be further improved by applying selection criteria on sensitive kinematic variables, as shown in the NNBAR experiment~\cite{Santoro:2024lvc}. Key observables include final state invariant mass and event sphericity, which help distinguish signal events from background. Unlike the ILL experiment, the WASA detector provides neutral pion tagging and reconstruction, serving as a primary selection tool. Additionally, the TPC in this search offers precision 3D tracking, expected to significantly surpass the vertex resolution achieved at ILL, which relied on limited streamer tubes. While the full signal-to-background optimization for the WASA-based detector is still ongoing, the advanced capabilities of the various HIBEAM detector subsystems strongly suggest superior performance compared to previous neutron oscillation searches.

The primary factors of interest concerning cosmic rays are the expected particle fluxes and, ultimately, the detector's ability to reject these events. The expected number of particles per $m^2$ has been estimated in the studies for the NNBAR case (Table 49 of Ref.~\cite{Santoro:2024lvc}), and these values are the same for HIBEAM. However, the total area of the WASA detector is approximately 30 times less than that of the NNBAR detector and 15 times smaller than that of ILL's. Therefore, the final particle fluxes from cosmics, when adjusted for area, are correspondingly smaller by the same factor. This is significant because if one assumes the ILL experiment achieved a zero-background selection, and assuming three times the running time, HIBEAM could tolerate a background rejection efficiency 5 times worse than ILL and still reach the same level of background rejection. There is no reason to believe that WASA would be any less effective than ILL in rejecting cosmic ray background, particularly given WASA's ability to reconstruct neutral pions with very high precision, a capability the ILL detector lacked.

Regarding cosmic ray event readout rates, taking into account estimated fluxes, the effectiveness of the passive shielding in stopping neutral particles, and the active cosmic veto in suppressing charged particles, the expected cosmic ray readout rate should be on the order of a few kHz. The signal trigger rate would be a few Hz, primarily caused by cosmic muons traversing the detector without triggering the veto, as was observed in the ILL experiment and is expected for NNBAR as well.

\subsubsection{Low energy spallation background}
An isotropically produced flux of MeV-scale photons per second is expected from thermal and cold neutron capture on the foil. The photons may also lead to Compton electrons entering the TPC. In signal candidates, tracks will be matched to a calorimeter signal which has a timing resolution of $\mathcal{O}$($10$)~ns. 

As was done in the ILL experiment, capture in the beampipe and other beam-related infrastructure can be suppressed with $^6$Li neutron poison from which low-energy alpha particles are emitted which are stopped in the beampipe. 

\begin{figure}[bt!] 
\centering
\includegraphics[width=0.8\textwidth]{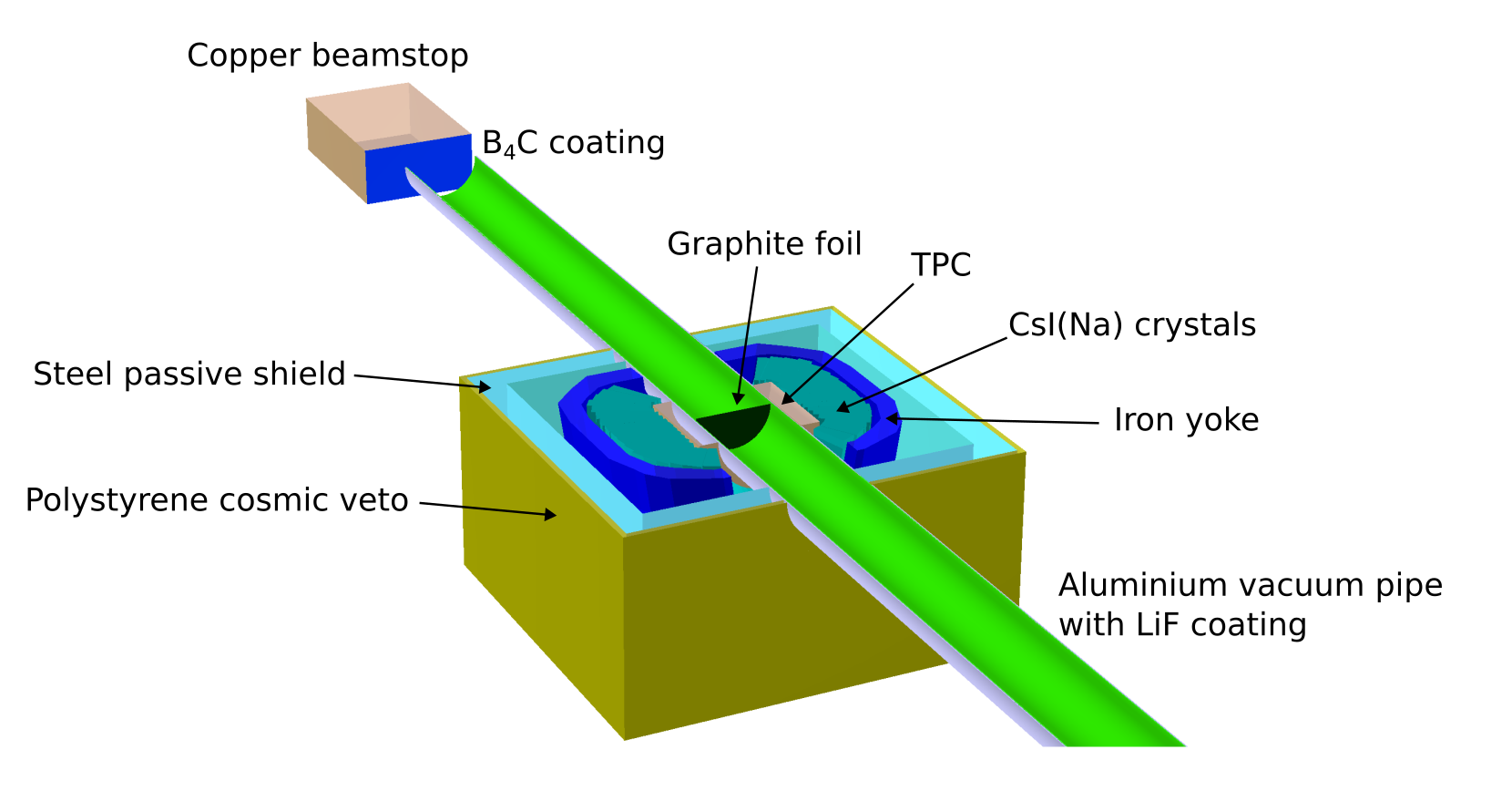}
\caption{The \software{Geant4} detector model used to estimate the spallation background.}
\label{G4bar} 
\end{figure}

The beamline simulations (see Section~\ref{sec:beamline}) generated a \software{MCPL}  file that recorded the particles entering the experimental cave, approximately 61~m from the moderator. This \software{MCPL} file was used as input for a detailed \texttt{Geant4} detector simulation. The \texttt{Geant4} model, converted from \texttt{ROOT}~\cite{BRUN199781} geometry using the \texttt{VGM} conversion tool~\cite{VGM}, is illustrated in Figure \ref{G4bar}. The physics list used is \texttt{G4HadronPhysicsINCLXX} including the Neutron-HP package. As for the beamline simulations using \software{PHITS}, this uses INCL at high energies ($> \SI{20}{MeV}$) while nuclear data libraries (G4NDL4.7) are applied below 20 MeV. To treat electromagnetic interactions, \texttt{G4EmStandardPhysics\_option4} list is used, with \texttt{G4EmExtraPhysics} for photonuclear reactions. 

\begin{figure}[bt!] 
\centering
\includegraphics[width=0.8\textwidth]{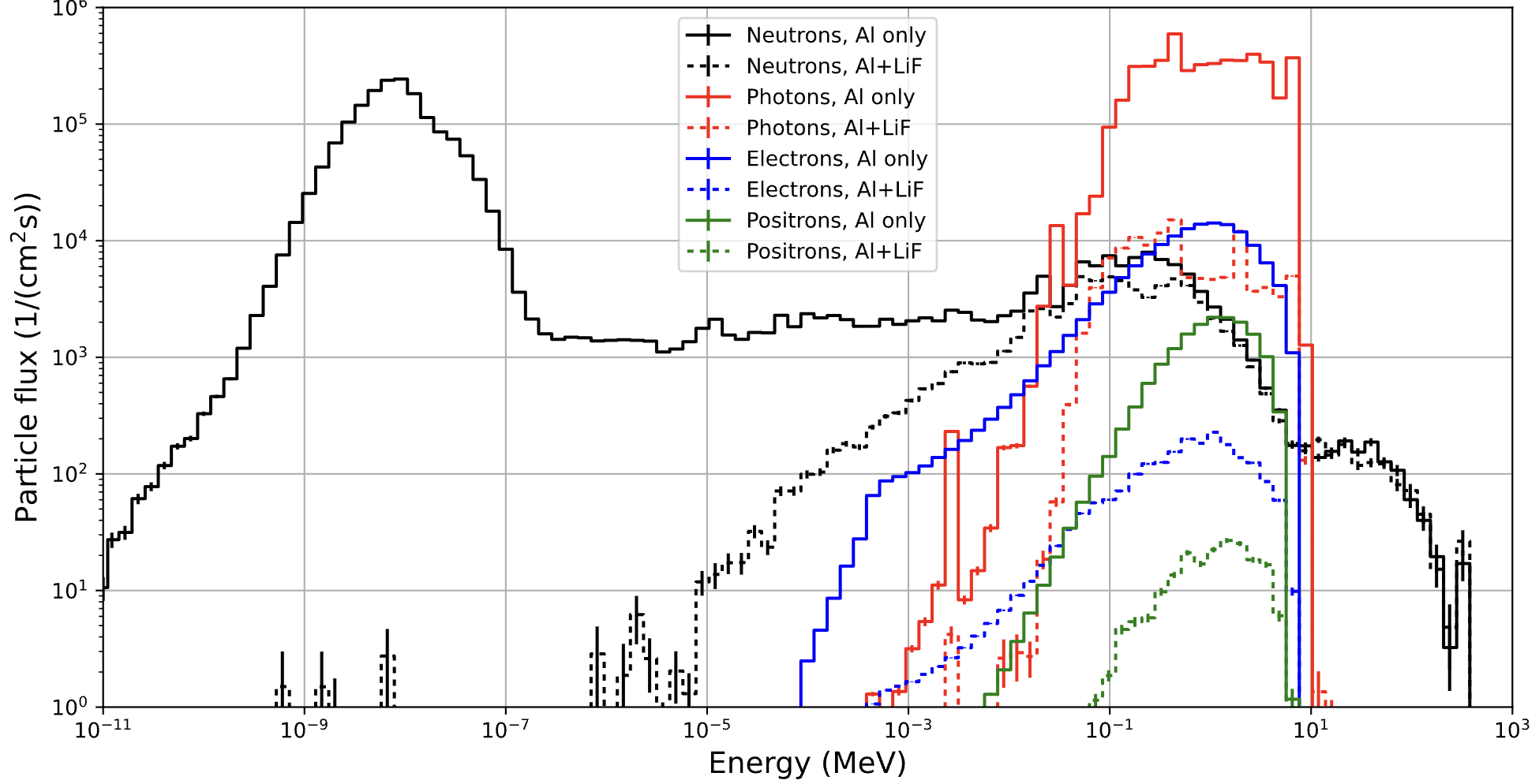}
\caption{Particle intensity (particle/cm$^2$s$^{-1}$) in the TPC as a function of particle energy, as predicted by \texttt{Geant4}.}
\label{backgrounds} 
\end{figure}

Using the \texttt{Geant4} model, the properties of the spallation background can be determined. As an example, the solid lines in Figure \ref{backgrounds} show the energy distribution of the neutron, photon, electron and positron volumetric flux in the TPC. The fluxes are calculated using the particle track lengths. Since an excessive flux of photons and electrons in the TPC complicates event reconstruction, a configuration where 5 mm of $^6$LiF cladding is added on the inside of the vacuum vessel is also considered in Figure \ref{backgrounds} as indicated by dashed lines. The $^6$LiF~ essentially eliminates the cold neutron flux and reduces the electron and positron flux by around two orders of magnitude. 

The dominant background in the TPC is due to electrons from the Compton scattering of photons produced by capture in the foil and elsewhere in the beampipe and other material. The rate of electrons entering the TPC is estimated to be $\sim 5 \times 10^6$ s$^{-1}$.  Energy depositions in associated WASA cells can be used to distinguish between Compton electron tracks and those from charged pions produced in an annihilation event. Furthermore, the electron tracks would not typically appear to come from a single vertex at the foil. The rate of photons entering the 1000-cell WASA calorimeter is $\sim 3 \times 10^9$ s$^{-1}$. The implies a possible mild contribution of several MeV-scale photons for, e.g., a 2$\micro$s signal integration time. Signal photons from neutral pion decays have energies that are typically at the 100-MeV scale.     

\subsubsection{Detector Prototypes}

Small-scale prototypes of the detector systems listed above have been built and are currently being tested. The prototype TPC is a cylindrical detector with a projected track length of 10~cm and a maximum drift length of 23~cm. These dimensions are close to the ones of the planned full detector, resulting in similarly short drift times and small diffusion effects. For charge collection, the detector is equipped with a $10\times10$~cm$^2$ three-layer GEM stack readout with 256 zigzag shaped pads. It is operational and has been tested with cosmic rays. As a future upgrade, a field cage with a $10\times10$~cm$^2$ cross section will be installed, allowing to investigate the effect of distortions at the field edges.

Prototype scintillator staves have been obtained from the FNAL-NICADD Extrusion Line Facility \cite{fermilab}. Originally intended for the hadronic range detector of the NNBAR experiment \cite{Santoro:2024lvc,Dunne:2021}, the same technology is now being considered for the HIBEAM cosmic ray veto and developments are running in parallel. The staves have dimensions of $50\times5\times2$~cm$^3$. They consist of polystyrene and have a reflective TiO$_2$ coating. Each stave is read out using two Kuraray wavelength-shifting fibers equipped with Hamamatsu S14160 SiPMs at each end. Dedicated frontend boards for the SiPMs have been produced and an FPGA-based digitization and readout system is currently under development. At the same time, detector tests with cosmic rays and radioactive sources are ongoing. First in-beam tests of these prototypes are currently scheduled for fall 2025.

\section{Summary and future plans}\label{sec:summary}
The ESS offers a unique opportunity to address fundamental open questions in particle physics, including the origin of the matter-antimatter asymmetry and the nature of dark matter. The HIBEAM collaboration has developed a dedicated beamline and search program to fully exploit the scientific potential of ESS.

The HIBEAM program includes searches for neutron-to-antineutron and neutron-to-sterile neutron conversions, as well as searches for ALPs and a nonzero electric charge of the neutron.  Improvements in discovery sensitivity of at least an order of magnitude compared with current or previous work can be achieved.  

Future plans include the tests and characterisation of magnetic control and detector prototypes and the further development of the HIBEAM program. Further activities at HIBEAM can include measurements of the neutron EDM and neutron decay. The potential of HIBEAM, including required infrastructure modifications and optimized configurations, will continue to be explored. A key objective of HIBEAM is to establish a flexible beamline concept, ensuring that the instrument remains scientifically valuable for decades.




\section{Acknowledgements }

D.M., V.S., B.M., T.N., S.S., and M.W. gratefully acknowledge support from the Council for Swedish Research Infrastructure at the Swedish Research Council for the grant “HIBEAM pre-studies.” V.S. gratefully acknowledges support from the Swedish Research Council for the grants: “Development of an innovative neutron detector for the European Spallation Source (ESS),” and “The First Particle Physics Experiment at the ESS: Search for Axion-like Particles at the HIBEAM Beamline.” V.~S. also acknowledges support from Stiftelsen för Strategisk Forskning for the grant "Development of a magnetic control beamline for fundamental physics and condensed matter science at the European Spallation Source." Additionally, V.~S. acknowledges support from the Crafoord Foundation. This work was also supported by the Italian Ministry of Foreign Affairs and International Cooperation, grant number PGR SE24GR04. The work of Y.~V.~S. was supported by the Australian Research Council under the Discovery Early Career Researcher Award No.~DE210101593. B.~M., A.~N., J.~A., and T.~Q. gratefully acknowledge support from STINT, The Swedish Foundation for International Cooperation in Research and Higher Education. L.~J.~B. acknowledges support from the U.S. Department of Energy (DOE), Office of Science, Office of Nuclear Physics under contract DE-AC05-00OR22725. A.~N. gratefully acknowledges support from the National Council for Scientific and Technological Development (CNPq) [403291/2023-2] and Fundação Carlos Chagas Filho de Amparo à Pesquisa do Estado do Rio de Janeiro (FAPERJ) [210.355/2024]. D.~M. and A.~B. gratefully acknowledge support from Olle Engkvists Stiftelse. Detector simulation computations were enabled by resources provided by LUNARC, The Centre for Scientific and Technical Computing at Lund University.

\pagebreak

\typeout{}

\end{document}